\documentclass[%
 reprint,
 amsmath,amssymb,
 aps,
]{revtex4-2}
\usepackage[version=4]{mhchem}
\usepackage{ragged2e} 
\usepackage{hyperref}

\usepackage[font=small,justification=Justified,format=plain]{caption}
\usepackage{graphicx}
\usepackage{dcolumn}
\usepackage{bm}
\usepackage{color}
\usepackage{xr}
\usepackage{comment}
\renewcommand{\vec}[1]{\ensuremath{\boldsymbol{#1}}}

\newcommand{\eff}{\ensuremath{\mbox{\tiny eff}}}
\newcommand{\friction}{\ensuremath{\eta}}
\newcommand{\tens}[1]{\ensuremath{\boldsymbol{\mathsf{#1}}}}
\usepackage{subcaption}
\newcommand{\CR}{\ensuremath{\mbox{\tiny CR}}}
\newcommand{\CH}{\ensuremath{\mbox{\tiny CH}}}

\usepackage{booktabs}
\usepackage{cleveref}
\usepackage{soul}
\usepackage[mathlines]{lineno}

\usepackage{xcolor}

\newcommand{\RB}[1]{\textcolor{blue}{#1}}

\begin{document}

\title{The diffusive dynamics and electrochemical regulation of weak polyelectrolytes across liquid interfaces}

\author{Giulia L Celora}
\email[]{giulia.celora@maths.ox.ac.uk}

\affiliation{University College London, Department of Mathematics, 25 Gordon Street, London, WC1H 0AY, UK}
\affiliation{Mathematical Institute, University of Oxford, Andrew Wiles Building, Woodstock Road, Oxford OX2 6GG, UK}

\author{Ralf Blossey}
\email[]{ralf.blossey@univ-lille.fr}

\affiliation{University of Lille, Unité de Glycobiologie Structurale et Fonctionnelle (UGSF), CNRS UMR8576, F-59000 Lille, France}

\author{Andreas M\"unch}
\email[]{muench@maths.ox.ac.uk}

\affiliation{Mathematical Institute, University of Oxford, Andrew Wiles Building, Woodstock Road, Oxford OX2 6GG, UK}

\author{Barbara Wagner}
\email[]{barbara.wagner@wias-berlin.de}

\affiliation{Weierstrass Institute, Mohrenstr. 39, 10117 Berlin, Germany}

\date{\today}

\begin{abstract}
We propose a framework to study the spatio-temporal evolution of liquid-liquid phase separation of weak polyelectrolytes in ionic solutions. Unlike strong polyelectrolytes, which carry a fixed charge, the charge state of weak polyelectrolytes is modulated by the electrochemical environment through protonation and deprotonation processes. Leveraging numerical simulations and analysis, our work reveals how solution acidity (pH) influences the formation, interactions, and structural properties of phase-separated coacervates. We find that pH gradients can be maintained across coacervate interfaces resulting in a clear distinction in the electro-chemical properties within and outside the coacervate. By regulating the charge state of weak polyelectrolytes, pH gradients interact and modulate the electric double layer forming at coacervate interfaces eventually determining how they interact. Further linear and nonlinear analyses of stationary localised solutions reveal a rich spectrum of behaviours that significantly distinguish weak from strong polyelectrolytes. Overall, our results demonstrate the importance of charge regulation on phase-separating solutions of charge-bearing molecules and the possibility of harnessing charge-regulated mechanisms to control coacervates and shape their stability and spatial organisation.

\end{abstract}

\maketitle


\section{Introduction}

In recent years, the understanding of liquid-liquid phase separation (LLPS) has gained enormous interest because of its putative role in the assembly of macromolecules (mostly proteins and nucleic acids) into membrane-less organelles (also known as biomolecular condensates) in cells~\cite{hyman_liquid-liquid_2014,Riback2020,Villegas2022}. This phenomenon has also been suggested as a primordial mechanism for compartmentalization in prebiotic cells, making it highly relevant to understanding the origin of life~\cite{oparin_origin_1965,bartolucci_sequence_2023,fox_evolutionary_1976}. However, there are still open challenges in understanding how different molecular mechanisms affect the formation, regulation and properties of biomolecular condensates in cells due to the complex and dynamic nature of the cellular environment and its constituents~\cite{Villegas2022}. 

Grounded in the seminal work by Flory and Huggins and experimental evidence, the balance between enthalpic and entropic interactions is considered to be the driving force of LLPS. While this is often the case, theoretical and computational studies have highlighted the important role that electrostatic interactions, mediated by the presence of ionizable groups, play in the process. By competing with entropic effects, electrostatic forces can slow down, and even arrest coarsening, potentially resulting in stable microphase separation~\cite{CeloraPL1,grzetic_electrostatic_2021,gavish_spatially_2017,hennessy_breakdown_2024}. 

In physical chemistry, the process by which liquid droplets composed of charged polymers and counterions form is known as coacervation. Detailed reviews and discussions of recent advances in the experimental and theoretical studies of coacervation can be found in~\cite{muthukumar_50th_2017,sing_recent_2020}. Since the first coacervation theory proposed by Voorn and Overbeck (VO) in 1957~\cite{VO_theory}, significant efforts have been invested in extending this framework to overcome its limitations. These include field-theoretic~\cite{grzetic_electrostatic_2021,gavish_spatially_2017,hennessy_breakdown_2024,rumyantsev_microphase_2018,CeloraPL1,majee_charge_2023} and liquid-state theories, scaling arguments, and coarse-grained and (more recently) atomistic simulations that have also been vital to validate theoretical predictions. Collectively, these approaches have helped elucidate the connection between the properties of coacervates, the properties of their components and/or environmental conditions, which aided the design of smart materials-- such as nano/bioreactors, drug delivery, sensors~\cite{calvert_hydrogels_2009,li_smart_2009}, and understanding of cellular physiology -- from the electrical excitability of cells~\cite{kolan_propagation_2025,wnek_bio-mimicking_2022,tasaki_spread_2002} to the adaptive compartmentalization of the intracellular space~\cite{feng_formation_2019,berry_physical_2018}. For instance, applications of random phase approximation have been used to study the impact of charges-connectivity and salt ions on the coacervation of polyelectrolyte systems~\cite{grzetic_electrostatic_2021,rumyantsev_microphase_2018}, findings that helped to explain experimental evidence about the role of intrinsically disordered domain sequences on protein LLPS ~\cite{nott2015phase,an_effects_2024,Lin2016,Meca2023}. 

Despite the significant advances in understanding coacervates and the role of electrostatics in liquid-liquid phase separation (LLPS), most of our understanding originates from the study of strong polyelectrolytes (\emph{i.e.}, polymers with approximately fixed charge). However, the majority of biomacromolecules are weak polyelectrolytes, whose charge is dynamically regulated by the local pH via protonation/deprotonation of ionizable groups. Recently, experiments have shown that coacervates of biomolecules can sustain stable pH gradients across coacervate phases, as it minimises the overall electrostatic repulsion~\cite{ausserwoger_biomolecular_2024} by modulating ionisation asymmetry across coacervate phases. These findings further support the putative role of biomolecular condensates in shaping pH gradients intra-cellularly and highlight the importance of charge regulation mechanisms in a theory of LLPS.
There have been a few attempts at characterising the phase behaviour of weak polyelectrolytes and their pH-dependent coacervation, including experimental~\cite{chollakup2010,stano2024}, computational~\cite{meneses2019,stano2024} and theoretical~\cite{Adame-Arana2020,majee2020,knoerdel2021,celora2023,zheng2021,avni2019,Avni2020} studies. These have revealed the complexity and interesting behaviour of weak polyelectrolyte coarcevates, which include asymmetric phase diagrams under pH-dependent conditions~\cite{knoerdel2021}, phase separation driven by charge symmetry breaking~\cite{majee2020} and reentrant phase behaviour arising from pH-controlled protonation equilibrium shifts \cite{celora2023}. Despite this progress, unexplored questions remain on how pH and electrostatics regulate the dynamics of liquid-liquid phase separation of weak, rather than strong, polyelectrolytes. This is the focus of this work.

In particular, the (diffusive) dynamics of charge-regulated macroions have only recently been addressed \cite{zheng2024}, following earlier work by the same authors on the equilibrium properties of
these systems \cite{avni2019,zheng2021}. In
a similar spirit, 
in the present paper, we extend our previous work on the equilibrium properties of weak polyelectrolytes~\cite{celora2023} to capture the dynamics and spatial arrangements of phase-separated domains near equilibrium. In our work, special
emphasis is given to the role of electrostatic interactions, captured within a mean-field description, and pH gradients in the formation and properties of liquid interfaces emerging via phase separation of weak polyelectrolyte solutions. In~\Cref{sec:model}, we start by presenting the underlying statistical physics description of the charge regulation polymer solution, where we explicitly consider hydronium ions and hence have access to pH. 
In~\Cref{sec:results}, the dynamics of phase separation are studied both numerically and analytically, providing us with a quantitative handle on the emergence of patterns via different types of instabilities. As for strong polyelectrolytes, we find that electrostatic forces can significantly affect phase separation; in particular, we find regimes in which micro-phase separation and arrests of Ostwald ripening can occur. Interestingly, we also identify non-linear effects that, to the best of our knowledge, are unique to weak polyelectrolytes and arise from the nucleation of nano-structured bulk solutions. In~\Cref{sec:conclusion}, we conclude by summarising our main findings and future extension of this work; in particular, we discuss how our results highlight differences in the phase behaviour of strong and weak polyelectrolytes and possible ways forward to extend our framework to LLPS in biology.

\begin{figure}[htb]
\begin{center}
\includegraphics[width=0.4\textwidth]{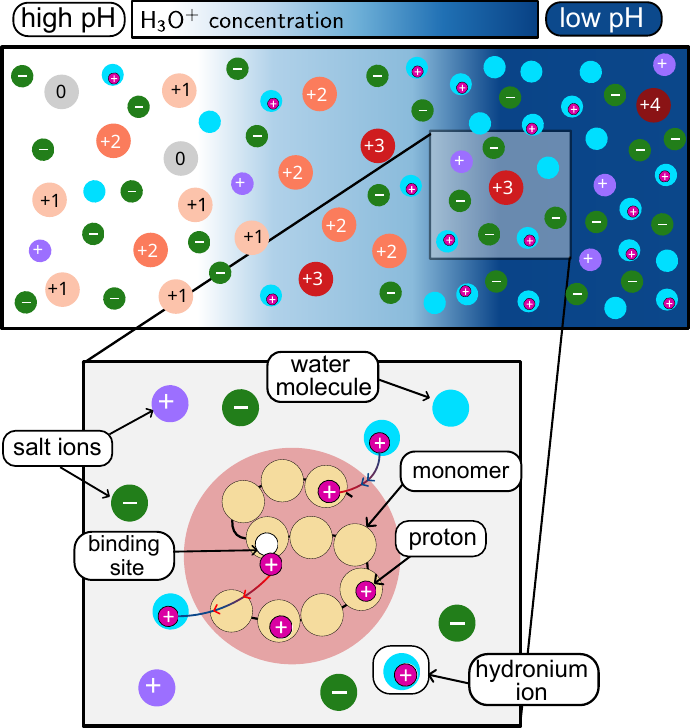}
\end{center}
\captionof{figure}{Schematic of the proposed model of phase-separation in ionic mixture with charge-regulated (CR) polymers. The protonation of binding sites on the polymer chains is regulated by the local pH -- or, equivalently, the local concentration of hydronium ions. Upon phase separation, asymmetric distribution of charges, including hydronium (\ce{H3O+}) ions, across the liquid interface results in the spatial organisation of polymer chains depending on their effective charge, here spanning from 0 to 4.}
\label{fig:schematic}
\end{figure}

\section{\label{sec:model}Theoretical Model}

\subsection{The free energy}

We consider a solution consisting of $4$ different mobile species, which we list here together with the
variable for their corresponding local number density:
water molecules ($c_s(\vec{x},t)$), positive and negative ions from a dissociated monovalent salt
($c_\pm(\vec{x},t)$), and weak polyelectrolytes ($c_p(\vec{x},t)$). We assume that \ce{H+} exists in the mixture but only bound to either water molecules or polymer monomers. Consequently, water molecules can exist in either a neutral (\ce{H2O}) and protonated (\ce{H3O+}) state and we denote by $\pi_s^q(\vec{x},t)$, the fraction of the water molecules in charge state $q=0,1$ at location $\vec{x}$ at time $t$. We further assume that the polyelectrolyte chains consist of $N$ monomers of which $Q_p$ have a group whose charge state is regulated via protonation/deprotonation, that is, absorption/desorption, respectively, of
\ce{H+}, resulting in $Q_p+1$ different charge states of the macromolecule; see~\Cref{fig:schematic} for a schematic illustration. We denote by $\pi^q_{p}(\vec{x},t)$, the fraction of the macromolecules in charge state $q=0,\ldots,Q_p$ at location $\vec{x}$ at time $t$.

We assume that all species in the mixture are incompressible; for simplicity, we assume that solvent, counter-ions and monomers have the same molecular volume $\nu$, independently of their charge state. The molecular volume of \ce{H+} ions has indeed a negligible contribution to the molecular volume of the macromolecules and water molecules. 
Assuming there are no voids in the mixture, the concentration of the different species must locally satisfy
\begin{equation}
    \nu\left(\sum_{i\in\mathcal{I}\setminus\left\{p\right\}}c_i(\vec{x},t)+N c_{p}(\vec{x},t)\right)=1.\label{eq:no_void}
\end{equation}
where the set $\mathcal{I}$ collects the subscript indicating all the species in the solution $\mathcal{I}=\left\{p,\pm,s\right\}$.

We assume that the mixture forms a dielectric medium with permittivity $\epsilon$; while this is, in general, a function of the mixture composition ~\cite{liese2022dielectric,Meca2023}, here for simplicity, we take it to be constant. The presence of charges in the solution gives rise to an electric potential $\psi$, which satisfies the Poisson equation:
\begin{subequations}\label{eq:poisson}
    \begin{align}
    -\nabla\cdot\left(\epsilon \nabla\psi(\vec{x},t)\right) &=e\varrho(\vec{x},t),\\
    \varrho(\vec{x},t)&=\sum_{i\in\mathcal{I}}q_i(\vec{x},t) c_{i},\label{eq:def_c_q}
\end{align}
\end{subequations}
where $e$ is the elementary charge, $\varrho$ is the local charge density of the mixture (in units of elementary charge) and $q_i(\vec{x},t)$ is the molecular charge of the $i$-th (in units of elementary charge). While the ions have a constant charge, $q_\pm\equiv \pm 1$, the charge in the solvent and on the polymer can change both spatially and temporally because of charge regulation: 
\begin{align*}
q_s(\vec{x},t)&=\sum\limits_{q=0}^1q\pi_s^{q}(\vec{x},t)
,\\ 
q_p(\vec{x},t)&=\sum\limits_{q=0}^{Q_p}q\pi_p^{q}(\vec{x},t).
\end{align*} 
This contrasts with other models where the charge on macromolecules is taken to be fixed~\cite{majee_charge_2023}. Eq.~(\ref{eq:poisson}) relates the electric potential to the local charge density and naturally introduces a length scale $\ell_B=e^2/(4\pi\epsilon k_BT)$, also known as Bjerrum length, which describes the length scale at which the energy due to the Coulomb force between two charges and the thermal energy, $k_B T$, balance.
 
We assume that the mixture is maintained at constant temperature $T$ and is locally at equilibrium. Consequently, the thermodynamics of the mixture is given by the mean-field Helmholtz free energy density, $f$, which is composed of 4 terms:
\begin{equation}
\begin{aligned}\label{eq:f}
f[c_m,\psi]=\bar{f}(c_m,\pi_p^{q})+\frac{\kappa}{2}\left|\nabla c_p\right|^{2}-\frac{\epsilon}{2} \left|\nabla \psi\right|^2  +\psi  e\varrho.
 \end{aligned}
\end{equation}
The first term, $\bar{f}$, corresponds to the bulk free-energy for a homogeneous mixture, see, e.g., ~\cite{celora2023,Avni2020}:
\begin{widetext}
\begin{equation}
\begin{aligned}\label{fbar}
    \frac{\bar{f}}{k_BT} =c_+\ln \nu c_++c_-\ln \nu c_-+c_s\sum\limits_{q=0}^1\pi_s^{q}\ln[\nu \pi_s^{q} c_s] +c_p\sum\limits_{q=0}^{Q_p}\pi_p^{q}\ln[N\nu \pi_p^{q} c_p]+\chi
N\nu c_p c_s+c_p\sum\limits_{q=0}^{Q_p}\pi_p^{q}\frac{u^q_{\CR,p}}{k_B T}\\+c_s\sum\limits_{q=0}^1\pi_s^q\frac{u^q_{\CR,s}}{k_B T}
-\lambda_p\left(1-\sum_{q=0}^{Q_p}\pi_p^{q}\right)-\lambda_s\left(1-\sum_{q=0}^1\pi_s^{q}\right)-\lambda_{\CR} \left(c_H-q_pc_p-q_sc_s\right). \end{aligned}
\end{equation}
\end{widetext}
In Eq.~(\ref{fbar}) the first 3 terms correspond to the standard Flory-Huggins contribution containing the entropy of mixing associated with each component of the mixture, the mean-field interactions between the macromolecules and water with strength $\chi$ and the internal free energies associated with different charge states $u_{\CR,m}^q$. We keep this expression of $q$ general in the derivation of the model, but choose a particular form later for the simulation results in Section~\ref{sec:results}. For simplicity, we have assumed that $\chi$ is independent of both the macromolecule and the water charge state. In Eq.~(\ref{fbar}), $\lambda_p$ and $\lambda_s$ and $\lambda_{\CR}$ are Lagrange multipliers associated with the normalisation condition for the charge state distributions, $\pi_p^{q}$ and $\pi_s^{q}$, and the local conservation of \ce{H+} ions by the charge regulation mechanisms. Indeed, protonation/deprotonation reactions only result in the exchange of charge between species, leading to the local number density of \ce{H+} ions 
\begin{equation}
c_{H}(\vec{x},t)=q_s(\vec{x},t) c_{s}(\vec{x},t)+q_p(\vec{x},t)c_p(\vec{x},t),\label{def:CH}
\end{equation}
to be conserved. The second term in Eq.~(\ref{eq:f}) accounts for the energy cost associated with the phase-separated polymer-rich and polymer-poor regions, which scales with the interfacial tension $\kappa$. For simplicity, we assume the latter to be constant, \emph{i.e.}, independent of the mixture composition. Note that in writing Eq.~(\ref{eq:f}) we have neglected the contribution due to the gradient cost associated with the salt, which has been investigated in~\cite{majee_charge_2023}. The last two terms on the right-hand side of Eq.~(\ref{eq:f}) account for the electrostatic energy of the mixture~\cite{markovichChargedMembranesPoisson2021}. 

\subsection{The CR-mechanisms}
For simplicity, we assume that charge regulation mechanisms occur at a much faster time scale than molecule transport. As a result, the local polymer and solvent charge distribution $\left\{\pi_p^{q}\right\}_{q=0}^{Q_p}$ and  $\left\{\pi_s^{q}\right\}_{q=0}^{1}$ can be obtained by minimising the free-energy density $\bar{f}$ (see Eq.~(\ref{fbar})) with respect to $\pi_p^{q}$. This procedure is equivalent to imposing the detailed balance for the charge regulation mechanisms (see \cite{celora2023} for more details). Completing the minimization yields a similar result as described in \cite{celora2023} for the polymer and water charge distributions 
\begin{subequations}\label{sys:CR_mechanism}
\begin{align}
    \pi_s^{q} &= \frac{\exp\left(-\beta (u^q_{\CR,s}+\tilde\alpha q)\right)}{\sum\limits_{\zeta=0}^1\exp\left(-\beta(u^q_{\CR,s}+\tilde\alpha \zeta)\right)}, \quad q=0,1,\label{eq:charge_distribution_solvent} \\
    \pi_p^{q} &= \frac{\exp\left(-\beta (u^q_{\CR,p}+\tilde\alpha q)\right)}{\sum\limits_{\zeta=0}^{Q_p}\exp\left(-\beta (u^\zeta_{\CR,p}+\tilde\alpha\zeta)\right)}, \quad q=0,\ldots,Q_p,\label{eq:discrete_charge_distribution} 
    \end{align}
where $\beta=(k_BT)^{-1}$ and the function 
\begin{align}
\tilde\alpha=e\psi+k_BT \lambda_{\CR}\end{align} 
\end{subequations}
captures the contribution of the local environmental conditions in regulating protonation/deprotonation reactions. Note that the multiplier $\lambda_{\CR}$ is implicitly defined in terms of the local number density of \ce{H+} ions by Eq.~(\ref{def:CH}). This is apparent when introducing the partition function $\mathcal{Z}_{m}(\tilde{\alpha})= \sum\limits_{q=0}^{Q_m}\exp\left(-\beta (u^q_{m,\CR}+\tilde\alpha q)\right)$, which contains the necessary information on the distribution $\pi_m^{q}$ as the $j$-th cumulant of $\pi_m^{q}$, $\mathcal{Q}^{(j)}_{m}$, which
is given by 
\begin{equation}\label{eq:moments charge distribution}
\mathcal{Q}^{(j)}_m=-k_BT\frac{d^j}{d\tilde\alpha^j} \ln \mathcal{Z}_{m}(\tilde{\alpha})\, .
\end{equation} 
\\

\subsection{Diffusive dynamics}

The motion of the different species in the solution yields changes to their local number densities $c_p$, $c_s$, $c_\pm$, but can also drive changes in the local number density of the \ce{H+} ions, $c_H$. The conservation of mass associated with the different species is expressed by:
\begin{subequations}\label{sys:time_dependent}
\begin{align}
    \partial_t c_m +\nabla \cdot (c_m\vec{v})+\nabla \cdot \vec{j}_m=0,\label{eq:cons_mass}
\end{align}
for $m\in\mathcal{I}_H=\left\{p,\pm,H,s\right\}$. In Eq.~(\ref{eq:cons_mass}), $\vec{v}$ is the mixture velocity and $\vec{j}_m$ are the thermodynamic (or diffusive) fluxes corresponding to each component. For simplicity, we here neglect hydrodynamic effects and, subsequently, take the mixture velocity $\vec{v}\equiv 0$. As a result, we only have diffusion-type fluxes in \eqref{eq:cons_mass}. Note that the fluxes are not independent, rather they must satisfy the following constraint $\vec{j}_s+N\vec{j}_p+\vec{j}_++\vec{j}_-=0$ to be consistent with the no-void condition~(\ref{eq:no_void}). This constraint can be weakly enforced by introducing the thermodynamic pressure $P$.

The form of the thermodynamic fluxes and pressure are derived using the Maxwell-Stefan approach (details are given in~\Cref{app:thermodynamic_fluxes}). Under the assumption that the friction coefficients between the different components are proportional to their molecular volumes and independent of their charge states, the Maxwell-Stefan conditions are easily invertible and we obtain an explicit definition of the fluxes: 
\begin{align}
    \vec{j}_m=-\sum_{K\in\mathcal{I}_H} \mathcal{M}_{mK}\nabla\mu^{\text{el}}_K, \quad m\in\mathcal{I}_H,\label{eq:fluxes_general} 
\end{align}
where $\mathcal{M}_{mK}$ is the mobility matrix and the variables $\mu^{\text{el}}_m$ are the electro-chemical potentials for the different components of the mixture. Based on our assumption $\mathcal{M}$ is almost diagonal (see Eq.~(\ref{eq:motility_matrix})) and cross-terms only arise to account for the couplings between the movement of proton-carrying molecules (\emph{i.e.}, solvent and polymer) and \ce{H+} ions (see~\Cref{app:thermodynamic_fluxes} for more details). 

The electrochemical potentials for the components $ i\in\left\{\pm,s,p\right\}$ of the mixture have the standard form:
\begin{align}
    \mu^{\text{el}}_i = P\nu_i+q_ie\psi + \frac{\partial \bar{f}_{\eff}}{\partial c_i} -\kappa_i \nabla^2 c_i,\quad i\in\mathcal{I},\label{eq:mu_m}
\end{align} 
while we find that the chemical potential for the \ce{H+} ions is set by the Lagrange multiplier associated with the charge regulation, 
$\mu^{\text{el}}_H=-k_BT\lambda_{\CR}$.
In Eq.~(\ref{eq:mu_m}), $\nu_i$ indicates the molecular volume of the $m$-th component ($\nu_p=N\nu$ and $\nu_{s,\pm}=\nu$ otherwise), $P$ is the thermodynamic pressure (see Eq.~(\ref{eq:thermodynamic_pressure})), $q_i$ is the molecular charge associated with each component, $\bar{f}_{\eff}$ is the effective free energy and the interfacial tension parameter $\kappa_i$ is non-zero only for $i=p$. The form of the effective free energy is obtained by substituting Eqs.~(\ref{eq:discrete_charge_distribution}) into Eq.~(\ref{fbar})
\begin{widetext}
    \begin{align}
\begin{aligned}
     \frac{\bar{f}_{\eff}}{k_BT} &= c_s\ln[\nu c_s]+c_+\ln\left[\nu c_+\right]+c_-\ln\left[\nu c_-\right] +c_p\ln\left[\nu Nc_p\right]+\chi
N\nu c_p c_s -c_H\lambda_{\CR}\\[2pt]\qquad&-c_p\left(\ln\left[\mathcal{Z}_{p}(\tilde{\alpha}(\psi,\lambda_{\CR}))\right]+\frac{eq_p\psi}{k_BT}\right)-c_s\left(\ln[\mathcal{Z}_{s}(\tilde{\alpha}(\psi,\lambda_{\CR}))]+\frac{eq_s\psi}{k_BT}\right).
\end{aligned}\label{eq:f_bar_eff}
\end{align}
\end{widetext}
\end{subequations}
To summarise, the system~(\ref{sys:time_dependent}) governs the dynamics of the mixture composed of charged components, some of which can locally regulate their charge via the presence of ionizable groups. Electrothermodynamics forces (\emph{i.e.}, gradients in the chemical potentials) drive the dynamics of the system towards its thermodynamic equilibrium and are influenced by the electrostatics of the mixture via the electrostatic potential $\psi$ defined by the Poisson equation, Eq.~(\ref{eq:poisson}), and charge regulation. 
The CR mechanisms enter the problem via the multiplier $\lambda_{\CR}$ -- corresponding to the electrochemical potential of \ce{H+} ions -- and the moments of the charge distributions associated with the components that present ionizable groups, which are locally defined by the non-linear algebraic system given by Eqs.~(\ref{def:CH})-(\ref{sys:CR_mechanism}). 

The non-linearities in the model prevent its analytic solution and we therefore solve the dynamic equations numerically. We consider the simplest case of an isolated system consisting of a domain $\Omega$ and impose fluxes and stresses to vanish at the boundary of the domain:
\begin{subequations}\label{sys:BC}
   \begin{align}
   \vec{j}_m \cdot \vec{n}&=0,\quad\vec{x}\in\partial\Omega,\quad m\in\mathcal{I}_H\label{eq:cond_flux}\\
   \nabla\phi_p\cdot\vec{n}&=0,\quad\vec{x}\in\partial\Omega,\\
   \quad \nabla\psi\cdot \vec{n}&=0, \quad\vec{x}\in\partial\Omega.\label{eq:cond_stress}
   \end{align}
The no-flux conditions imply that the total concentration of all species in the domain is conserved over time, while the condition~(\ref{eq:cond_stress}) implies that there is no net electric field across the solutions and requires the solution to be electroneutral:
\[\int_\Omega \varrho(\vec{x},t) d\vec{x}\equiv 0.\] However, this condition only imposes $\psi$ up to a constant. We therefore need to impose an additional constrained on its value; without loss of generality we require:
\begin{align}
    \psi(0,t)=0.\label{eq:norm_electric_field}
\end{align}%
\end{subequations}
For the problem to be well-posed the initial conditions on the number densities ($c_H$, $c_p$, $c_s$, $c_\pm$) must satisfy global electroneutrality of the solution and the no-void condition, \emph{i.e.}, Eq.~(\ref{eq:no_void}), locally.

\section{Results}\label{sec:results}

We now employ the derived quantitative theory and a combination of numerical and analytical approaches to study the time-evolution and long-term behaviour of phase-separated CR polymer solutions. For simplicity, we focus on flat interfaces and consider a 1D problem defined on the interval $\Omega=\left[0,L\right]$, where $L$ is the size of the domain. We present the non-dimensional 1D form of the modelling equations in the Supplementary Material (SM). Following the literature on dilute electrolyte solutions~\cite{gavish2017PRE}, we scaled space by the Debye-like scale $L_\psi=\sqrt{\epsilon \nu k_BT}/e$, even though this deviates from the typical screening length for non-dilute weak polyelectolytes; more details in~\Cref{sec:localised domains}. Details on the numerical approach to solving the time-dependent problem and parameter values used in the simulations are given in the SM. For the numerical simulations, a specific model for the charge regulation mechanisms must be specified. Here, we use the standard linear-quadratic form for the charge regulation energy as in \cite{celora2023,Avni2020} (see SM).

\begin{figure*}[htb]
    \centering
    \begin{subfigure}{0.001\textwidth}
      \centering
      \captionlistentry{}
      \label{fig2A}
     \end{subfigure}
     \begin{subfigure}{0.001\textwidth}
      \centering
      \captionlistentry{}
      \label{fig2B}
     \end{subfigure}
          \begin{subfigure}{0.002\textwidth}
      \centering
      \captionlistentry{}
      \label{fig2C}
     \end{subfigure}
    \begin{subfigure}{0.99\textwidth}
    \includegraphics[width=\linewidth]{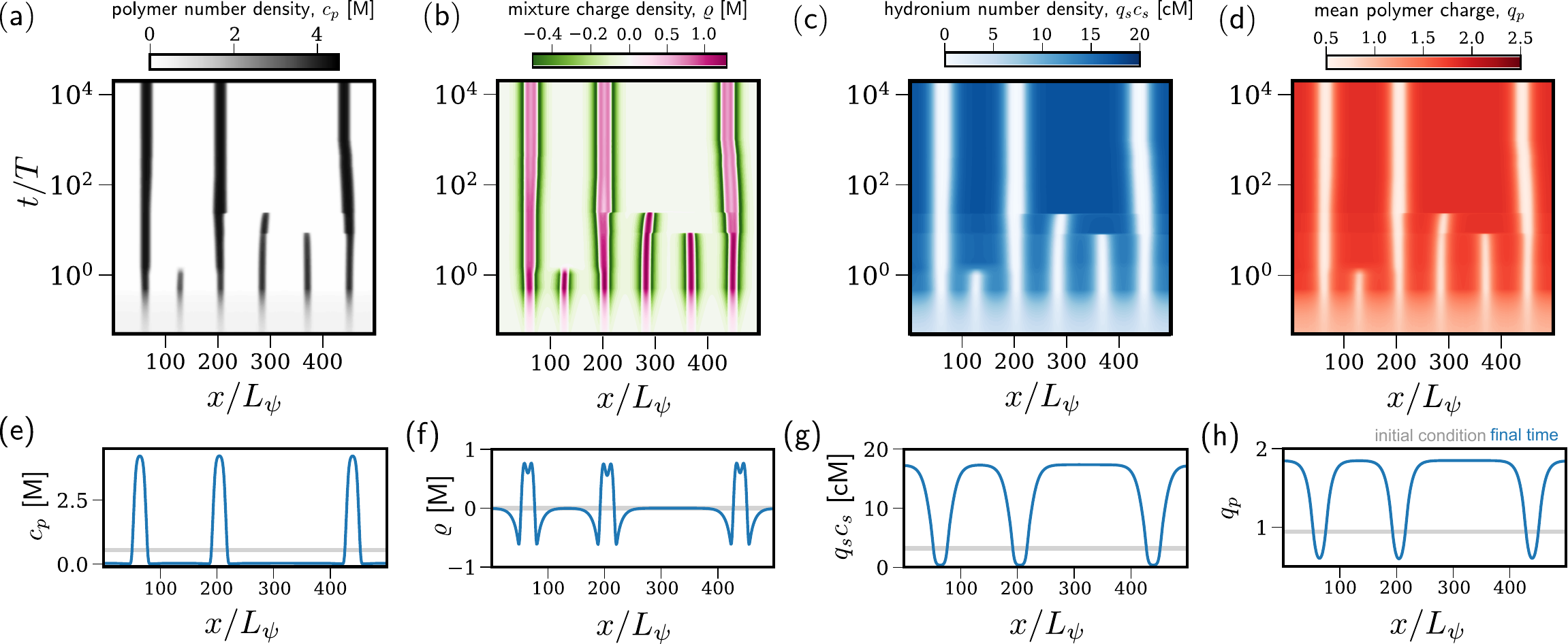}
                \captionlistentry{}
                \label{fig2D}
    \end{subfigure}
    \begin{subfigure}{0.001\textwidth}
          \centering
                \captionlistentry{}
                     \label{fig2E}
    \end{subfigure}
        \begin{subfigure}{0.001\textwidth}
          \centering
                \captionlistentry{}
                     \label{fig2F}
    \end{subfigure}
        \begin{subfigure}{0.001\textwidth}
          \centering
                \captionlistentry{}
                     \label{fig2G}
    \end{subfigure}
        \begin{subfigure}{0.001\textwidth}
          \centering
                \captionlistentry{}
                     \label{fig2H}
    \end{subfigure}
    \vspace{-5mm}
    \caption{Charge and phase-separation dynamics predicted by the model. Top: plots illustrate the time evolution of (a) the local polymer number density $c_p$, in units of molar; (b) the local charge density of the mixture $\varrho$, in units of molar (defined in Eq.~(\ref{eq:poisson})); (c) the local number density of hydronium ions $q_sc_s$, in units of centi-molar; (d) the local mean charge on the polymer macromolecules $q_p$. Bottom: plots illustrate the spatial distribution at the initial ($t/T=0$) and final ($t/T=20000$) simulation times for (e) $c_p$, (f) $\varrho$, (g) $q_sc_s$, and (h) $q_p$. Model and scaling parameters are listed in~\Cref{tab:table_par} in the Supplementary Material.}
    \label{fig2:charge dynamics}
\end{figure*}

\subsection{Coacervation dynamics of weak polyelectrolytes }
\label{sec:dynamics1}
We start by discussing the dynamics of phase separation of an initially homogeneous mixture when perturbed by introducing noise in the initial concentration of polymer macromolecules in a regime where demixing occurs. The stability of homogeneous mixtures of weak polyelectrolytes will be discussed in~\Cref{sec:spinodal}.

The initial perturbations in the polymer number density rapidly lead to the formation of regions of higher liquid density (\Cref{fig2A,fig2E}). 
As in recent works~\cite{CeloraPL1,hennessy_breakdown_2024,majee_charge_2023,luo2024condensatesizecontrolcharge}, we observe the separation of the charges across the condensate liquid interface which results in the local breaking of charge electroneutrality at the condensate interface. While the mixture is neutral in the dilute homogeneous regions of the domain (see~\Cref{fig2B,fig2F}), an electric double-layer forms across the condensate interface due to the asymmetric distribution of the charges in the mixture that jumps from being slightly negative outside the condensed regions to being positive in their interior. In the initial transient, we observe standard Ostwald ripening dynamics whereby 
smaller condensates transfer mass towards the larger domains (see~\Cref{fig2A}) and eventually disappear. While the smaller condensates present at earlier times are highly positively charged, as their number decreases and size increases via coarsening, their interior tends to be less positively charged (see~\Cref{fig2B,fig2F}). At longer times ($t\approx 100 T$), we observe mass is transferred from the larger (middle) to the smaller (left) domain. This inverse transfer of material stabilises the coexistence of the three equally-sized domains that persists until the end of the simulations ($t=20000 T$). The stabilisation of the microscopic domains at long times is well-understood and arises from the competition between two mechanisms~\cite{grzetic_electrostatic_2021}. On the one hand, the electrostatic forces favour the asymmetric distribution of co-ions and counter-ions across the domain interface to balance the accumulation of the charged polymer phase in the condensed region and restore electroneutrality. On the other hand, the salt ion entropy favours homogeneous salt distribution by driving diffusive fluxes across the domain interface.

An interesting aspect resulting from the inclusion of the CR mechanism is the significant redistribution of free \ce{H3O+} ions in the solution during phase separation (\Cref{fig2C,fig2G}), leading to spatial regulation of the local pH of the mixture. In particular, we find that \ce{H3O+} ions are excluded from the condensed domains and accumulate in the more dilute regions. This is in line with predictions of asymmetric proton partitioning in weak simple coacervates~\cite{celora2023}. The local modulation of the mixture acidity via phase separation results in the self-organised arrangement of the polymer molecules across the liquid interphase depending on their charge state (\Cref{fig2D,fig2H}). In particular, we find that the more highly charged polymer chains are located in the dilute area of the domains, where \ce{H3O+} ions are most abundant, while the condensed regions are occupied by the poorly charged macromolecules. 

We now study how the dynamics of phase separation are affected by the average concentration of \ce{H+} ions in solution -- which is a conserved quantity over time: 
\[c^0_H=\frac{1}{L}\int_0^L c_H(x,0)\,dx.\]
The following simulation results were generated with the same random initial condition for the polymer volume fraction to ensure an appropriate comparison between the different conditions.

\begin{figure*}[htb]
    \centering
    \begin{subfigure}{0.001\textwidth}
      \centering
      \captionlistentry{}
      \label{fig1A}
     \end{subfigure}
         \begin{subfigure}{0.001\textwidth}
      \centering
      \captionlistentry{}
      \label{fig1B}
     \end{subfigure}
    \begin{subfigure}{0.99\textwidth}
    \includegraphics[width=\linewidth]{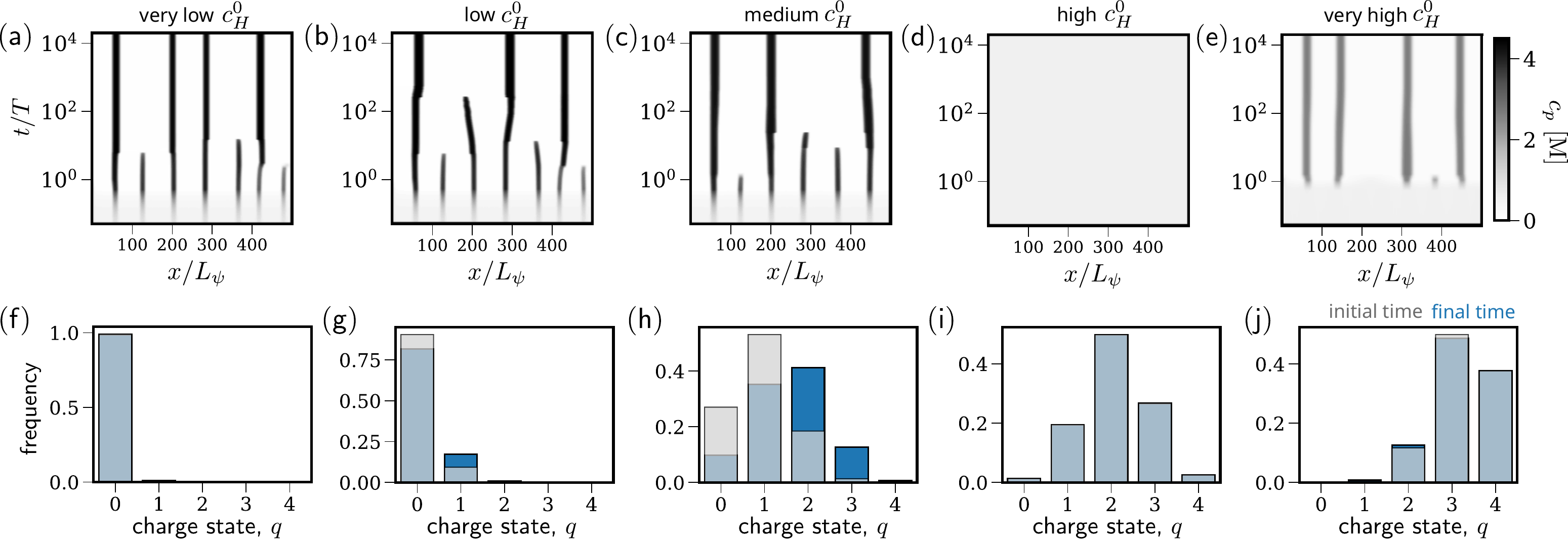}
                \captionlistentry{}
                \label{fig1C}
    \end{subfigure}
    \begin{subfigure}{0.001\textwidth}
          \centering
                \captionlistentry{}
                     \label{fig1D}
    \end{subfigure}
        \begin{subfigure}{0.001\textwidth}
      \centering
      \captionlistentry{}
      \label{fig1E}
     \end{subfigure}

     \begin{subfigure}{0.001\textwidth}
      \centering
      \captionlistentry{}
      \label{fig1F}
     \end{subfigure}
         \begin{subfigure}{0.001\textwidth}
      \centering
      \captionlistentry{}
      \label{fig1G}
     \end{subfigure}
    \begin{subfigure}{0.99\textwidth}
                \captionlistentry{}
                \label{fig1H}
    \end{subfigure}
    \begin{subfigure}{0.001\textwidth}
          \centering
                \captionlistentry{}
                     \label{fig1I}
    \end{subfigure}
        \begin{subfigure}{0.001\textwidth}
      \centering
      \captionlistentry{}
      \label{fig1J}
     \end{subfigure}
     \vspace{-8mm}
    \caption{Simulation of the dynamics of phase separation for increasing value of the control parameter $c_H^0$. Top: plots illustrate the spatio-temporal evolution of the polymer concentration for (a) $\nu c_H^0=0.0001$; (b) $\nu c_H^0=0.001$ and (c) $\nu c_H^0=0.01$; (d) $\nu c_H^0=0.032$; (e) $\nu c_H^0=0.063$. Bottom: plots compare the spatially-average polymer charge distribution, $\bar{\pi}^q_p(t)=(\int_0^L\pi_p^q(x,t)dx)/L$ at the initial and final time of the simulations for (f) $\nu c_H^0=0.0001$; (g) $\nu c_H^0=0.001$ and (h) $\nu c_H^0=0.01$; (i) $\nu c_H^0=0.032$; (j) $\nu c_H^0=0.063$. Note that~\Cref{fig1C} and~\Cref{fig2A} are identical. Model parameters are as in~\Cref{fig2:charge dynamics}. }
    \label{fig1:dynamics}
\end{figure*}

In Figure~\ref{fig1:dynamics}, we compare the coarsening dynamics (top panels) and the impact of phase separation on the overall polymer charge distribution (bottom panels) for increasing values of $c_H^0$. 
We find that the propensity of the mixture to phase separate depends non-monotonically on $c_H^0$. In particular, we find that while the homogeneous state is unstable at low and at very high levels of acidity, intermediate levels of acidity can favour mixing, see~\Cref{fig1D}. This is in line with previous analysis of the equilibrium behaviour of CR mixtures~\cite{celora2023}, which identified non-linear dependence of demixing on solution acidity as a key feature of this system. In either very low or very high $c_H^0$ values, the polymer phase exhibits minimal charge heterogeneity in the homogeneous initial state: most polymers are uncharged at very low $c_H^0$, and highly charged at very high $c_H^0$.  under these limiting conditions, we find that the homogeneous states are unstable, and phase separation has a negligible impact on the overall charge distribution. Charge-heterogeneity in the initial homogeneous state peaks for intermediate $c_H^0$ values (\Cref{fig1H,fig1I}). Due to the convexity of the entropic term $\sum_{z}\pi_z\ln (\pi_z)$ in Eq.~(\ref{fbar}), homogeneous states characterised by larger polymer charge heterogeneity are associated with higher entropy. At medium $c_H^0$ values, the homogeneous state remains unstable and we observe a significant increase in the overall charge-heterogeneity upon spatial segregation of polymers in micro-domains. In contrast, for a high value of $c_H^0$ (\Cref{fig1I}) the homogeneous state stabilises to noisy initial conditions. This corresponds to the initial state of maximal charge heterogeneity suggesting that the contribution of charge heterogeneity to the entropy is sufficiently strong (at least for the parameter value considered) to compete with short-range interactions to counteract the tendency of the polymers to phase separate and stabilise the homogeneous state.

For conditions under which demixing occurs, in line with previous observations, phase separation initially follows a standard Ostwald ripening dynamics, whereby smaller condensates form and then dissipate to transfer mass towards the larger domains. The timescale at which condensed domains first appear as well as their initial number is controlled by $c_H^0$. This number decreases with $c_H^0$ in the low-medium acidity regime,  then increases again upon re-entrance in the demixing region at very high acidity. It turns out that the same trend holds for the number of long-lasting condensed domains. Interestingly, for low values of $c_H^0$ -- corresponding to an initially weakly charged polymer solution, we observe that condensed regions of different sizes persist. In contrast, in the medium and very high acidity case the remaining condensates have similar sizes (see Figures~\ref{fig1:dynamics} and SM1). 
This suggests that medium values of $c_H^0$ stabilise the structure of long-lasting micro-domains. We validate this hypothesis further by running ensemble simulations of the model for the same environmental condition but different noisy initial conditions and characterising the micro-domains size distributions at later times -- $t/T=1000$ (Figure SM1);
the results are detailed in the SM. 
We find that $c_H^0$ modulates the level of heterogeneity in micro-domain size, with medium levels of acidity reducing variability. On the one hand, this is due to the long-term coexistence of micro-domains of distinct size in low acidity conditions (\Cref{fig1A}); additionally, this reflects variability in the number of micro-domains that persist at later times. For example, for very low $c_H^0$, out of 48 simulations, we observe three or four micro-domains with approximately equal probability. In contrast, for medium $c_H^0$, the three micro-domains are highly likely; we observed them in 44 out of 48 randomly initialised simulations. These results suggest the system admits multiple spatially heterogeneous equilibria, whose stability and existence are regulated by the total proton concentration, $c_H^0$. 

\subsection{Emergence and coexistence of localised domains}
\label{sec:localised domains}
\label{sec:spinodal}
\begin{figure*}[t]
    \centering
    \begin{subfigure}{0.001\textwidth}
        \captionlistentry{}
        \label{fig3A}
    \end{subfigure}
    \begin{subfigure}{0.001\textwidth}
        \captionlistentry{}
        \label{fig3B}
    \end{subfigure}
    \begin{subfigure}{0.98\textwidth}
    \includegraphics[width=\linewidth]{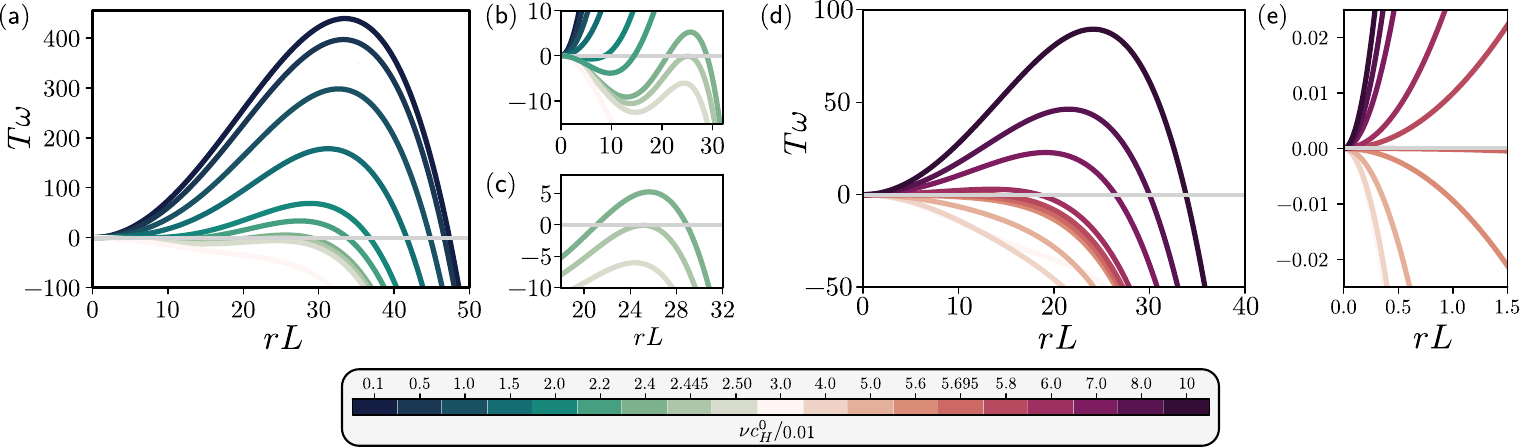}
        \captionlistentry{}
        \label{fig3C}
    \end{subfigure}
    \begin{subfigure}{0.001\textwidth}
        \captionlistentry{}
        \label{fig3D}
    \end{subfigure}
    \begin{subfigure}{0.001\textwidth}
        \captionlistentry{}
        \label{fig3E}
    \end{subfigure}
    \vspace{-5mm}
    \caption{Dispersion relation, $\hat{\omega}_r=\max\left\{\omega_r\in\mathbb{R} \mbox{ s.t. Eq.~(\ref{eq:dispertion_relation}) holds}\right\}$, computed for for the parameter values used in~\Cref{fig1:dynamics} as a function of the solution acidity (a)-(c) $\nu c_H^0\in[0.001,0.03]$ and (d)-(e) $\nu c_H^0\in[0.03,0.1]$. (b)-(c) Insets zooming into the region near the onset of low-$c_H^0$ instability from panel (a). (d) Insets zooming into the region near the onset of high-$c_H^0$ instability from (e). Model parameters are as in~\Cref{fig2:charge dynamics}. }
    \label{fig:dispersion_relation}
\end{figure*}
Our previous results have shown that polymer solubility, phase separation dynamics and the length-scale selection of the long-term patterns are controlled by the total amount of \ce{H+}ions; further investigation of the charge distribution hints at the role that charge-heterogeneity plays in shaping the phase behaviour of polymers under variable pH. Inspired by the presented dynamical simulations, we now focus on investigating the mechanisms driving the emergence of spatial patterns.

\subsubsection{Linear stability of homogeneous weak polyelectrolytes}
Firstly, we analyse the linear stability of well-mixed electroneutral solutions; details of the analysis are presented in the SM. We consider spatio-temporal perturbations of the solution of the form:
\begin{equation}
    c_i=c^0_i+\delta u_i(\vec{x},t), \quad \psi=\delta u_\psi(\vec{x},t), \label{eq:ansatz_linear_spacetime}
\end{equation}
where $m\in\left\{p,\pm,s,H\right\}$ $\delta \ll1$. We again consider the 1D problem: $\vec{x}=x\vec{e}_x$, $x\in[0,L]$. Substituting Eq.~(\ref{eq:ansatz_linear_spacetime}) in the original modelling equations, keeping only the linear terms in $\delta$, we obtain a linear set of coupled PDE, which we solve using standard separation of variable techniques by writing \begin{equation}
u_i=U^r_i \cos\left(r \pi x\right)e^{\omega t}, \quad  i\in\left\{p,\pm,s,H,\psi\right\}.\label{eq:ansatz_linearisation}
\end{equation} 
where $\omega$ is the growth rate of the perturbation associated with the wavenumber $r$. 
In what follows, we consider a semi-infinite domain so that the wavenumber $r$ can take values on the positive real line, \emph{i.e.}, $r\in (0,\infty)$. This approximation, which neglects the second no flux conditions at $x=L$, is reasonable as long as the characteristic length scale of the observed pattern is significantly smaller than $L$. 
The growth rates $\omega_r$ associated with the wave number $r$ are given by the solution of a generalised eigenvalue problem,
\begin{equation}
\det\left[\tens{A}_r- \tens{B}\omega_r\right]=0.\label{eq:dispertion_relation}
\end{equation}
The matrices $\tens{B}\in\mathbb{R}^{4\times 4}$ and $\tens{A}_r\in\mathbb{R}^{4\times 4}$ are symmetric matrices which are derived in SM. The matrix $\tens{B}$ is independent of the wavenumber $r$ and is totally positive (hence positive definite) provided $c_j^0>0$. Both $\tens{A}_{r}$ and $\tens{B}$ are symmetric, hence the growth rates $\omega_r$ are real for all wavenumbers $r$. This rules out the possibility of oscillatory instabilities. In the linear regime, the onset of patterns can therefore only result from stationary instabilities and the boundary of the stability region for homogeneous solutions can be studied by looking at the reduced problem of finding $r^*>0$ such that $\det \tens{A}_{r^*}=0$.

\Cref{fig:dispersion_relation} illustrates how the dispersion relation changes as a function of $c_H^0$. When $c_H^0$ is either very small or sufficiently large (dark curves in either~\Cref{fig3A} or~\Cref{fig3D}), we recover the same qualitative behaviour as for a standard conservative Cahn-Hilliard model. In this case, the growth rate $\omega$ monotonically increases with $r$ up to a local maximum, corresponding to the fastest growing spatial mode -- which determines the evolution of the system in the very early stages of spinoidal decomposition, and then decreases monotonically to $-\infty$. As a result, all eigenmodes below a critical value of $r_{\rm{II}}$ have a positive growth rate $\omega$. This implies that the homogeneous state is unstable to both short- and long-wave perturbations. 

As the solution acidity increases from $\nu c_H^0=0.001$ to $\nu c_H^0=0.03$, that is, from dark to light green colour curves in~\Cref{fig3A}, we observe a significant change in the shape of the dispersion curve which now presents two local extremal points in $r\in(0,\infty)$ (\Cref{fig3B}). 
Despite the gradient structure of the modelling equation, the growth rate associated with small wavenumbers becomes negative, while larger wavenumbers remain unstable. As a result, the homogeneous steady state is unstable to short-scale perturbations while it remains stable to large-scale perturbations (corresponding to $r\rightarrow 0$). The damping of the large-scale instabilities is common in system coupling gradient systems with the Poisson equation~\cite{gavish_spatially_2017,gavish2017PRE,gavish_theory_2016}. As shown in the inset (see panel (c)), the instability transition due to dilution of $c_H^0$ below the critical value $\nu c^{0}_H=0.02445$ is associated with the typical length scale, $(r^*)^{-1}\approx 20L_\psi$. Interestingly, we observe a different behaviour for the re-entrant instability observed for even larger solution acidity (see~\Cref{fig3D}-\ref{fig3E}). In this case, the dispersion relation presents at most one extremum for $r>0$. Upon increase in the solution acidity past the critical value $\nu c_H^0\approx 0.0568$, we find that the homogeneous state first becomes unstable to long-range perturbations and only subsequently to short-range perturbations. This corresponds to the standard picture of spinodal decomposition in Cahn-Hilliard models which is associated with macroscopic phase separation. 

\subsubsection{Microphase separation of weak polyelectrolytes}

The presence of a short-scale instability and its connection with microphase separation is a characteristic feature of models of electrically controlled phase separation \cite{CeloraPL1,hennessy_breakdown_2024,gavish_spatially_2017,grzetic_electrostatic_2021} and it is exemplified by the canonical Ohta-Kawasaki (OK) model~\cite{shiwa_amplitude_1997}. In particular, short-scale instabilities are controlled by the relative strength of short-range and long-range (\emph{i.e.}, Coulombic) interactions, which is captured by the non-dimensional parameter \[\ell_\kappa=\frac{e}{k_BT\nu N}\sqrt{\frac{\kappa}{\epsilon}}.\] When $\ell_\kappa\gg1$, the cost associated with interfacial tension dominates over the electrostatic forces and the only driver of phase separation is the polymer hydrophobicity -- captured by the Flory-Huggings term and phase separation is not sufficient to drive a breaking of electro-neutrality across the liquid interface~\cite{hennessy_breakdown_2024}. In contrast, when $\ell_\kappa\lesssim1$ short- and long-range interactions can compete; consequently, electrostatic forces can drive demixing in the presence of short-range charge fluctuations. As shown in~\cite{majee_charge_2023}, this can also result in the formation of charge layers near liquid interfaces. However, the analysis is more complicated in the case of concentrated solutions~\cite{gavish2017PRE,majee_charge_2023}, where composition-dependent effects come into play. For weak polyelectrolytes, the competition amongst entropy, long- and short-range interactions is further modulated by the CR mechanisms. This is apparent in the computation of the stability of homogeneous states, detailed in~\Cref{app:stability analysis}. In particular, we find that the competition between short- and long-range interactions is captured by the effective parameter
\begin{subequations}\label{eq:moder_T}
\begin{align}
    \tilde{\ell}_\kappa=\ell_\kappa\sqrt{\frac{b}{a}},
\end{align}
where
\begin{align}
a&=\chi^2c^0_s-\frac{(\nu^{-1}-\chi c^0_s)^2}{\nu^{-1}- Nc_p^0}-\frac{1}{(N\nu)^2 c_p^0},\\
b&=2 c_+^0+2 c_H^0+ c_p^0\sigma_p^0-q_p^0 c_p^0\left(1+\frac{q^0_p c_p^0}{\nu^{-1}-N c^0_p}\right),
\end{align}
\end{subequations}
and $\sigma_p^0=\langle q^2\rangle_p-q_p^2$ is the standard deviation of the charge distribution in the homogeneous steady state. Note that in defining $\tilde{\ell}_k$, we are assuming that the homogeneous solution is such that $a>0$, which is the most interesting case since $a<0$ corresponds to the entropy-dominated regime in which the steady state is stable. By looking at Eq.~(\ref{eq:moder_T}), we see that the CR mechanisms come into play via the coefficients $b$, as the latter depends on both the mean and standard deviation of the charge distribution. By entering positively into the factor $b$, $\sigma_p^0$ effectively increases $\ell_\kappa$, favouring the contribution from short-range interaction over electrostatics. In the linear problem, this appears by taking the variation of the mean polymer charge $q_p$ with respect to the electric field; therefore larger values of $\sigma^0_p$ imply the charge-state of the polymer phase is more sensitive to the electrostatic environment. In this regime, upon macrophase separation, the exclusion of protons from the condensed region can contribute to reducing the repulsion between polymers within the condensed domain and the need for charge separation across the interface. 
In contrast, increasing the mean charge on the polymer $q_p^0$ reduces $\tilde{\ell}_\kappa$ by increasing the strength of electrostatic interactions. In our model $q_p^0$ and $\sigma_0$ are not independent variables, but are rather set by the properties of the internal free energies of different polymer charge states $u^q_{\CR,p}$.

\begin{figure*}
\centering
\begin{subfigure}{0.002\textwidth}
    \captionlistentry{}
    \label{fig4A}
\end{subfigure}
\begin{subfigure}{0.001\textwidth}
    \captionlistentry{}
    \label{fig4B}
\end{subfigure}
\begin{subfigure}{0.7\textwidth}
    \includegraphics[width=\textwidth]{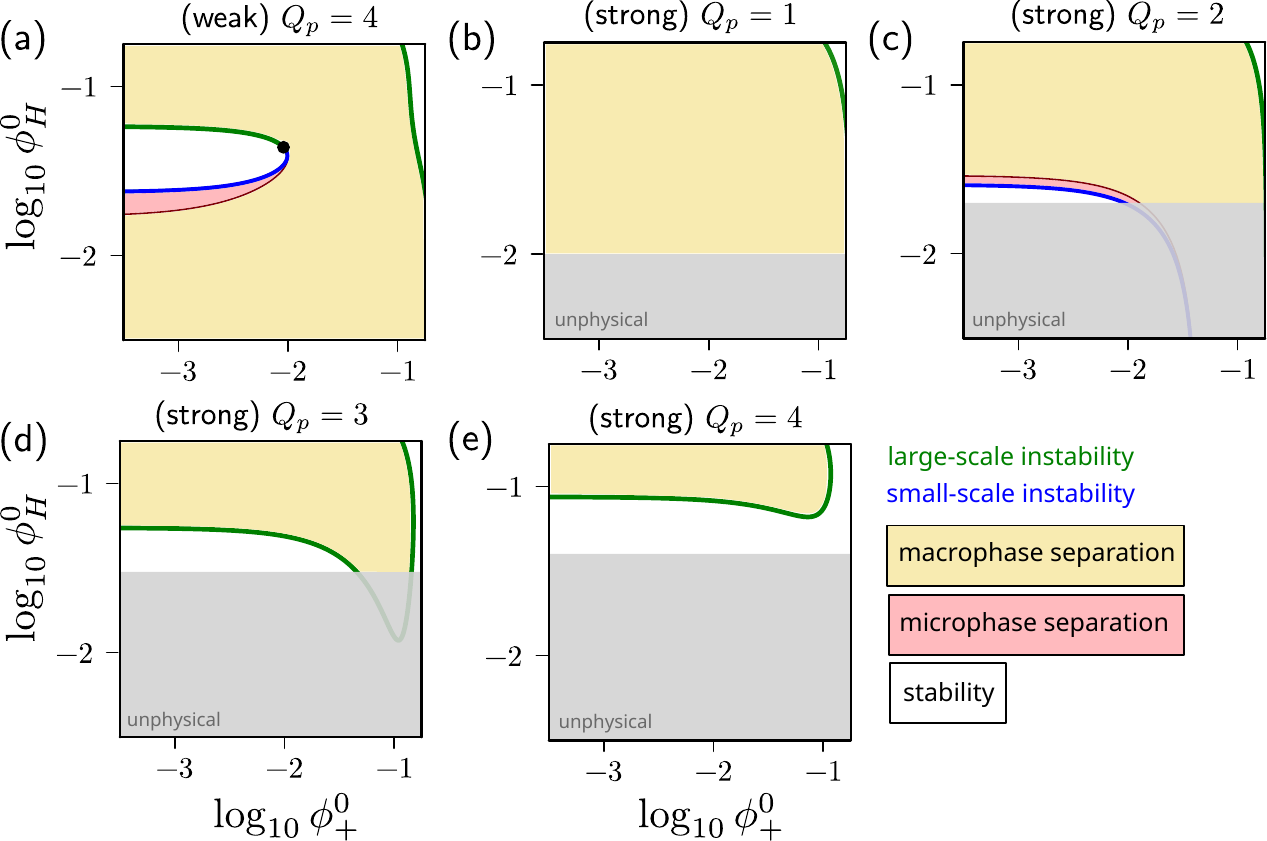}
    \captionlistentry{}
    \label{fig4C}
\end{subfigure}
\begin{subfigure}{0.001\textwidth}
    \captionlistentry{}
    \label{fig4D}
\end{subfigure}
\begin{subfigure}{0.001\textwidth}
    \captionlistentry{}
    \label{fig4E}
\end{subfigure}
    \caption{Spinoidal phase diagram in the $(c_+^0,c_H^0)$-coordinates. (a) Phase diagram for weak polyelectrolytes; same parameter values as in~\Cref{fig1:dynamics}. (b)-(e) Phase diagram for strong polyelectrolytes with increasing fixed charge $Q_p$. Black dots indicate a tricritical point (or Lifshitz point) where the microphase separation spinoidal vanishes. Microphase separation region: homogeneous state only unstable to short-range perturbations; Macrophase separation region: homogeneous state unstable to long-range perturbations; White region: homogeneous state stable to both long- and short-range perturbations. Model parameters are as in~\Cref{fig2:charge dynamics}.
    }
    \label{fig:spinodals}
\end{figure*}

We now compare the phase behaviour of weak and strong polyelectrolytes by computing the spinodal curves associated with both macro- and micro-phase separaton; or equivalently short- and long-range instabilities.
In particular, we investigate the phase behaviour of weak and strong polyelectrolytes in response to the average proton concentration $c_H^0$ and the average concentration of added salt $c_+^0$. The results are shown in~\Cref{fig4A}, where coloured and white areas indicate phase space regions where homogeneous solutions are unstable and stable, respectively. Spinodal curves in the plot indicate codimension-one bifurcations, while the black dot indicates codimension-two bifurcations, specifically tricritical Lifshitz points, \emph{i.e.}, points in the phase diagram where homogeneous, micro- and macrophase separation can coexist. We find that the stability of weak and strong polyelectrolytes is controlled by both axes considered, even if in a different manner. A stabilisation of homogenous states at intermediate levels of acidity is only possible at sufficiently low salinity (the reference value used in our simulations is $\nu c_+^0=10^{-3}$). 

The corresponding phase diagrams for strong polyelectrolytes are given in~\Cref{fig4B,fig4E} (details on the computation for the strong polyelectrolyte case can be found in SM. Since in the weak polyelectrolyte case, molecules can exist in any charge state $q_p\in[1,Q_p]$, we compute the phase diagrams for strong polyelectrolyte with fixed charges in the same range (see~\Cref{fig4B}-\ref{fig4E}). Note that because of the presence of fixed charges, we have the additional constraint that $c_H^0\geq Q_pc_p^0$ which limits the area of the phase space which corresponds to physical states. As we increase the fixed charge $Q_p$ on the molecules, we find that the phase space region, in which demixing occurs, reduces. Microphase separation only occurs for medium-charged polymers with $Q_p=2$, while in all other cases, only macrophase instabilities are observed. When comparing the weak and strong phase diagrams, we find that the first appears like a combination of the others but also presents unique features such as the codimension-two bifurcations.
This suggests the phase behaviour of weak polyelectrolytes can deviate significantly from that of strong polyelectrolytes due to the coupling between local environmental conditions and the polymer's charge state, which allows for electro-chemical regulation of short- and long-range interactions during phase separation.

\subsubsection{Analysis of localised states: the role of non-linear mechanisms.}
\label{sec:bifurcation}
In the previous section, we investigated the phase behaviour of weak polyelectrolytes under small amplitude perturbations, which allowed us to neglect non-linear effects. Even in the linear regime, we identify the possibility of microphase separation, \emph{i.e.}, the coexistence of spatially localized states. However, the evolution and interactions between such domains are eventually regulated by non-linear interactions between Coulombic and interfacial forces. Similar to the case of strong polyelectrolytes, as discussed in~\Cref{sec:dynamics1}, these non-linear effects influence the dynamics following macrophase separation, potentially arresting or reversing standard Ostwald ripening to stabilize the coexistence of micro-domains. Here, we discuss additional features that are unique to weak polyelectrolytes as they arise from the interaction of CR mechanisms, microphase separation and the re-entrant behaviour of the phase diagram.

\begin{figure*}
    \centering
    \begin{subfigure}{0.002\textwidth}
    \captionlistentry{}
    \label{fig6A}
\end{subfigure}
\begin{subfigure}{0.001\textwidth}
    \captionlistentry{}
    \label{fig6B}
\end{subfigure}
\begin{subfigure}{0.9\textwidth}
    \includegraphics[width=\textwidth]{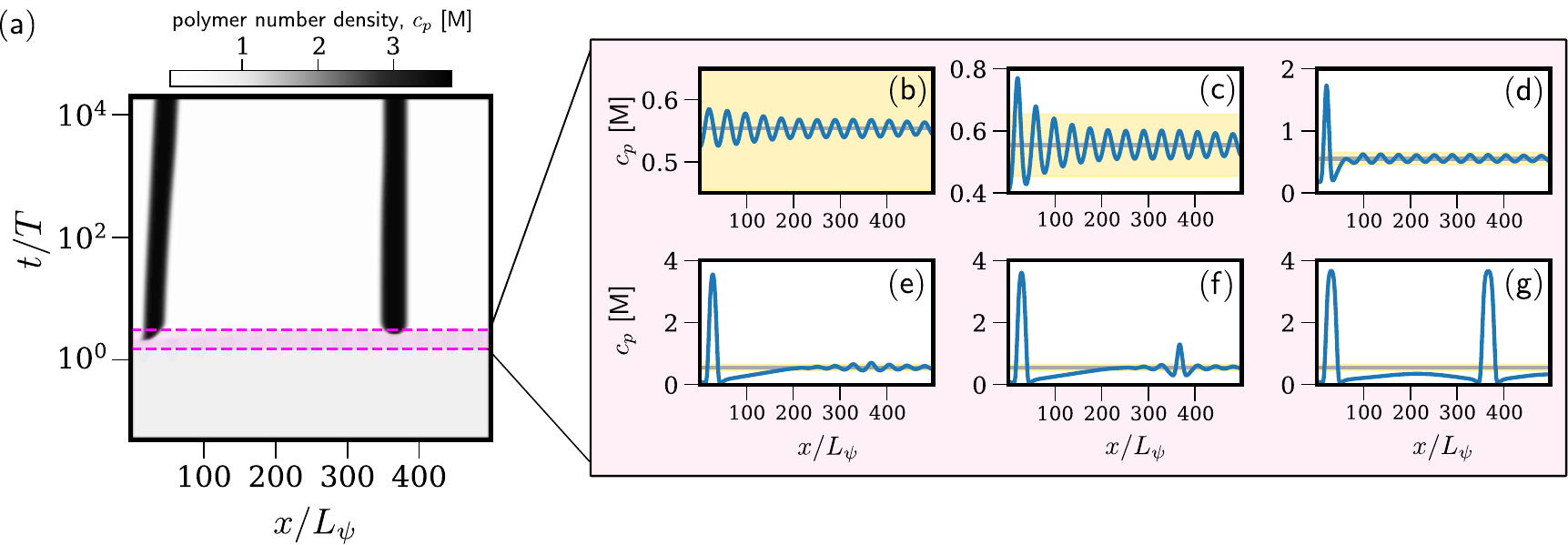}
    \captionlistentry{}
    \label{fig6C}
\end{subfigure}
\begin{subfigure}{0.001\textwidth}
    \captionlistentry{}
    \label{fig6D}
\end{subfigure}
\begin{subfigure}{0.001\textwidth}
    \captionlistentry{}
    \label{fig6E}
\end{subfigure}
\begin{subfigure}{0.001\textwidth}
    \captionlistentry{}
    \label{fig6F}
\end{subfigure}
\begin{subfigure}{0.001\textwidth}
    \captionlistentry{}
    \label{fig6G}
\end{subfigure}
    \caption{Simulation of the dynamics of phase separation for $\nu c_H^0=0.02425$ and $\nu c_+^0=0.001$, which lies in the region of micro-phase separation in~\Cref{fig4A}. (b)-(g) Plot illustrating the spatial polymer concentration at different time points within the pink-shaded area in panel (a). Note that in panels (b)-(f) the y-axis is changed. To help comparison, the grey line indicates the initial polymer concentration ($c_p(x,0)\approx c_p^0$) and the shaded yellow area indicates the interval $|c_p-c_p^0|<0.1$. Values of model parameters are as in~\Cref{fig2:charge dynamics}. }
    \label{fig:microphase_separation_dynamics}
\end{figure*}

In~\Cref{fig:microphase_separation_dynamics}, we illustrate the phase separation dynamics for values of $c_H^0$ near the onset of a multiscale instability. In this regime, the linear stability analysis predicts microphase separation and selection of a persistent
typical length $1/r^*\approx 20 L_\psi$. We find this is what occurs in the initial transient, until $t\approx T$. Unlike in~\Cref{fig2:charge dynamics}, Ostwald ripening does not occur. On the contrary, the system quickly converges towards a layered configuration, in which small amplitude micro-domains coexist. However, after this transient, the dynamics significantly change due to a nucleation event on the left-hand side of the domain, which is driven by a small asymmetry in the initial condition (\Cref{fig6B,fig6D}). Interestingly, nucleation does not lead to the dissipation of the layered domains, which occupy the right-hand side of the domain (\Cref{fig6D}). Eventually, a second nucleation event breaks the layered structure and drives to the formation of two condensed domains that stabilise over time.

The phase separation dynamics illustrated in~\Cref{fig:microphase_separation_dynamics} are unexpected compared to standard micro-phase separation. Although nonlinear instabilities have been reported also in models of strong polyelectrolytes, they usually preserve the layered structures and the associated length scales~\cite{gavish2017PRE}. This indicates that the interplay between short- and long-range Coulombic interactions alone can not fully explain the observed dynamics. Instead, it suggests that this behaviour may be unique to weak polyelectrolytes and is driven by charge-regulation (CR) mechanisms. 

We study the structure of localised states by looking at stationary solutions of our model, which can be obtained by setting all temporal derivatives in Eq.~(\ref{eq:cons_mass}) to zero. As detailed in the SM, the corresponding system of 5 coupled boundary value problems can be reduced to a system of two coupled boundary value problems for the stationary (rescaled) electric potential $\Psi=\Psi(x)$ and the polymer volume fraction $\Phi_p=\Phi_p(x)$:
\begin{subequations}
\begin{align}   
-\Psi''&=\tilde{F}_\psi(\Phi_p,\Psi;\vec{C}_0),\\
    \ell^2_\kappa\Phi_p''&=\tilde{F}_\phi (\Phi_p,\Psi;\vec{C}_0), 
\end{align}\label{eq:equilibria_spatial_dynamics}%
with the boundary conditions:
\begin{align}
\Psi(0)=\Psi'(\ell)=\Phi_p'(0)=\Phi_p'(\ell)=0,
\end{align}
where  $\tilde{F}_\phi$, $\tilde{F}_\psi$ denote variations of the function
\begin{align}\label{eq:energ_hamiltonian}
\begin{aligned}
    \tilde{F}(\phi,\psi;\vec{C}_0)= (1-\phi)\ln\left[\gamma_s(\phi,\psi;\vec{C}_0)(1-\phi)\right] \\\qquad+ \chi \phi(1-\phi)
    +\frac{\phi}{N} \ln\left[\gamma_\phi(\psi;\vec{C}_0)\phi\right],
    \end{aligned}
\end{align}
with respect to its first two scalar arguments. Because of the choice of no-flux boundary conditions, the dynamics of the number densities $c_m$ preserve both the local and global mass of each component $m\in\mathcal{I}_H$. While the no-void and electroneutrality conditions constrain two of the five degrees of freedom associated with mass conservation, three additional constraints have to be specified for the computation of stationary solutions. Here we decide to impose the average number density of polymer, co-ions and protons, which emerges in Eq.~(\ref{eq:equilibria_spatial_dynamics}) by the dependency of the functional $\tilde{F}$ on the constant vector $\vec{C}_0=[c_p^0,c^0_+,c_H^0]$. 
\end{subequations}
Here $\Phi_p$ corresponds to the equilibrium polymer volume fraction so that $\Phi_p=1$ corresponds to the space being fully occupied by polymer molecules, while $\Phi_p=0$ to the space being fully occupied by the solution phase, which includes solvent, salt and hydronium ions. 
The system energy $\tilde{F}$ has the standard form of the energy describing the mixing of a binary mixture of non-ideal immiscible fluid phases, where the function $\gamma_s$ and $\gamma_p$ are the activity coefficients, a unitless correctional factor that describes the extent to which the component of the mixture deviates from the ideal behaviour~\cite{lewis_outlines_1907}.

\begin{figure*}[htb]
    \centering
      \begin{subfigure}{0.002\textwidth}
    \captionlistentry{}
    \label{fig7A}
\end{subfigure}
\begin{subfigure}{0.85\textwidth}
\includegraphics[width=\linewidth]{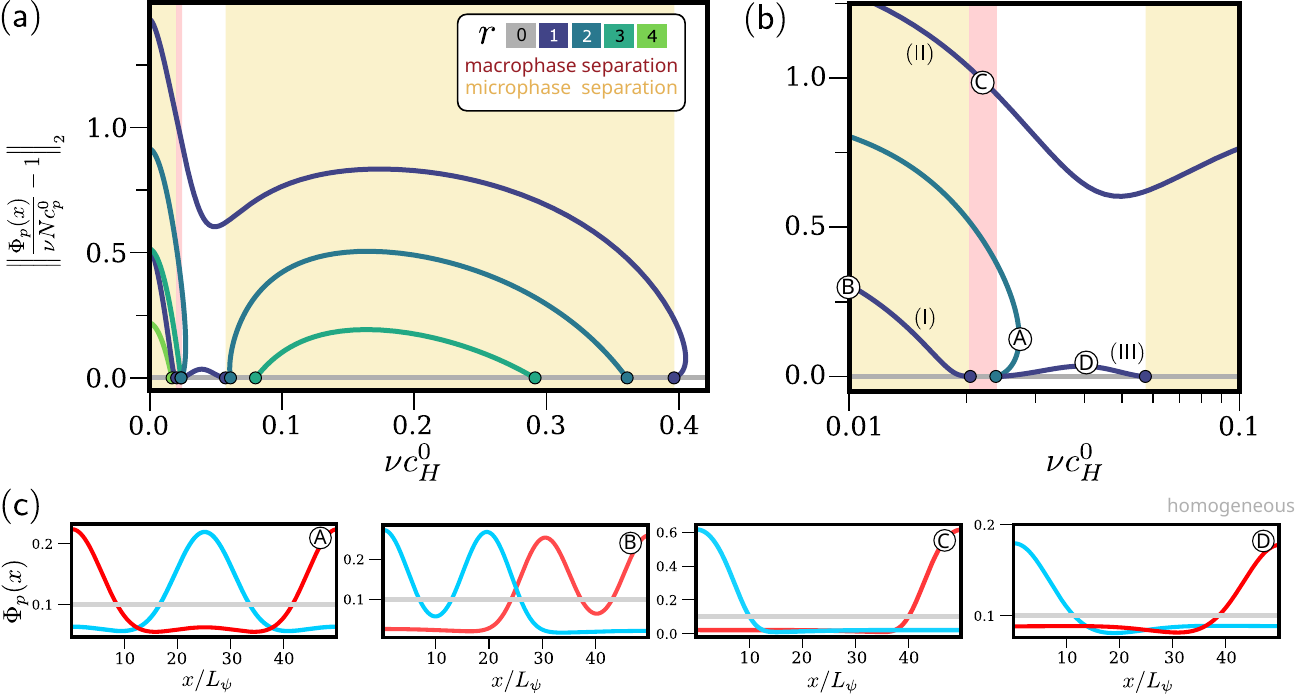}
    \captionlistentry{}
    \label{fig7B}
\end{subfigure}
\begin{subfigure}{0.002\textwidth}
    \captionlistentry{}
    \label{fig7C}
\end{subfigure}
\vspace{-4mm}
    \caption{(a) Characteristic bifurcation diagram for Eqs.~(\ref{eq:equilibria_spatial_dynamics}) using numerical continuation with $c_H^0$ as a control parameter. Here the focus is on the steady phase-separated solutions that branch from the homogeneous steady state. Shown is the normalised $L^2$-norm as a function of the average proton number density $c_H^0$. Different colours are used to indicate the eigenmode from which the branches first appear. Dots identify the bifurcation points of the homogeneous states, where the colour indicates the corresponding spatial modes. As in~\Cref{fig:spinodals}, the yellow and red shaded area indicates the region in which the linear stability analysis predicts macro- and micro-phase separation, respectively. (b) A close-up look at the bifurcation diagram in the interval $\nu c_0\in[0.01,0.1]$. (c) Example profiles of the polymer spatial distribution $\Phi_p(x)$ corresponding to the points A-D in panel (b). Note that because of the symmetries of Eqs.~(\ref{eq:equilibria_spatial_dynamics}), each branch in panels (a)-(b) corresponds to two solutions, see cyan and red curves in panel (c). Model parameters and scalings are as listed in~\Cref{fig2:charge dynamics}, except for the domain size which is reduced to $L=50L_\psi$.}
    \label{fig:bifurcation_diagram}
\end{figure*} 

Eqs.~(\ref{eq:equilibria_spatial_dynamics}) are highly non-linear. We solve them using numerical continuation techniques which allow us to follow how solutions change as a function of a given model parameters and track local bifurcations. We unfold the structure of model solutions as a function of the control parameter $c_H^0$ (\Cref{fig:bifurcation_diagram}). Since the number of solutions increases significantly with the domain size $\ell$, we consider solutions on a domain that is 10 times smaller than in the dynamical simulations ($L=50L_\psi$ rather than $L=500L_\psi$). While this simplifies the structure of the bifurcation diagram, it retains all of the interesting features that can explain the non-linear instabilities in the dynamical simulation presented in~\Cref{fig:microphase_separation_dynamics}.

While we illustrate the full bifurcation diagram for completeness (\Cref{fig7A}), we focus on discussing the subpart of the diagram presented in~\Cref{fig7B}, that zooms in near the region of mixing for intermediate values of $c_H^0$ (\Cref{fig7B}). Interestingly, we find two different types of non-linear instabilities. The first is associated with the appearance of the spatial mode $r=2$, which is characterised via a fold (or type-II) bifurcation~\cite{alma991025286663307026} (see point A in~\Cref{fig7B,fig7C}). Dynamical simulations reveal that of the two branches originating from (A), the one that persists at low $c_H^0$ is stable, while the one that eventually crosses the trivial solution via a subcritical pitchfork bifurcation is unstable. Similar bifurcations are observed in strong polyelectrolyte systems
and can be studied analytically via weakly non-linear analysis~\cite{gavish2017PRE,gavish_spatially_2017}. Significantly different instead is the behaviour of the eigenmode $r=1$, for which we find the coexistence of multiple branches. Branch (III), which bridges the two boundaries of the stability region, is unstable and plays an important role only in determining the critical value $c_H^0$ for re-entrant phase separation. Of the two that exist in dilute conditions (see (I) and (II) in~\Cref{fig7B}), branch (II) corresponds to standard macrophase separation (see point C), while branch (I) is modulated by the non-linear interactions interaction of mode $r=1$ and mode $r=3$ (see point B). Dynamical simulations reveal that branch (I) is unstable to numerical noise while (II) is stable. Eventually, branch (I) disappears by crossing the homogeneous solution; this is the point at which the linear stability analysis predicted microphase separation. However, the top branch persists beyond the end of the demixing region leading to a regime of multistability. First, a region of bistability between the micro- and macro-phase separated states (see shaded red area); then a region of tristability between the former two and the homogenous state; and finally, for even larger value of $c_H^0$, a region of bistability between the mixed and macro-phase separated solutions. This explains the possibility of nucleation of micro-phase separated states observed in~\Cref{fig:microphase_separation_dynamics}. 

\section{Conclusion}
\label{sec:conclusion}
Building on previous studies~\cite{celora2023,zheng2024}, we presented a theoretical framework to investigate the phase separation dynamics of weak polyelectrolytes in solutions containing protons and salt. This framework extends standard kinetic models for the coacervation of strong polyelectrolytes by incorporating charge regulation mechanisms; specifically, we account for the rapid protonation and deprotonation dynamics of binding sites along polymer chains, which regulate their charge state in response to the local solution acidity. The movement of components in the solution is guided by diffusive dynamics, via thermodynamics fluxes, and the electric field, which is accounted for via \RB{a} mean-field description.

We use our framework to characterize both the kinetic and equilibrium phase behaviour of weak polyelectrolytes. To do so, we integrate dynamical simulations, linear stability analysis, and numerical bifurcation analysis. These reveal the rich behaviour of weak polyelectrolytes and interesting features that distinguish their behaviour from the one of strong polyelectrolytes. Our results reveal the formation of condensed domains and the emergence of persistent pH gradients across liquid interfaces, which remain stable over long timescales (\Cref{fig2:charge dynamics}). These sustained gradients lead to a continuous variation in the charge state of the polymer across the interface and align with experimental observations~\cite{ausserwoger_biomolecular_2024}.
We show that the protons concentration shapes the demixing of weak polyelectrolytes in a non-linear manner, and, in the case of phase separation, it can impact the coarsening dynamics and the long-term shape and number of condensed regions. Interestingly, we observe non-standard effects-- such as reverse Ostwald ripening-- that lead to the stabilization of states in which multiple condensed regions coexist. This is due to the effect of long-range Coloumb forces that are driven by the asymmetric distribution of charges across the diffusive interfaces between condensed and dilute phases. 

Linear stability analysis around homogeneous states confirms the non-linear dependence of homogeneous solution stability on solution acidity and further reveals how acidity modulates the type of transition (or bifurcation) that leads to spontaneous demixing of weak polyelectrolytes. On the one hand, destabilisation of the homogeneous state at a high level of acidity is modulated by a large-scale instability which is typical in models of macrophase separation. In contrast, the destabilisation of the homogeneous solutions at low levels of acidity is associated with a small-scale (or finite length) instability, which results in stable microphase separation in which multiple condensed domains of the same characteristic size coexist and are stabilised by the electric-field driven separation of charges across the liquid interface. While micro-phase separation can occur also in solutions of strong polyelectrolytes, our results highlight how, for weak polyelectrolytes, the existence of metastable microphase states can be controlled via modulation of the acid environment, which impact electrostatic forces via the regulation of charges on the polymer. Interestingly, full non-linear analysis of stationary solutions reveals the possibility of metastable regions in the acidity space, where the macrophase, microphase and homogeneous states are all stable introducing additional complexity that, to the best of our knowledge, has not been reported to strong polyelectrolyte systems.  

The present work represents an initial investigation of the complex phase behaviour of weak polyelectrolyte solutions and their ability to self-organise into localised coacervates in an environment-dependent manner. While the bifurcation analysis in~\Cref{sec:bifurcation} reveals interesting transitions in which acidity regulates the coacervation of weak polyelectrolyte system, this study is not exhaustive and motivates further analytical and numerical exploration to fully characterize the patterns predicted by Eqs.~(\ref{eq:equilibria_spatial_dynamics}). This may be achieved by leveraging the variational structure of the problem. As we show in~\Cref{app:spatial dynamics}, Eqs.~(\ref{eq:equilibria_spatial_dynamics}) can be reformulated as a 4D reversible Hamiltonian system,  a class of systems known for exhibiting rich homoclinic unfolding, as exemplified in the well-known Swift–Hohenberg model for pattern formation~\cite{champneys_homoclinic_1998}. From this point of view, a generalisation of the standard Maxwell construction~\cite{greve2024coexistenceuniformoscillatorystates} could be employed to construct such homoclinic (\emph{i.e.}, localised) solutions to better characterise the phase behaviour of weak polyelectrolytes in response to different control parameters (such as salt content and solution acidity). From a pattern formation perspective, it would be interesting to extend our study to non-planar interfaces to reveal the structure, interaction and regulation of weak polyelectrolyte coacervates via full 2D simulations.

The proposed theoretical framework reveals a rich and interesting range of physical behaviours that highlight the critical role of charge regulation mechanisms in shaping the formation and interactions of coacervates. However, further theoretical developments are needed to generalise our findings to liquid-liquid phase separation of biomolecules. In particular, this would require extending our framework to account for the complexity of biopolymers and of the cytoplasm. For example, we have here assumed binding sites along the polymer chain to be identical; however, biomolecules -- such as protein -- are often heterogeneous and consist of distinct amino acids, which can differ both in their binding affinity and/or charge. From this point of view, a natural extension of our work is to consider weak polyampholytes, which are a better theoretical model of proteins. Furthermore, while mean-field theory are extremely useful improving our understanding they are limited as they assume polyelectrolytes are macromolecules. In contrast, proteins often present intrinsically disordered regions, which have been shown to play a significant role in determining intracellular LLPS. The local electrochemical environment and charge state can have a substantial influence on the conformation of such intrinsically disordered regions, hence contributing to the formation and mesoscopic organisation of biomolecular condensates~\cite{davis2025mesoscopicheterogeneitybiomolecularcondensates,Lin2016}. Such conformation-dependency can not be easily captured using mean-field descriptions of phase separation dynamics. It would, therefore, be interesting to complement our results with the study of these effects using molecular simulations~\cite{curk_accelerated_2022,luo2024condensatesizecontrolcharge,davis2025mesoscopicheterogeneitybiomolecularcondensates}.

Overall, our work lays the foundation for a continuum theory of pH-mediated coacervation of weak polyelectrolytes and emphasise the role of charge regulation mechanisms in mediating the dynamics, interactions and structure of these assemblies.


\begin{acknowledgments}
The authors thank Dr Matthew Hennessy and Prof. Sarah Waters for the helpful discussions in the initial phase of this project. GLC acknowledges funding from the UK Engineering and Physical Sciences Research Council (EPSRC), grant number EP/W524335/1, and support from the Mathematical Institute through a Hooke fellowship.  
\end{acknowledgments}

\appendix

\section{Computation of the thermodynamic fluxes.}
\label{app:thermodynamic_fluxes}
To define the fluxes we adopt the Maxwell-Stephan approach~\cite{leonardi_maxwellstefan_2010}, which is based on balancing the friction forces between relative phases with the thermodynamic forces. An important aspect to consider in our system is the presence of heterogeneous phases that require a careful derivation of appropriate effective Maxwell-Stephan conditions. Here we assume that the friction coefficient for heterogeneous components is independent of their charge state. Then, the balance of forces acting on the $i$-th component takes the standard form:
\begin{equation}
    \vec{\mathcal{T}}_i + \sum_{i\neq j} \friction_{ij}(c_j\vec{j}_i-c_i\vec{j}_j)=0, \quad i,j\ \in \left\{\pm,p,s\right\}^2, \label{eq:fluxes}
\end{equation}
where $\friction_{ij}>0$ are the friction coefficients  satisfying the symmetry condition $\friction_{ij}=\friction_{ji}$, $\vec{j}_i$ is the thermodynamic flux for the component $i$ and $\vec{\mathcal{T}}_i$ is the thermodynamic force acting on the component $i$. For homogeneous components, such as salt ions, the thermodynamic force takes the standard form
\begin{equation}
    \vec{\mathcal{T}}_\pm =  c_\pm\nabla \mu^{\text{el}}_\pm.\label{eq:thermodynamic_forces_salt}
 \end{equation}
where $\mu^{\text{el}}_\pm$ is the electrochemical potential (see Eq.~(\ref{eq:mu_m})). For heterogeneous components such as the solvent and the polymer macromolecules, the thermodynamic forces include an additional contribution associated with the presence of variable charges:
\begin{equation}
    \vec{\mathcal{T}}_{m} = \sum_{q=0}^{Q_m}\vec{\mathcal{T}}^z_{m}=c_m \nabla  \mu^{\text{el}}_m + q_mc_m\nabla\mu^{\text{el}}_H.\label{eq:thermodynamic_forces_solvent_and_polymer}
\end{equation}%
Summing over $i$ in Eq.~(\ref{eq:fluxes}), we conclude that the total force $\sum_{i}\mathcal{\vec{Q}}_i$ must vanish. This defines the value of thermodynamic pressure $P$ up to an arbitrary constant $\bar{P}$:
\begin{equation}	
\begin{aligned}
P&=\bar{P}+\bar{f}_{\eff}-\sum_{m}c_m\frac{\partial \bar{f}_{\eff}}{\partial c_m}+\frac{\epsilon}{2}|\nabla \psi|^2 \\
&\qquad\qquad+\sum_{m} \kappa_m  \left(c_m\nabla^2 c_m -\frac{|\nabla c_m|^2}{2}\right),\end{aligned}\label{eq:thermodynamic_pressure}
\end{equation}
Here we assume that the dissipation between all components scales with their molecular volume $\friction_{ij}=\eta\nu_i\nu_j \neq0$.  Then the flux of each component is simply proportional to the thermodynamic force they experience; namely:
\begin{equation}
\vec{j}_m=-\frac{\vec{\mathcal{T}}_m\nu}{\eta\nu_m}.\label{eq:flux_components_except_cH}
 \end{equation} 
The flux of the \ce{H+} ions can be estimated by summing over the flux of \ce{H+} ions transported by both polymer and solvent macromolecules and reads:
\begin{equation}
\begin{aligned}
    \eta \vec{j}_H&=\frac{c_pq_p}{N}\nabla\mu^{\text{el}}_{p}+c_sq_s\nabla\mu^{\text{el}}_s \\&\qquad+ \left(\langle q^2\rangle_s c_s +\langle q^2\rangle_p \frac{c_p}{N}\right)\nabla\mu_H^{\text{el}},
    \end{aligned}\label{eq:flux_cH}
    \end{equation}
where $\langle q^2\rangle_i$ indicates the second moment of the charge distribution associated with the component $i$; namely $\langle q^2\rangle_i=\sum_{q=0}^{Q_i}q^2\pi^q_i$. Eqs.~(\ref{eq:flux_components_except_cH})-(\ref{eq:flux_cH}) can be written in compact form given in the main text \[\vec{j}_m=\sum_{K\in\left\{p,s,H,+,-\right\}}\mathcal{M}_{mK}\nabla\mu_K\] by defining the mobility matrix $\mathcal{M}$ as
\begin{equation}
    \mathcal{M}=D\beta \begin{bmatrix} \dfrac{c_p}{N} &0 & \dfrac{q_pc_p}{N} &0 &0\\[4pt]
    0 &c_s& q_sc_s &0 &0\\[2pt]
     \dfrac{q_pc_p}{N}  & q_sc_s & \langle q^2\rangle_s c_s +  \dfrac{\langle q^2\rangle_p c_p}{N} &0 &0\\[4pt]
     0 & 0 &0 & c_+ &0\\[2pt]
    0 & 0 &0 & 0 &c_-\\
\end{bmatrix}\label{eq:motility_matrix}
\end{equation}
based on the following order for the indices: $p$, $s$, $H$, $+$ and $-$. In writing Eq.~(\ref{eq:motility_matrix}), $\beta=(k_BT)^{-1}$ is a constant and $D$ is the diffusion coefficient, which is related to the friction coefficient $\eta$ in Eqs.~(\ref{eq:flux_components_except_cH})-(\ref{eq:flux_cH}) via $D=k_BT/\eta$.

\section{Stability analysis}
\label{app:stability analysis}
The analytical steps used in deriving the conditions for the linear stability of homogenous, electroneutral, weak polyelectrolyte mixtures are detailed in SM. We note that in the derivation we use the non-dimensional form of the modelling equations (see SM).

Ultimately, the stability of homogeneous and electroneutral states \[\begin{pmatrix}c_p\\c_s\\c_+\\c_-\\c_H\\\psi\end{pmatrix}=\begin{pmatrix}
    c_p^0\\
    1-c_p^0-2c_+^0-c_H^0\\
    c^0_++c^0_H\\
    c_H^0\\
    0
\end{pmatrix}\] is connected to the existence and number of real roots of the following fourth-order polynomial:
\begin{align}
    \ell^2_\kappa (\pi \hat{r})^4 - 2\left[\frac{\ell^2_\kappa \tilde{F}_{\psi\psi}^0-\tilde{F}^0_{\phi\phi}}{2}\right](\pi \hat{r})^2 - \text{det}(\mathcal{H}^0_{\tilde{F}})= 0,\label{eq:polynomial_r}
\end{align}
where $\text{det}(\mathcal{H}^0_{\tilde{F}})=\tilde{F}^0_{\psi\psi}\tilde{F}^0_{\phi\phi}-(\tilde{F}^0_{\psi\phi})^2$ is the determinant of the Hessian of the functional $\tilde{F}$ (see Eq.~(\ref{eq:energ_hamiltonian})) evaluated at the equilibrium point $(\phi,\psi;\vec{C}_0)=(\nu N c_p^0,0;[\nu Nc_p^0,c_+^0,c_H^0])$
\begin{subequations}
\begin{align}
F_{\psi\psi}&=q_p^0 c_p^0\nu\left(1+\frac{\nu q^0_p c_p^0}{1-N \nu c^0_p}\right)-2 \nu c_+^0-2 \nu c_H^0- \nu c_p^0\sigma_p^0,\\
    F_{\phi\phi}&=\frac{(1-\chi \nu c^0_s)^2}{1- \nu Nc_p^0}+\frac{\nu}{(N\nu)^2 c_p^0}-\chi^2\nu c^0_s,\\
    F_{\psi\phi}&=\frac{q^0_p}{N}\frac{1-\chi\nu^2 Nc_p^0c_s^0}{1-N\nu c_p^0}-\chi\nu(c_H-q_pc_p^0).
\end{align}\label{eq:coeff_free_energy_hessian}%
\end{subequations}

If we denote by \[\Delta=\left(\frac{\ell^2_\kappa \tilde{F}_{\psi\psi}^0-\tilde{F}^0_{\phi\phi}}{2}\right)^2+\ell_\kappa^2\text{det}(\mathcal{H}_{\tilde{F}}),\] then we distinguish three regimes:
\begin{itemize}
    \item[(\rm{I})] When $\det\mathcal{H}_{\tilde{F}}^0>0$, Eq.~(\ref{eq:polynomial_r}) has two real roots $\left\{\pm \hat{r}_{\CH}\right\}$,
    \[\hat{r}_{\CH}=\frac{1}{\pi\ell_\kappa}\sqrt{\left(\frac{\ell^2_\kappa \tilde{F}_{\psi\psi}^0-\tilde{F}^0_{\phi\phi}}{2}\right)+\sqrt{\Delta}},\]
    the homogenous solution is unstable in time to spatial perturbations with modes in the interval $rL_\psi\in(0,\hat{r}_{\text{CH}})$. 
  \item[(\rm{II})] When $\det\mathcal{H}_{\tilde{F}}^0<0$, $\Delta>0$ and $\ell^2_\kappa \tilde{F}_{\psi\psi}^0-\tilde{F}^0_{\phi\phi}>0$, Eq.~(\ref{eq:polynomial_r}) has four real roots $\left\{\pm \hat{r}^+_{T},\pm \hat{r}^-_T\right\}$
  \[\hat{r}^\pm_T=\frac{1}{\pi\ell_\kappa}\sqrt{\left(\frac{\ell^2_\kappa \tilde{F}_{\psi\psi}^0-\tilde{F}^0_{\phi\phi}}{2}\right)\pm\sqrt{\Delta}},\]the homogenous solution $\left\{\phi_m^0\right\}$ is unstable in time to spatial perturbations with modes in the interval $rL_\psi\in(\hat{r}_T^-,\hat{r}^+_{T})$. 
    \item[(\rm{III})] Otherwise, Eq.~(\ref{eq:polynomial_r}) has no real roots, the homogenous solution is linearly stable in time to spatial perturbations. 
\end{itemize}

While regimes (\rm{I}) and (\rm{III}) are common to models of phase separation (such as classical Cahn-Hilliard models), the regime (\rm{II}), where the homogeneous state is unstable to small-scale but not long-scale perturbations, is associated with microphase separation.
We note that the condition $\ell^2_\kappa \tilde{F}_{\psi\psi}^0-\tilde{F}^0_{\phi\phi}>0$ necessary to be in regime (\rm{II}) can be expressed in terms of the characteristic length scale $\tilde{\ell}_\kappa=\ell_\kappa\sqrt{F_{\psi\psi}/F_{\phi\phi}}$ and the coefficient $a=-F_{\phi\phi}$ defined by Eq.~(\ref{eq:moder_T}) as
\[a(1-\tilde{\ell}^2_\kappa)>0.\]
Taking $a>0$, this highlights how the length scale $\tilde{\ell}_\kappa$ mediates the transition to the regime in which microphase separation is possible.

\paragraph{Spinodal curves.}
Spinodal curves are identified as the curves that determine the onset of the stability of the homogeneous solution. This corresponds to transitions from the regime (\rm{III}) described above to either regime (\rm{I}) or (\rm{II}). While the former corresponds to a standard Cahn-Hilliard (or large-scale) instability the second case leads to a Turing-like (or short-scale) instability~\cite{frohoff-hulsmann_non-reciprocity_2023}. In light of this, the boundary of the Cahn-Hilliard and Turing spinodals are defined respectively by the conditions
\begin{equation}
\begin{cases}
\text{det}(\mathcal{H}_{\tilde{F}})=0,\\
\ell_\kappa^2\tilde{F}_{\psi\psi}^0<\tilde{F}_{\phi\phi}^0,
\end{cases}
\end{equation}
and
\begin{equation}\label{cond:Turing}
\begin{cases}
\left(\ell_\kappa^2\tilde{F}^0_{\psi\psi}-\tilde{F}^0_{\phi\phi}\right)^2+4\ell_\kappa^2\text{det}(\mathcal{H}_{\tilde{F}})=0,\\[2pt]
\ell_\kappa^2\tilde{F}_{\psi\psi}^0>\tilde{F}_{\phi\phi}^0.
\end{cases}
\end{equation}
We can further characterise the critical wavenumber $r_T^*$ selected at the onset of the Turing bifurcation, by substituting condition~(\ref{cond:Turing}) into the definition of $\hat{r}_T$ given above:
\begin{equation}
r^*_T = \frac{1}{L_\psi\pi\ell_\kappa}\sqrt{\frac{\ell_\kappa^2\tilde{F}_{\psi\psi}^0-\tilde{F}_{\phi\phi}^0}{2}}.\label{eq:mode_unstable}
\end{equation}
Note that $r^*_T$ can also be expressed in terms of the characteristic length scale $\tilde{\ell}_\kappa$ introduced in~\Cref{sec:spinodal} 

\begin{equation}
r_T^*=\frac{1}{L_\psi\pi\ell_\kappa}\sqrt{a \frac{1-\tilde{\ell}_\kappa}{2}},
\end{equation}%

\noindent where $a$ is as defined in Eq.~(\ref{eq:moder_T}).

A similar result holds for the case of strong polyelectrolyte solutions. Assuming that the polymer is characterised by fixed charges $Q$, then the spinoidals are still defined by the conditions given above provided that $q_p^0\rightarrow Q$ and $\sigma_p^0\rightarrow 0$ in Eqs.~(\ref{eq:coeff_free_energy_hessian}); furthermore, the possible homogeneous states are now constrained by electroneutrality which requires $\nu c_H^0\in [Q\nu c_p^0,1)$.

\section{Spatial Dynamics.}
\label{app:spatial dynamics}
Following the steps detailed in the SM, we reduce the problem of computing stationary solutions of the model presented in~\Cref{sec:model} to a system of two coupled, non-linear, boundary value problems~(\ref{eq:equilibria_spatial_dynamics}). Here we show how Eqs.~(\ref{eq:equilibria_spatial_dynamics}) can be rewritten as a 4D reversible Hamiltonian system so that localised solutions can be described as homoclinic orbits of such a system~\cite{champneys_homoclinic_1998} 
in terms of the stationary (rescaled) electric potential $\Psi=\Psi(x)$ and the order parameter $\Phi=\Phi(x)$. To do so, we introduce the moments variables, $q_\psi$ and $q_\phi$ and rewrite Eq.~(\ref{eq:equilibria_spatial_dynamics}) as:
\begin{subequations}\label{eq:spatial_hamiltonian_system}
\begin{align}
  \begin{cases} 
   \Psi'&=-q_\psi,\\
    \ell^2_\kappa\Phi'&=q_\phi,\\
     q'_\psi &= \tilde{F}_\psi(\Phi,\Psi;\vec{C}_0),\\
 q'_\phi &=  \tilde{F}_\phi (\Phi,\Psi;\vec{C}_0), 
\end{cases}
\end{align}
\end{subequations}%
where the prime is used to denote the spatial derivative and $\tilde{F}_\phi$, $\tilde{F}_\psi$ are the partial derivatives of a functional $\tilde{F}$ defined by Eq.~(\ref{eq:energ_hamiltonian}). As mentioned in the main text, it corresponds to the free energy of two immiscible non-ideal fluids, where the non-ideal nature is described by the activity coefficients:
\begin{widetext}
    \begin{subequations}\label{eq:activity}
    \begin{align}
    \gamma_s(\phi,\psi;\vec{C_0})&=\frac{1}{1+e^{-\psi+\mu^\Delta_+(\vec{C}_0)+\chi\phi}+e^{-\psi+\mu^\Delta_H(\vec{C}_0)}+e^{\psi+\mu^\Delta_-(\vec{C}_0)+\chi\phi}},\\
    \gamma_p(\psi)&=e^{N\mu^\Delta_p(\vec{C}_0)}\left(\sum_{q=0}^{Q_p} e^{-\beta u_{\CR}^q-q (\psi-\mu^\Delta_H(\vec{C}_0))}\right)^{-1},\label{eq:gamma_p}
\end{align}
\end{subequations}
\end{widetext}
where the constants $\mu_{\pm,p}^\Delta$ indicate the equilibrium chemical potential difference between salt ions and solvent, and polymer and solvent, while $\mu^\Delta_H$ is the equilibrium chemical potential of protons (both expressed in non-dimensional units). These are unknowns and can be determined by imposing electroneutrality and the spatially-averaged concentrations of each species; more details are given in the SM.
The Hamiltonian
\[\mathcal{H}=-\frac{1}{2} q_\psi^2+\frac{1}{2}\left(\frac{q_\phi}{\ell_\kappa}\right)^2-\tilde{F}(\Phi,\Psi;\vec{C}_0)\]
generates Eqs.~(\ref{eq:spatial_hamiltonian_system}) and it is a first integral along trajectories in the $(\Psi,\Phi,q_\psi,q_\phi)$ space. 


\bibliography{charge-regulation}

\begin{thebibliography}{52}%
\makeatletter
\providecommand \@ifxundefined [1]{%
 \@ifx{#1\undefined}
}%
\providecommand \@ifnum [1]{%
 \ifnum #1\expandafter \@firstoftwo
 \else \expandafter \@secondoftwo
 \fi
}%
\providecommand \@ifx [1]{%
 \ifx #1\expandafter \@firstoftwo
 \else \expandafter \@secondoftwo
 \fi
}%
\providecommand \natexlab [1]{#1}%
\providecommand \enquote  [1]{``#1''}%
\providecommand \bibnamefont  [1]{#1}%
\providecommand \bibfnamefont [1]{#1}%
\providecommand \citenamefont [1]{#1}%
\providecommand \href@noop [0]{\@secondoftwo}%
\providecommand \href [0]{\begingroup \@sanitize@url \@href}%
\providecommand \@href[1]{\@@startlink{#1}\@@href}%
\providecommand \@@href[1]{\endgroup#1\@@endlink}%
\providecommand \@sanitize@url [0]{\catcode `\\12\catcode `\$12\catcode
  `\&12\catcode `\#12\catcode `\^12\catcode `\_12\catcode `\%12\relax}%
\providecommand \@@startlink[1]{}%
\providecommand \@@endlink[0]{}%
\providecommand \url  [0]{\begingroup\@sanitize@url \@url }%
\providecommand \@url [1]{\endgroup\@href {#1}{\urlprefix }}%
\providecommand \urlprefix  [0]{URL }%
\providecommand \Eprint [0]{\href }%
\providecommand \doibase [0]{https://doi.org/}%
\providecommand \selectlanguage [0]{\@gobble}%
\providecommand \bibinfo  [0]{\@secondoftwo}%
\providecommand \bibfield  [0]{\@secondoftwo}%
\providecommand \translation [1]{[#1]}%
\providecommand \BibitemOpen [0]{}%
\providecommand \bibitemStop [0]{}%
\providecommand \bibitemNoStop [0]{.\EOS\space}%
\providecommand \EOS [0]{\spacefactor3000\relax}%
\providecommand \BibitemShut  [1]{\csname bibitem#1\endcsname}%
\let\auto@bib@innerbib\@empty
\bibitem [{\citenamefont {Hyman}\ \emph {et~al.}(2014)\citenamefont {Hyman},
  \citenamefont {Weber},\ and\ \citenamefont
  {Jülicher}}]{hyman_liquid-liquid_2014}%
  \BibitemOpen
  \bibfield  {author} {\bibinfo {author} {\bibfnamefont {A.~A.}\ \bibnamefont
  {Hyman}}, \bibinfo {author} {\bibfnamefont {C.~A.}\ \bibnamefont {Weber}},\
  and\ \bibinfo {author} {\bibfnamefont {F.}~\bibnamefont {Jülicher}},\
  }\bibfield  {title} {\bibinfo {title} {Liquid-{Liquid} {Phase} {Separation}
  in {Biology}},\ }\href
  {https://doi.org/10.1146/annurev-cellbio-100913-013325} {\bibfield  {journal}
  {\bibinfo  {journal} {Annual Review of Cell and Developmental Biology}\
  }\textbf {\bibinfo {volume} {30}},\ \bibinfo {pages} {39} (\bibinfo {year}
  {2014})}\BibitemShut {NoStop}%
\bibitem [{\citenamefont {Riback}\ \emph {et~al.}(2020)\citenamefont {Riback},
  \citenamefont {Zhu}, \citenamefont {Ferrolino}, \citenamefont {Tolbert},
  \citenamefont {Mitrea}, \citenamefont {Sanders}, \citenamefont {Wei},
  \citenamefont {Kriwacki},\ and\ \citenamefont {Brangwynne}}]{Riback2020}%
  \BibitemOpen
  \bibfield  {author} {\bibinfo {author} {\bibfnamefont {J.~A.}\ \bibnamefont
  {Riback}}, \bibinfo {author} {\bibfnamefont {L.}~\bibnamefont {Zhu}},
  \bibinfo {author} {\bibfnamefont {M.~C.}\ \bibnamefont {Ferrolino}}, \bibinfo
  {author} {\bibfnamefont {M.}~\bibnamefont {Tolbert}}, \bibinfo {author}
  {\bibfnamefont {D.~M.}\ \bibnamefont {Mitrea}}, \bibinfo {author}
  {\bibfnamefont {D.~W.}\ \bibnamefont {Sanders}}, \bibinfo {author}
  {\bibfnamefont {M.-T.}\ \bibnamefont {Wei}}, \bibinfo {author} {\bibfnamefont
  {R.~W.}\ \bibnamefont {Kriwacki}},\ and\ \bibinfo {author} {\bibfnamefont
  {C.~P.}\ \bibnamefont {Brangwynne}},\ }\bibfield  {title} {\bibinfo {title}
  {Composition-dependent thermodynamics of intracellular phase separation},\
  }\href {https://doi.org/10.1038/s41586-020-2256-2} {\bibfield  {journal}
  {\bibinfo  {journal} {Nature}\ }\textbf {\bibinfo {volume} {581}},\ \bibinfo
  {pages} {1476} (\bibinfo {year} {2020})}\BibitemShut {NoStop}%
\bibitem [{\citenamefont {Villegas}\ \emph {et~al.}(2022)\citenamefont
  {Villegas}, \citenamefont {Heidenreich},\ and\ \citenamefont
  {Levy}}]{Villegas2022}%
  \BibitemOpen
  \bibfield  {author} {\bibinfo {author} {\bibfnamefont {J.}~\bibnamefont
  {Villegas}}, \bibinfo {author} {\bibfnamefont {M.}~\bibnamefont
  {Heidenreich}},\ and\ \bibinfo {author} {\bibfnamefont {E.~D.}\ \bibnamefont
  {Levy}},\ }\bibfield  {title} {\bibinfo {title} {Molecular and environmental
  determinants of biomolecular condensate formation},\ }\href
  {https://doi.org/10.1038/s41589-022-01175-4} {\bibfield  {journal} {\bibinfo
  {journal} {Nature Chemical Biology}\ }\textbf {\bibinfo {volume} {18}},\
  \bibinfo {pages} {1319} (\bibinfo {year} {2022})}\BibitemShut {NoStop}%
\bibitem [{\citenamefont {Oparin}(1965)}]{oparin_origin_1965}%
  \BibitemOpen
  \bibfield  {author} {\bibinfo {author} {\bibfnamefont {A.}~\bibnamefont
  {Oparin}},\ }\href@noop {} {\emph {\bibinfo {title} {The {Origin} of
  {Life}}}},\ Dover books on biology and medicine\ (\bibinfo  {publisher}
  {Dover},\ \bibinfo {year} {1965})\BibitemShut {NoStop}%
\bibitem [{\citenamefont {{Bartolucci, Giacomo and Cala\c{c}a Serr\~{a}o,
  Adriana and Schwintek, Philipp and Kühnlein, Alexandra and Rana, Yash and
  Janto, Philipp and Hofer, Dorothea and Mast, Christof B. and Braun, Dieter
  and Weber, Christoph A.}}(2023)}]{bartolucci_sequence_2023}%
  \BibitemOpen
  \bibfield  {author} {\bibinfo {author} {\bibnamefont {{Bartolucci, Giacomo
  and Cala\c{c}a Serr\~{a}o, Adriana and Schwintek, Philipp and Kühnlein,
  Alexandra and Rana, Yash and Janto, Philipp and Hofer, Dorothea and Mast,
  Christof B. and Braun, Dieter and Weber, Christoph A.}}},\ }\bibfield
  {title} {\bibinfo {title} {Sequence self-selection by cyclic phase
  separation},\ }\href {https://doi.org/10.1073/pnas.2218876120} {\bibfield
  {journal} {\bibinfo  {journal} {Proceedings of the National Academy of
  Sciences}\ }\textbf {\bibinfo {volume} {120}},\ \bibinfo {pages}
  {e2218876120} (\bibinfo {year} {2023})}\BibitemShut {NoStop}%
\bibitem [{\citenamefont {Fox}(1976)}]{fox_evolutionary_1976}%
  \BibitemOpen
  \bibfield  {author} {\bibinfo {author} {\bibfnamefont {S.~W.}\ \bibnamefont
  {Fox}},\ }\bibfield  {title} {\bibinfo {title} {{The evolutionary
  significance of phase-separated microsystems}},\ }\href
  {https://doi.org/10.1007/BF01218513} {\bibfield  {journal} {\bibinfo
  {journal} {Origins of life}\ }\textbf {\bibinfo {volume} {7}},\ \bibinfo
  {pages} {49} (\bibinfo {year} {1976})}\BibitemShut {NoStop}%
\bibitem [{\citenamefont {{G. L. Celora, M. G. Hennessy, A. M\"{u}nch, B.
  Wagner, S. L. Waters}}(2022)}]{CeloraPL1}%
  \BibitemOpen
  \bibfield  {author} {\bibinfo {author} {\bibnamefont {{G. L. Celora, M. G.
  Hennessy, A. M\"{u}nch, B. Wagner, S. L. Waters}}},\ }\bibfield  {title}
  {\bibinfo {title} {A kinetic model of a polyelectrolyte gel undergoing phase
  separation},\ }\href
  {https://doi.org/https://doi.org/10.1016/j.jmps.2021.104771} {\bibfield
  {journal} {\bibinfo  {journal} {Journal of the Mechanics and Physics of
  Solids}\ }\textbf {\bibinfo {volume} {160}},\ \bibinfo {pages} {104771}
  (\bibinfo {year} {2022})}\BibitemShut {NoStop}%
\bibitem [{\citenamefont {Grzetic}\ \emph {et~al.}(2021)\citenamefont
  {Grzetic}, \citenamefont {Delaney},\ and\ \citenamefont
  {Fredrickson}}]{grzetic_electrostatic_2021}%
  \BibitemOpen
  \bibfield  {author} {\bibinfo {author} {\bibfnamefont {D.~J.}\ \bibnamefont
  {Grzetic}}, \bibinfo {author} {\bibfnamefont {K.~T.}\ \bibnamefont
  {Delaney}},\ and\ \bibinfo {author} {\bibfnamefont {G.~H.}\ \bibnamefont
  {Fredrickson}},\ }\bibfield  {title} {\bibinfo {title} {Electrostatic
  {Manipulation} of {Phase} {Behavior} in {Immiscible} {Charged} {Polymer}
  {Blends}},\ }\href {https://doi.org/10.1021/acs.macromol.1c00095} {\bibfield
  {journal} {\bibinfo  {journal} {Macromolecules}\ }\textbf {\bibinfo {volume}
  {54}},\ \bibinfo {pages} {2604} (\bibinfo {year} {2021})}\BibitemShut
  {NoStop}%
\bibitem [{\citenamefont {Gavish}\ \emph {et~al.}(2017)\citenamefont {Gavish},
  \citenamefont {Versano},\ and\ \citenamefont
  {Yochelis}}]{gavish_spatially_2017}%
  \BibitemOpen
  \bibfield  {author} {\bibinfo {author} {\bibfnamefont {N.}~\bibnamefont
  {Gavish}}, \bibinfo {author} {\bibfnamefont {I.}~\bibnamefont {Versano}},\
  and\ \bibinfo {author} {\bibfnamefont {A.}~\bibnamefont {Yochelis}},\
  }\bibfield  {title} {\bibinfo {title} {Spatially localized self-assembly
  driven by electrically charged phase separation},\ }\href
  {https://doi.org/10.1137/16M1105876} {\bibfield  {journal} {\bibinfo
  {journal} {{SIAM} Journal on Applied Dynamical Systems}\ }\textbf {\bibinfo
  {volume} {16}},\ \bibinfo {pages} {1946} (\bibinfo {year}
  {2017})}\BibitemShut {NoStop}%
\bibitem [{\citenamefont {Hennessy}\ \emph {et~al.}(2024)\citenamefont
  {Hennessy}, \citenamefont {Celora}, \citenamefont {Waters}, \citenamefont
  {Münch},\ and\ \citenamefont {Wagner}}]{hennessy_breakdown_2024}%
  \BibitemOpen
  \bibfield  {author} {\bibinfo {author} {\bibfnamefont {M.~G.}\ \bibnamefont
  {Hennessy}}, \bibinfo {author} {\bibfnamefont {G.~L.}\ \bibnamefont
  {Celora}}, \bibinfo {author} {\bibfnamefont {S.~L.}\ \bibnamefont {Waters}},
  \bibinfo {author} {\bibfnamefont {A.}~\bibnamefont {Münch}},\ and\ \bibinfo
  {author} {\bibfnamefont {B.}~\bibnamefont {Wagner}},\ }\bibfield  {title}
  {\bibinfo {title} {Breakdown of electroneutrality in polyelectrolyte gels},\
  }\href {https://doi.org/10.1017/S0956792523000244} {\bibfield  {journal}
  {\bibinfo  {journal} {European Journal of Applied Mathematics}\ }\textbf
  {\bibinfo {volume} {35}},\ \bibinfo {pages} {359} (\bibinfo {year}
  {2024})}\BibitemShut {NoStop}%
\bibitem [{\citenamefont {Muthukumar}(2017)}]{muthukumar_50th_2017}%
  \BibitemOpen
  \bibfield  {author} {\bibinfo {author} {\bibfnamefont {M.}~\bibnamefont
  {Muthukumar}},\ }\bibfield  {title} {\bibinfo {title} {50th {Anniversary}
  {Perspective}: {A} {Perspective} on {Polyelectrolyte} {Solutions}},\ }\href
  {https://doi.org/10.1021/acs.macromol.7b01929} {\bibfield  {journal}
  {\bibinfo  {journal} {Macromolecules}\ }\textbf {\bibinfo {volume} {50}},\
  \bibinfo {pages} {9528} (\bibinfo {year} {2017})}\BibitemShut {NoStop}%
\bibitem [{\citenamefont {Sing}\ and\ \citenamefont
  {Perry}(2020)}]{sing_recent_2020}%
  \BibitemOpen
  \bibfield  {author} {\bibinfo {author} {\bibfnamefont {C.~E.}\ \bibnamefont
  {Sing}}\ and\ \bibinfo {author} {\bibfnamefont {S.~L.}\ \bibnamefont
  {Perry}},\ }\bibfield  {title} {\bibinfo {title} {Recent progress in the
  science of complex coacervation},\ }\href
  {https://doi.org/10.1039/D0SM00001A} {\bibfield  {journal} {\bibinfo
  {journal} {Soft Matter}\ }\textbf {\bibinfo {volume} {16}},\ \bibinfo {pages}
  {2885} (\bibinfo {year} {2020})}\BibitemShut {NoStop}%
\bibitem [{\citenamefont {Overbeek}\ and\ \citenamefont
  {Voorn}(1957)}]{VO_theory}%
  \BibitemOpen
  \bibfield  {author} {\bibinfo {author} {\bibfnamefont {J.~T.~G.}\
  \bibnamefont {Overbeek}}\ and\ \bibinfo {author} {\bibfnamefont {M.~J.}\
  \bibnamefont {Voorn}},\ }\bibfield  {title} {\bibinfo {title} {Phase
  separation in polyelectrolyte solutions. theory of complex coacervation},\
  }\href {https://doi.org/https://doi.org/10.1002/jcp.1030490404} {\bibfield
  {journal} {\bibinfo  {journal} {Journal of Cellular and Comparative
  Physiology}\ }\textbf {\bibinfo {volume} {49}},\ \bibinfo {pages} {7}
  (\bibinfo {year} {1957})}\BibitemShut {NoStop}%
\bibitem [{\citenamefont {Rumyantsev}\ \emph {et~al.}(2018)\citenamefont
  {Rumyantsev}, \citenamefont {Kramarenko},\ and\ \citenamefont
  {Borisov}}]{rumyantsev_microphase_2018}%
  \BibitemOpen
  \bibfield  {author} {\bibinfo {author} {\bibfnamefont {A.~M.}\ \bibnamefont
  {Rumyantsev}}, \bibinfo {author} {\bibfnamefont {E.~Y.}\ \bibnamefont
  {Kramarenko}},\ and\ \bibinfo {author} {\bibfnamefont {O.~V.}\ \bibnamefont
  {Borisov}},\ }\bibfield  {title} {\bibinfo {title} {Microphase {Separation}
  in {Complex} {Coacervate} {Due} to {Incompatibility} between {Polyanion} and
  {Polycation}},\ }\href {https://doi.org/10.1021/acs.macromol.8b00721}
  {\bibfield  {journal} {\bibinfo  {journal} {Macromolecules}\ }\textbf
  {\bibinfo {volume} {51}},\ \bibinfo {pages} {6587} (\bibinfo {year}
  {2018})}\BibitemShut {NoStop}%
\bibitem [{\citenamefont {Majee}\ \emph {et~al.}(2024)\citenamefont {Majee},
  \citenamefont {Weber},\ and\ \citenamefont {J\"ulicher}}]{majee_charge_2023}%
  \BibitemOpen
  \bibfield  {author} {\bibinfo {author} {\bibfnamefont {A.}~\bibnamefont
  {Majee}}, \bibinfo {author} {\bibfnamefont {C.~A.}\ \bibnamefont {Weber}},\
  and\ \bibinfo {author} {\bibfnamefont {F.}~\bibnamefont {J\"ulicher}},\
  }\bibfield  {title} {\bibinfo {title} {Charge separation at liquid
  interfaces},\ }\href {https://doi.org/10.1103/PhysRevResearch.6.033138}
  {\bibfield  {journal} {\bibinfo  {journal} {Phys. Rev. Res.}\ }\textbf
  {\bibinfo {volume} {6}},\ \bibinfo {pages} {033138} (\bibinfo {year}
  {2024})}\BibitemShut {NoStop}%
\bibitem [{\citenamefont {Calvert}(2009)}]{calvert_hydrogels_2009}%
  \BibitemOpen
  \bibfield  {author} {\bibinfo {author} {\bibfnamefont {P.}~\bibnamefont
  {Calvert}},\ }\bibfield  {title} {\bibinfo {title} {Hydrogels for {Soft}
  {Machines}},\ }\href {https://doi.org/10.1002/adma.200800534} {\bibfield
  {journal} {\bibinfo  {journal} {Advanced Materials}\ }\textbf {\bibinfo
  {volume} {21}},\ \bibinfo {pages} {743} (\bibinfo {year} {2009})}\BibitemShut
  {NoStop}%
\bibitem [{\citenamefont {Li}(2009)}]{li_smart_2009}%
  \BibitemOpen
  \bibfield  {author} {\bibinfo {author} {\bibfnamefont {H.}~\bibnamefont
  {Li}},\ }\href {https://doi.org/10.1007/978-3-642-02368-2} {\emph {\bibinfo
  {title} {Smart {Hydrogel} {Modelling}}}}\ (\bibinfo  {publisher} {Springer},\
  \bibinfo {address} {Berlin, Heidelberg},\ \bibinfo {year} {2009})\BibitemShut
  {NoStop}%
\bibitem [{\citenamefont {Kolan}\ \emph {et~al.}(2025)\citenamefont {Kolan},
  \citenamefont {Kozawa}, \citenamefont {Weitzer}, \citenamefont {Wnek},\ and\
  \citenamefont {Mussel}}]{kolan_propagation_2025}%
  \BibitemOpen
  \bibfield  {author} {\bibinfo {author} {\bibfnamefont {D.}~\bibnamefont
  {Kolan}}, \bibinfo {author} {\bibfnamefont {S.}~\bibnamefont {Kozawa}},
  \bibinfo {author} {\bibfnamefont {D.}~\bibnamefont {Weitzer}}, \bibinfo
  {author} {\bibfnamefont {G.}~\bibnamefont {Wnek}},\ and\ \bibinfo {author}
  {\bibfnamefont {M.}~\bibnamefont {Mussel}},\ }\bibfield  {title} {\bibinfo
  {title} {Propagation of a {Chemo}-{Mechanical} {Phase} {Boundary} in
  {Polyacrylate} {Gels}},\ }\href
  {https://doi.org/10.1016/j.polymer.2025.128039} {\bibfield  {journal}
  {\bibinfo  {journal} {Polymer}\ ,\ \bibinfo {pages} {128039}} (\bibinfo
  {year} {2025})}\BibitemShut {NoStop}%
\bibitem [{\citenamefont {Wnek}\ \emph {et~al.}(2022)\citenamefont {Wnek},
  \citenamefont {Costa},\ and\ \citenamefont
  {Kozawa}}]{wnek_bio-mimicking_2022}%
  \BibitemOpen
  \bibfield  {author} {\bibinfo {author} {\bibfnamefont {G.~E.}\ \bibnamefont
  {Wnek}}, \bibinfo {author} {\bibfnamefont {A.~C.~S.}\ \bibnamefont {Costa}},\
  and\ \bibinfo {author} {\bibfnamefont {S.~K.}\ \bibnamefont {Kozawa}},\
  }\bibfield  {title} {\bibinfo {title} {Bio-{Mimicking}, {Electrical}
  {Excitability} {Phenomena} {Associated} {With} {Synthetic} {Macromolecular}
  {Systems}: {A} {Brief} {Review} {With} {Connections} to the {Cytoskeleton}
  and {Membraneless} {Organelles}},\ }\bibfield  {journal} {\bibinfo  {journal}
  {Frontiers in Molecular Neuroscience}\ }\textbf {\bibinfo {volume} {15}},\
  \href {https://doi.org/10.3389/fnmol.2022.830892} {10.3389/fnmol.2022.830892}
  (\bibinfo {year} {2022})\BibitemShut {NoStop}%
\bibitem [{\citenamefont {Tasaki}(2002)}]{tasaki_spread_2002}%
  \BibitemOpen
  \bibfield  {author} {\bibinfo {author} {\bibfnamefont {I.}~\bibnamefont
  {Tasaki}},\ }\bibfield  {title} {\bibinfo {title} {Spread of {Discrete}
  {Structural} {Changes} in {Synthetic} {Polyanionic} {Gel}: {A} {Model} of
  {Propagation} of a {Nerve} {Impulse}},\ }\href
  {https://doi.org/10.1006/jtbi.2002.3095} {\bibfield  {journal} {\bibinfo
  {journal} {Journal of Theoretical Biology}\ }\textbf {\bibinfo {volume}
  {218}},\ \bibinfo {pages} {497} (\bibinfo {year} {2002})}\BibitemShut
  {NoStop}%
\bibitem [{\citenamefont {Feng}\ \emph {et~al.}(2019)\citenamefont {Feng},
  \citenamefont {Chen}, \citenamefont {Wu},\ and\ \citenamefont
  {Zhang}}]{feng_formation_2019}%
  \BibitemOpen
  \bibfield  {author} {\bibinfo {author} {\bibfnamefont {Z.}~\bibnamefont
  {Feng}}, \bibinfo {author} {\bibfnamefont {X.}~\bibnamefont {Chen}}, \bibinfo
  {author} {\bibfnamefont {X.}~\bibnamefont {Wu}},\ and\ \bibinfo {author}
  {\bibfnamefont {M.}~\bibnamefont {Zhang}},\ }\bibfield  {title} {\bibinfo
  {title} {Formation of biological condensates via phase separation:
  {Characteristics}, analytical methods, and physiological implications},\
  }\href {https://doi.org/10.1074/jbc.REV119.007895} {\bibfield  {journal}
  {\bibinfo  {journal} {Journal of Biological Chemistry}\ }\textbf {\bibinfo
  {volume} {294}},\ \bibinfo {pages} {14823} (\bibinfo {year}
  {2019})}\BibitemShut {NoStop}%
\bibitem [{\citenamefont {Berry}\ \emph {et~al.}(2018)\citenamefont {Berry},
  \citenamefont {Brangwynne},\ and\ \citenamefont
  {Haataja}}]{berry_physical_2018}%
  \BibitemOpen
  \bibfield  {author} {\bibinfo {author} {\bibfnamefont {J.}~\bibnamefont
  {Berry}}, \bibinfo {author} {\bibfnamefont {C.~P.}\ \bibnamefont
  {Brangwynne}},\ and\ \bibinfo {author} {\bibfnamefont {M.}~\bibnamefont
  {Haataja}},\ }\bibfield  {title} {\bibinfo {title} {Physical principles of
  intracellular organization via active and passive phase transitions},\ }\href
  {https://doi.org/10.1088/1361-6633/aaa61e} {\bibfield  {journal} {\bibinfo
  {journal} {Reports on Progress in Physics}\ }\textbf {\bibinfo {volume}
  {81}},\ \bibinfo {pages} {046601} (\bibinfo {year} {2018})}\BibitemShut
  {NoStop}%
\bibitem [{\citenamefont {Nott}\ \emph {et~al.}(2015)\citenamefont {Nott},
  \citenamefont {Petsalaki}, \citenamefont {Farber}, \citenamefont {Jervis},
  \citenamefont {Fussner}, \citenamefont {Plochowietz}, \citenamefont {Craggs},
  \citenamefont {Bazett-Jones}, \citenamefont {Pawson}, \citenamefont
  {Forman-Kay} \emph {et~al.}}]{nott2015phase}%
  \BibitemOpen
  \bibfield  {author} {\bibinfo {author} {\bibfnamefont {T.~J.}\ \bibnamefont
  {Nott}}, \bibinfo {author} {\bibfnamefont {E.}~\bibnamefont {Petsalaki}},
  \bibinfo {author} {\bibfnamefont {P.}~\bibnamefont {Farber}}, \bibinfo
  {author} {\bibfnamefont {D.}~\bibnamefont {Jervis}}, \bibinfo {author}
  {\bibfnamefont {E.}~\bibnamefont {Fussner}}, \bibinfo {author} {\bibfnamefont
  {A.}~\bibnamefont {Plochowietz}}, \bibinfo {author} {\bibfnamefont {T.~D.}\
  \bibnamefont {Craggs}}, \bibinfo {author} {\bibfnamefont {D.~P.}\
  \bibnamefont {Bazett-Jones}}, \bibinfo {author} {\bibfnamefont
  {T.}~\bibnamefont {Pawson}}, \bibinfo {author} {\bibfnamefont {J.~D.}\
  \bibnamefont {Forman-Kay}}, \emph {et~al.},\ }\bibfield  {title} {\bibinfo
  {title} {Phase transition of a disordered nuage protein generates
  environmentally responsive membraneless organelles},\ }\href
  {https://doi.org/10.1016/j.molcel.2015.01.013} {\bibfield  {journal}
  {\bibinfo  {journal} {Molecular cell}\ }\textbf {\bibinfo {volume} {57}},\
  \bibinfo {pages} {936} (\bibinfo {year} {2015})}\BibitemShut {NoStop}%
\bibitem [{\citenamefont {An}\ \emph {et~al.}(2024)\citenamefont {An},
  \citenamefont {Gao}, \citenamefont {Wang}, \citenamefont {Zhang},\ and\
  \citenamefont {Bharti}}]{an_effects_2024}%
  \BibitemOpen
  \bibfield  {author} {\bibinfo {author} {\bibfnamefont {Y.}~\bibnamefont
  {An}}, \bibinfo {author} {\bibfnamefont {T.}~\bibnamefont {Gao}}, \bibinfo
  {author} {\bibfnamefont {T.}~\bibnamefont {Wang}}, \bibinfo {author}
  {\bibfnamefont {D.}~\bibnamefont {Zhang}},\ and\ \bibinfo {author}
  {\bibfnamefont {B.}~\bibnamefont {Bharti}},\ }\bibfield  {title} {\bibinfo
  {title} {Effects of charge asymmetry on the liquid–liquid phase separation
  of polyampholytes and their condensate properties},\ }\href
  {https://doi.org/10.1039/D4SM00532E} {\bibfield  {journal} {\bibinfo
  {journal} {Soft Matter}\ }\textbf {\bibinfo {volume} {20}},\ \bibinfo {pages}
  {6150} (\bibinfo {year} {2024})}\BibitemShut {NoStop}%
\bibitem [{\citenamefont {Lin}\ \emph {et~al.}(2016)\citenamefont {Lin},
  \citenamefont {Forman-Kay},\ and\ \citenamefont {Chan}}]{Lin2016}%
  \BibitemOpen
  \bibfield  {author} {\bibinfo {author} {\bibfnamefont {Y.-H.}\ \bibnamefont
  {Lin}}, \bibinfo {author} {\bibfnamefont {J.~D.}\ \bibnamefont
  {Forman-Kay}},\ and\ \bibinfo {author} {\bibfnamefont {H.~S.}\ \bibnamefont
  {Chan}},\ }\bibfield  {title} {\bibinfo {title} {Sequence-specific
  polyampholyte phase separation in membraneless organelles},\ }\href
  {https://doi.org/10.1103/PhysRevLett.117.178101} {\bibfield  {journal}
  {\bibinfo  {journal} {Phys. Rev. Lett.}\ }\textbf {\bibinfo {volume} {117}},\
  \bibinfo {pages} {178101} (\bibinfo {year} {2016})}\BibitemShut {NoStop}%
\bibitem [{\citenamefont {Meca}\ \emph {et~al.}(2023)\citenamefont {Meca},
  \citenamefont {Fritsch}, \citenamefont {Iglesias-Artola}, \citenamefont
  {Reber},\ and\ \citenamefont {Wagner}}]{Meca2023}%
  \BibitemOpen
  \bibfield  {author} {\bibinfo {author} {\bibfnamefont {E.}~\bibnamefont
  {Meca}}, \bibinfo {author} {\bibfnamefont {A.~W.}\ \bibnamefont {Fritsch}},
  \bibinfo {author} {\bibfnamefont {J.~M.}\ \bibnamefont {Iglesias-Artola}},
  \bibinfo {author} {\bibfnamefont {S.}~\bibnamefont {Reber}},\ and\ \bibinfo
  {author} {\bibfnamefont {B.}~\bibnamefont {Wagner}},\ }\bibfield  {title}
  {\bibinfo {title} {Predicting disordered regions driving phase separation of
  proteins under variable salt concentration},\ }\bibfield  {journal} {\bibinfo
   {journal} {Frontiers in Physics}\ }\textbf {\bibinfo {volume} {11}},\ \href
  {https://doi.org/10.3389/fphy.2023.1213304} {10.3389/fphy.2023.1213304}
  (\bibinfo {year} {2023})\BibitemShut {NoStop}%
\bibitem [{\citenamefont {Ausserwöger}\ \emph {et~al.}(2024)\citenamefont
  {Ausserwöger}, \citenamefont {Scrutton}, \citenamefont {Sneideris},
  \citenamefont {Fischer}, \citenamefont {Qian}, \citenamefont {Csilléry},
  \citenamefont {Saar}, \citenamefont {Białek}, \citenamefont {Oeller},
  \citenamefont {Krainer}, \citenamefont {Franzmann}, \citenamefont {Wittmann},
  \citenamefont {Iglesias-Artola}, \citenamefont {Invernizzi}, \citenamefont
  {Hyman}, \citenamefont {Alberti}, \citenamefont {Lorenzen},\ and\
  \citenamefont {Knowles}}]{ausserwoger_biomolecular_2024}%
  \BibitemOpen
  \bibfield  {author} {\bibinfo {author} {\bibfnamefont {H.}~\bibnamefont
  {Ausserwöger}}, \bibinfo {author} {\bibfnamefont {R.}~\bibnamefont
  {Scrutton}}, \bibinfo {author} {\bibfnamefont {T.}~\bibnamefont {Sneideris}},
  \bibinfo {author} {\bibfnamefont {C.~M.}\ \bibnamefont {Fischer}}, \bibinfo
  {author} {\bibfnamefont {D.}~\bibnamefont {Qian}}, \bibinfo {author}
  {\bibfnamefont {E.~d.}\ \bibnamefont {Csilléry}}, \bibinfo {author}
  {\bibfnamefont {K.~L.}\ \bibnamefont {Saar}}, \bibinfo {author}
  {\bibfnamefont {A.~Z.}\ \bibnamefont {Białek}}, \bibinfo {author}
  {\bibfnamefont {M.}~\bibnamefont {Oeller}}, \bibinfo {author} {\bibfnamefont
  {G.}~\bibnamefont {Krainer}}, \bibinfo {author} {\bibfnamefont {T.~M.}\
  \bibnamefont {Franzmann}}, \bibinfo {author} {\bibfnamefont {S.}~\bibnamefont
  {Wittmann}}, \bibinfo {author} {\bibfnamefont {J.~M.}\ \bibnamefont
  {Iglesias-Artola}}, \bibinfo {author} {\bibfnamefont {G.}~\bibnamefont
  {Invernizzi}}, \bibinfo {author} {\bibfnamefont {A.~A.}\ \bibnamefont
  {Hyman}}, \bibinfo {author} {\bibfnamefont {S.}~\bibnamefont {Alberti}},
  \bibinfo {author} {\bibfnamefont {N.}~\bibnamefont {Lorenzen}},\ and\
  \bibinfo {author} {\bibfnamefont {T.~P.~J.}\ \bibnamefont {Knowles}},\ }\href
  {https://doi.org/10.1101/2024.05.23.595321} {\bibinfo {title} {Biomolecular
  condensates sustain {pH} gradients at equilibrium driven by charge
  neutralisation}} (\bibinfo {year} {2024})\BibitemShut {NoStop}%
\bibitem [{\citenamefont {Chollakup}\ \emph {et~al.}(2010)\citenamefont
  {Chollakup}, \citenamefont {Smitthipong}, \citenamefont {Eisenbach},\ and\
  \citenamefont {Tirrell}}]{chollakup2010}%
  \BibitemOpen
  \bibfield  {author} {\bibinfo {author} {\bibfnamefont {R.}~\bibnamefont
  {Chollakup}}, \bibinfo {author} {\bibfnamefont {W.}~\bibnamefont
  {Smitthipong}}, \bibinfo {author} {\bibfnamefont {C.}~\bibnamefont
  {Eisenbach}},\ and\ \bibinfo {author} {\bibfnamefont {M.}~\bibnamefont
  {Tirrell}},\ }\bibfield  {title} {\bibinfo {title} {{Phase Behavior and
  Coacervation of Aqueous Poly(acrylic acid)-Poly(allylamine) Solutions}},\
  }\href {https://doi.org/10.1021/MA902144K} {\bibfield  {journal} {\bibinfo
  {journal} {Macromolecules}\ }\textbf {\bibinfo {volume} {43}},\ \bibinfo
  {pages} {2518} (\bibinfo {year} {2010})}\BibitemShut {NoStop}%
\bibitem [{\citenamefont {Sta\v{n}o}\ \emph {et~al.}(2024)\citenamefont
  {Sta\v{n}o}, \citenamefont {van Lente}, \citenamefont {Lindhoud},\ and\
  \citenamefont {Ko\v{s}ovan}}]{stano2024}%
  \BibitemOpen
  \bibfield  {author} {\bibinfo {author} {\bibfnamefont {R.}~\bibnamefont
  {Sta\v{n}o}}, \bibinfo {author} {\bibfnamefont {J.~J.}\ \bibnamefont {van
  Lente}}, \bibinfo {author} {\bibfnamefont {S.}~\bibnamefont {Lindhoud}},\
  and\ \bibinfo {author} {\bibfnamefont {P.}~\bibnamefont {Ko\v{s}ovan}},\
  }\bibfield  {title} {\bibinfo {title} {Sequestration of small ions and weak
  acids and bases by a polyelectrolyte complex studied by simulation and
  experiment},\ }\href {https://doi.org/10.1021/acs.macromol.3c01209}
  {\bibfield  {journal} {\bibinfo  {journal} {Macromolecules}\ }\textbf
  {\bibinfo {volume} {57}},\ \bibinfo {pages} {1383 } (\bibinfo {year}
  {2024})}\BibitemShut {NoStop}%
\bibitem [{\citenamefont {Meneses-Ju\'arez}\ \emph {et~al.}(2019)\citenamefont
  {Meneses-Ju\'arez}, \citenamefont {M\'arquez-Beltr\'an},\ and\ \citenamefont
  {Gonz\'alez-Melchor}}]{meneses2019}%
  \BibitemOpen
  \bibfield  {author} {\bibinfo {author} {\bibfnamefont {E.}~\bibnamefont
  {Meneses-Ju\'arez}}, \bibinfo {author} {\bibfnamefont {C.}~\bibnamefont
  {M\'arquez-Beltr\'an}},\ and\ \bibinfo {author} {\bibfnamefont
  {M.}~\bibnamefont {Gonz\'alez-Melchor}},\ }\bibfield  {title} {\bibinfo
  {title} {Influence of $p\mathrm{H}$ on the formation of a polyelectrolyte
  complex by dissipative particle dynamics simulation: From an extended to a
  compact shape},\ }\href {https://doi.org/10.1103/PhysRevE.100.012505}
  {\bibfield  {journal} {\bibinfo  {journal} {Phys. Rev. E}\ }\textbf {\bibinfo
  {volume} {100}},\ \bibinfo {pages} {012505} (\bibinfo {year}
  {2019})}\BibitemShut {NoStop}%
\bibitem [{\citenamefont {Adame-Arana}\ \emph {et~al.}(2020)\citenamefont
  {Adame-Arana}, \citenamefont {Weber}, \citenamefont {Zaburdaev},
  \citenamefont {Prost},\ and\ \citenamefont
  {J{\"{u}}licher}}]{Adame-Arana2020}%
  \BibitemOpen
  \bibfield  {author} {\bibinfo {author} {\bibfnamefont {O.}~\bibnamefont
  {Adame-Arana}}, \bibinfo {author} {\bibfnamefont {C.~A.}\ \bibnamefont
  {Weber}}, \bibinfo {author} {\bibfnamefont {V.}~\bibnamefont {Zaburdaev}},
  \bibinfo {author} {\bibfnamefont {J.}~\bibnamefont {Prost}},\ and\ \bibinfo
  {author} {\bibfnamefont {F.}~\bibnamefont {J{\"{u}}licher}},\ }\bibfield
  {title} {\bibinfo {title} {{Liquid Phase Separation Controlled by pH}},\
  }\href {https://doi.org/10.1016/J.BPJ.2020.07.044} {\bibfield  {journal}
  {\bibinfo  {journal} {Biophysical Journal}\ }\textbf {\bibinfo {volume}
  {119}},\ \bibinfo {pages} {1590} (\bibinfo {year} {2020})}\BibitemShut
  {NoStop}%
\bibitem [{\citenamefont {Majee}\ \emph {et~al.}(2020)\citenamefont {Majee},
  \citenamefont {Bier}, \citenamefont {Blossey},\ and\ \citenamefont
  {Podgornik}}]{majee2020}%
  \BibitemOpen
  \bibfield  {author} {\bibinfo {author} {\bibfnamefont {A.}~\bibnamefont
  {Majee}}, \bibinfo {author} {\bibfnamefont {M.}~\bibnamefont {Bier}},
  \bibinfo {author} {\bibfnamefont {R.}~\bibnamefont {Blossey}},\ and\ \bibinfo
  {author} {\bibfnamefont {R.}~\bibnamefont {Podgornik}},\ }\bibfield  {title}
  {\bibinfo {title} {Charge symmetry broken complex coacervation},\ }\href
  {https://doi.org/10.1103/PhysRevResearch.2.043417} {\bibfield  {journal}
  {\bibinfo  {journal} {Phys. Rev. Res.}\ }\textbf {\bibinfo {volume} {2}},\
  \bibinfo {pages} {043417} (\bibinfo {year} {2020})}\BibitemShut {NoStop}%
\bibitem [{\citenamefont {Knoerdel}\ \emph {et~al.}(2021)\citenamefont
  {Knoerdel}, \citenamefont {McTigue},\ and\ \citenamefont
  {Sing}}]{knoerdel2021}%
  \BibitemOpen
  \bibfield  {author} {\bibinfo {author} {\bibfnamefont {A.~R.}\ \bibnamefont
  {Knoerdel}}, \bibinfo {author} {\bibfnamefont {W.~C.~B.}\ \bibnamefont
  {McTigue}},\ and\ \bibinfo {author} {\bibfnamefont {C.}~\bibnamefont
  {Sing}},\ }\bibfield  {title} {\bibinfo {title} {Transfer matrix model of ph
  effects in polymeric complex coacervation},\ }\href
  {https://doi.org/10.1021/acs.jpcb.1c03065} {\bibfield  {journal} {\bibinfo
  {journal} {The Journal of Physical Chemistry B}\ }\textbf {\bibinfo {volume}
  {125}},\ \bibinfo {pages} {8965} (\bibinfo {year} {2021})}\BibitemShut
  {NoStop}%
\bibitem [{\citenamefont {Celora}\ \emph {et~al.}(2023)\citenamefont {Celora},
  \citenamefont {Blossey}, \citenamefont {M\"{u}nch},\ and\ \citenamefont
  {Wagner}}]{celora2023}%
  \BibitemOpen
  \bibfield  {author} {\bibinfo {author} {\bibfnamefont {G.~L.}\ \bibnamefont
  {Celora}}, \bibinfo {author} {\bibfnamefont {R.}~\bibnamefont {Blossey}},
  \bibinfo {author} {\bibfnamefont {A.}~\bibnamefont {M\"{u}nch}},\ and\
  \bibinfo {author} {\bibfnamefont {B.}~\bibnamefont {Wagner}},\ }\bibfield
  {title} {\bibinfo {title} {Counterion-controlled phase equilibria in a
  charge-regulated polymer solution},\ }\href
  {https://doi.org/10.1063/5.0169610} {\bibfield  {journal} {\bibinfo
  {journal} {The Journal of Chemical Physics}\ }\textbf {\bibinfo {volume}
  {159}},\ \bibinfo {pages} {184902} (\bibinfo {year} {2023})}\BibitemShut
  {NoStop}%
\bibitem [{\citenamefont {Zheng}\ \emph {et~al.}(2021)\citenamefont {Zheng},
  \citenamefont {Avni}, \citenamefont {Andelman},\ and\ \citenamefont
  {Podgornik}}]{zheng2021}%
  \BibitemOpen
  \bibfield  {author} {\bibinfo {author} {\bibfnamefont {B.}~\bibnamefont
  {Zheng}}, \bibinfo {author} {\bibfnamefont {Y.}~\bibnamefont {Avni}},
  \bibinfo {author} {\bibfnamefont {D.}~\bibnamefont {Andelman}},\ and\
  \bibinfo {author} {\bibfnamefont {R.}~\bibnamefont {Podgornik}},\ }\bibfield
  {title} {\bibinfo {title} {Phase separation of polyelectrolytes: The effect
  of charge regulation},\ }\href {https://doi.org/10.1021/acs.jpcb.1c01986}
  {\bibfield  {journal} {\bibinfo  {journal} {The Journal of Physical Chemistry
  B}\ }\textbf {\bibinfo {volume} {125}},\ \bibinfo {pages} {7863 } (\bibinfo
  {year} {2021})}\BibitemShut {NoStop}%
\bibitem [{\citenamefont {Avni}\ \emph {et~al.}(2019)\citenamefont {Avni},
  \citenamefont {Andelman},\ and\ \citenamefont {Podgornik}}]{avni2019}%
  \BibitemOpen
  \bibfield  {author} {\bibinfo {author} {\bibfnamefont {Y.}~\bibnamefont
  {Avni}}, \bibinfo {author} {\bibfnamefont {D.}~\bibnamefont {Andelman}},\
  and\ \bibinfo {author} {\bibfnamefont {R.}~\bibnamefont {Podgornik}},\
  }\bibfield  {title} {\bibinfo {title} {Charge regulation with fixed and
  mobile charged macromolecules},\ }\href
  {https://doi.org/https://doi.org/10.1016/j.coelec.2018.10.014} {\bibfield
  {journal} {\bibinfo  {journal} {Current Opinion in Electrochemistry}\
  }\textbf {\bibinfo {volume} {13}},\ \bibinfo {pages} {70} (\bibinfo {year}
  {2019})}\BibitemShut {NoStop}%
\bibitem [{\citenamefont {Avni}\ \emph {et~al.}(2020)\citenamefont {Avni},
  \citenamefont {Podgornik},\ and\ \citenamefont {Andelman}}]{Avni2020}%
  \BibitemOpen
  \bibfield  {author} {\bibinfo {author} {\bibfnamefont {Y.}~\bibnamefont
  {Avni}}, \bibinfo {author} {\bibfnamefont {R.}~\bibnamefont {Podgornik}},\
  and\ \bibinfo {author} {\bibfnamefont {D.}~\bibnamefont {Andelman}},\
  }\bibfield  {title} {\bibinfo {title} {Critical behavior of charge-regulated
  macro-ions},\ }\href {https://doi.org/10.1063/5.0011623} {\bibfield
  {journal} {\bibinfo  {journal} {The Journal of Chemical Physics}\ }\textbf
  {\bibinfo {volume} {153}},\ \bibinfo {pages} {024901} (\bibinfo {year}
  {2020})}\BibitemShut {NoStop}%
\bibitem [{\citenamefont {Zheng}\ \emph {et~al.}(2024)\citenamefont {Zheng},
  \citenamefont {Komura}, \citenamefont {Andelman},\ and\ \citenamefont
  {Podgornik}}]{zheng2024}%
  \BibitemOpen
  \bibfield  {author} {\bibinfo {author} {\bibfnamefont {B.}~\bibnamefont
  {Zheng}}, \bibinfo {author} {\bibfnamefont {S.}~\bibnamefont {Komura}},
  \bibinfo {author} {\bibfnamefont {D.}~\bibnamefont {Andelman}},\ and\
  \bibinfo {author} {\bibfnamefont {R.}~\bibnamefont {Podgornik}},\ }\href
  {https://arxiv.org/abs/2411.09448} {\bibinfo {title} {Diffusive dynamics of
  charge regulated macro-ion solutions}} (\bibinfo {year} {2024}),\ \Eprint
  {https://arxiv.org/abs/2411.09448} {arXiv:2411.09448} \BibitemShut {NoStop}%
\bibitem [{\citenamefont {Liese}\ \emph {et~al.}(2022)\citenamefont {Liese},
  \citenamefont {Schlaich},\ and\ \citenamefont {Netz}}]{liese2022dielectric}%
  \BibitemOpen
  \bibfield  {author} {\bibinfo {author} {\bibfnamefont {S.}~\bibnamefont
  {Liese}}, \bibinfo {author} {\bibfnamefont {A.}~\bibnamefont {Schlaich}},\
  and\ \bibinfo {author} {\bibfnamefont {R.~R.}\ \bibnamefont {Netz}},\
  }\bibfield  {title} {\bibinfo {title} {Dielectric constant of aqueous
  solutions of proteins and organic polymers from molecular dynamics
  simulations},\ }\href {https://doi.org/https://doi.org/10.1063/5.0089397}
  {\bibfield  {journal} {\bibinfo  {journal} {The Journal of Chemical Physics}\
  }\textbf {\bibinfo {volume} {156}},\ \bibinfo {pages} {224902} (\bibinfo
  {year} {2022})}\BibitemShut {NoStop}%
\bibitem [{\citenamefont {Markovich}\ \emph {et~al.}(2021)\citenamefont
  {Markovich}, \citenamefont {Andelman},\ and\ \citenamefont
  {Podgornik}}]{markovichChargedMembranesPoisson2021}%
  \BibitemOpen
  \bibfield  {author} {\bibinfo {author} {\bibfnamefont {T.}~\bibnamefont
  {Markovich}}, \bibinfo {author} {\bibfnamefont {D.}~\bibnamefont
  {Andelman}},\ and\ \bibinfo {author} {\bibfnamefont {R.}~\bibnamefont
  {Podgornik}},\ }\bibfield  {title} {\bibinfo {title} {Charged membranes:
  Poisson–boltzmann theory, the {DLVO} paradigm, and beyond},\ }in\
  \href@noop {} {\emph {\bibinfo {booktitle} {Handbook of Lipid Membranes}}}\
  (\bibinfo  {publisher} {{CRC} Press},\ \bibinfo {year} {2021})\BibitemShut
  {NoStop}%
\bibitem [{\citenamefont {Bier}\ \emph {et~al.}(2017)\citenamefont {Bier},
  \citenamefont {Gavish}, \citenamefont {Uecker},\ and\ \citenamefont
  {Yochelis}}]{gavish2017PRE}%
  \BibitemOpen
  \bibfield  {author} {\bibinfo {author} {\bibfnamefont {S.}~\bibnamefont
  {Bier}}, \bibinfo {author} {\bibfnamefont {N.}~\bibnamefont {Gavish}},
  \bibinfo {author} {\bibfnamefont {H.}~\bibnamefont {Uecker}},\ and\ \bibinfo
  {author} {\bibfnamefont {A.}~\bibnamefont {Yochelis}},\ }\bibfield  {title}
  {\bibinfo {title} {From bulk self-assembly to electrical diffuse layer in a
  continuum approach for ionic liquids: The impact of anion and cation size
  asymmetry},\ }\href {https://doi.org/10.1103/PhysRevE.95.060201} {\bibfield
  {journal} {\bibinfo  {journal} {Phys. Rev. E}\ }\textbf {\bibinfo {volume}
  {95}},\ \bibinfo {pages} {060201} (\bibinfo {year} {2017})}\BibitemShut
  {NoStop}%
\bibitem [{\citenamefont {Luo}\ \emph {et~al.}(2024)\citenamefont {Luo},
  \citenamefont {Hess}, \citenamefont {Aierken}, \citenamefont {Qiang},
  \citenamefont {Joseph},\ and\ \citenamefont
  {Zwicker}}]{luo2024condensatesizecontrolcharge}%
  \BibitemOpen
  \bibfield  {author} {\bibinfo {author} {\bibfnamefont {C.}~\bibnamefont
  {Luo}}, \bibinfo {author} {\bibfnamefont {N.}~\bibnamefont {Hess}}, \bibinfo
  {author} {\bibfnamefont {D.}~\bibnamefont {Aierken}}, \bibinfo {author}
  {\bibfnamefont {Y.}~\bibnamefont {Qiang}}, \bibinfo {author} {\bibfnamefont
  {J.~A.}\ \bibnamefont {Joseph}},\ and\ \bibinfo {author} {\bibfnamefont
  {D.}~\bibnamefont {Zwicker}},\ }\href@noop {} {\bibinfo {title} {Condensate
  size control by charge asymmetry}} (\bibinfo {year} {2024}),\ \Eprint
  {https://arxiv.org/abs/2409.15599} {arXiv:2409.15599} \BibitemShut {NoStop}%
\bibitem [{\citenamefont {Gavish}\ and\ \citenamefont
  {Yochelis}(2016)}]{gavish_theory_2016}%
  \BibitemOpen
  \bibfield  {author} {\bibinfo {author} {\bibfnamefont {N.}~\bibnamefont
  {Gavish}}\ and\ \bibinfo {author} {\bibfnamefont {A.}~\bibnamefont
  {Yochelis}},\ }\bibfield  {title} {\bibinfo {title} {Theory of {Phase}
  {Separation} and {Polarization} for {Pure} {Ionic} {Liquids}},\ }\href
  {https://doi.org/10.1021/acs.jpclett.6b00370} {\bibfield  {journal} {\bibinfo
   {journal} {The Journal of Physical Chemistry Letters}\ }\textbf {\bibinfo
  {volume} {7}},\ \bibinfo {pages} {1121} (\bibinfo {year} {2016})}\BibitemShut
  {NoStop}%
\bibitem [{\citenamefont {Shiwa}(1997)}]{shiwa_amplitude_1997}%
  \BibitemOpen
  \bibfield  {author} {\bibinfo {author} {\bibfnamefont {Y.}~\bibnamefont
  {Shiwa}},\ }\bibfield  {title} {\bibinfo {title} {The amplitude and
  phase-diffusion equations for lamellar patterns in block copolymers},\ }\href
  {https://doi.org/10.1016/S0375-9601(97)00128-X} {\bibfield  {journal}
  {\bibinfo  {journal} {Physics Letters A}\ }\textbf {\bibinfo {volume}
  {228}},\ \bibinfo {pages} {279} (\bibinfo {year} {1997})}\BibitemShut
  {NoStop}%
\bibitem [{\citenamefont {Lewis}(1907)}]{lewis_outlines_1907}%
  \BibitemOpen
  \bibfield  {author} {\bibinfo {author} {\bibfnamefont {G.~N.}\ \bibnamefont
  {Lewis}},\ }\bibfield  {title} {\bibinfo {title} {Outlines of a {New}
  {System} of {Thermodynamic} {Chemistry}},\ }\href
  {https://doi.org/10.2307/20022322} {\bibfield  {journal} {\bibinfo  {journal}
  {Proceedings of the American Academy of Arts and Sciences}\ }\textbf
  {\bibinfo {volume} {43}},\ \bibinfo {pages} {259} (\bibinfo {year}
  {1907})}\BibitemShut {NoStop}%
\bibitem [{\citenamefont {Pismen}(2023)}]{alma991025286663307026}%
  \BibitemOpen
  \bibfield  {author} {\bibinfo {author} {\bibfnamefont {L.~M.}\ \bibnamefont
  {Pismen}},\ }\href
  {https://doi.org/https://doi.org/10.1007/978-3-031-29579-9} {\emph {\bibinfo
  {title} {Patterns and Interfaces in Dissipative Dynamics}}},\ \bibinfo
  {edition} {2nd}\ ed.,\ Springer Series in Synergetics\ (\bibinfo  {publisher}
  {Springer Cham},\ \bibinfo {year} {2023})\BibitemShut {NoStop}%
\bibitem [{\citenamefont {Champneys}(1998)}]{champneys_homoclinic_1998}%
  \BibitemOpen
  \bibfield  {author} {\bibinfo {author} {\bibfnamefont {A.~R.}\ \bibnamefont
  {Champneys}},\ }\bibfield  {title} {\bibinfo {title} {Homoclinic orbits in
  reversible systems and their applications in mechanics, fluids and optics},\
  }\href {https://doi.org/10.1016/S0167-2789(97)00209-1} {\bibfield  {journal}
  {\bibinfo  {journal} {Physica D: Nonlinear Phenomena}\ }\bibinfo {series}
  {Proceedings of the {Workshop} on {Time}-{Reversal} {Symmetry} in {Dynamical}
  {Systems}},\ \textbf {\bibinfo {volume} {112}},\ \bibinfo {pages} {158}
  (\bibinfo {year} {1998})}\BibitemShut {NoStop}%
\bibitem [{\citenamefont {Greve}\ \emph {et~al.}(2024)\citenamefont {Greve},
  \citenamefont {Lovato}, \citenamefont {Frohoff-Hülsmann},\ and\
  \citenamefont {Thiele}}]{greve2024coexistenceuniformoscillatorystates}%
  \BibitemOpen
  \bibfield  {author} {\bibinfo {author} {\bibfnamefont {D.}~\bibnamefont
  {Greve}}, \bibinfo {author} {\bibfnamefont {G.}~\bibnamefont {Lovato}},
  \bibinfo {author} {\bibfnamefont {T.}~\bibnamefont {Frohoff-Hülsmann}},\
  and\ \bibinfo {author} {\bibfnamefont {U.}~\bibnamefont {Thiele}},\
  }\href@noop {} {\bibinfo {title} {Coexistence of uniform and oscillatory
  states resulting from nonreciprocity and conservation laws}} (\bibinfo {year}
  {2024}),\ \Eprint {https://arxiv.org/abs/2402.08634} {arXiv:2402.08634}
  \BibitemShut {NoStop}%
\bibitem [{\citenamefont {Davis}\ \emph {et~al.}(2025)\citenamefont {Davis},
  \citenamefont {Baldwin},\ and\ \citenamefont
  {Pearce}}]{davis2025mesoscopicheterogeneitybiomolecularcondensates}%
  \BibitemOpen
  \bibfield  {author} {\bibinfo {author} {\bibfnamefont {L.~K.}\ \bibnamefont
  {Davis}}, \bibinfo {author} {\bibfnamefont {A.~J.}\ \bibnamefont {Baldwin}},\
  and\ \bibinfo {author} {\bibfnamefont {P.}~\bibnamefont {Pearce}},\ }\href
  {https://doi.org/https://doi.org/10.48550/arXiv.2502.14587} {\bibinfo {title}
  {Mesoscopic heterogeneity in biomolecular condensates from sequence
  patterning}} (\bibinfo {year} {2025}),\ \Eprint
  {https://arxiv.org/abs/2502.14587} {arXiv:2502.14587} \BibitemShut {NoStop}%
\bibitem [{\citenamefont {Curk}\ \emph {et~al.}(2022)\citenamefont {Curk},
  \citenamefont {Yuan},\ and\ \citenamefont {Luijten}}]{curk_accelerated_2022}%
  \BibitemOpen
  \bibfield  {author} {\bibinfo {author} {\bibfnamefont {T.}~\bibnamefont
  {Curk}}, \bibinfo {author} {\bibfnamefont {J.}~\bibnamefont {Yuan}},\ and\
  \bibinfo {author} {\bibfnamefont {E.}~\bibnamefont {Luijten}},\ }\bibfield
  {title} {\bibinfo {title} {Accelerated simulation method for charge
  regulation effects},\ }\href {https://doi.org/10.1063/5.0066432} {\bibfield
  {journal} {\bibinfo  {journal} {The Journal of Chemical Physics}\ }\textbf
  {\bibinfo {volume} {156}},\ \bibinfo {pages} {044122} (\bibinfo {year}
  {2022})}\BibitemShut {NoStop}%
\bibitem [{\citenamefont {Leonardi}\ and\ \citenamefont
  {Angeli}(2010)}]{leonardi_maxwellstefan_2010}%
  \BibitemOpen
  \bibfield  {author} {\bibinfo {author} {\bibfnamefont {E.}~\bibnamefont
  {Leonardi}}\ and\ \bibinfo {author} {\bibfnamefont {C.}~\bibnamefont
  {Angeli}},\ }\bibfield  {title} {\bibinfo {title} {{On the Maxwell--Stefan
  Approach to Diffusion: A General Resolution in the Transient Regime for
  One-Dimensional Systems}},\ }\href {https://doi.org/10.1021/jp900760c}
  {\bibfield  {journal} {\bibinfo  {journal} {The Journal of Physical Chemistry
  B}\ }\textbf {\bibinfo {volume} {114}},\ \bibinfo {pages} {151} (\bibinfo
  {year} {2010})}\BibitemShut {NoStop}%
\bibitem [{\citenamefont {Frohoff-Hülsmann}\ \emph {et~al.}(2023)\citenamefont
  {Frohoff-Hülsmann}, \citenamefont {Thiele},\ and\ \citenamefont
  {Pismen}}]{frohoff-hulsmann_non-reciprocity_2023}%
  \BibitemOpen
  \bibfield  {author} {\bibinfo {author} {\bibfnamefont {T.}~\bibnamefont
  {Frohoff-Hülsmann}}, \bibinfo {author} {\bibfnamefont {U.}~\bibnamefont
  {Thiele}},\ and\ \bibinfo {author} {\bibfnamefont {L.~M.}\ \bibnamefont
  {Pismen}},\ }\bibfield  {title} {\bibinfo {title} {Non-reciprocity induces
  resonances in a two-field cahn–hilliard model},\ }\href
  {https://doi.org/10.1098/rsta.2022.0087} {\bibfield  {journal} {\bibinfo
  {journal} {Philosophical Transactions of the Royal Society A: Mathematical,
  Physical and Engineering Sciences}\ }\textbf {\bibinfo {volume} {381}},\
  \bibinfo {pages} {20220087} (\bibinfo {year} {2023})}\BibitemShut {NoStop}%
\end{thebibliography}%


\begin{thebibliography}{5}%
\makeatletter
\providecommand \@ifxundefined [1]{%
 \@ifx{#1\undefined}
}%
\providecommand \@ifnum [1]{%
 \ifnum #1\expandafter \@firstoftwo
 \else \expandafter \@secondoftwo
 \fi
}%
\providecommand \@ifx [1]{%
 \ifx #1\expandafter \@firstoftwo
 \else \expandafter \@secondoftwo
 \fi
}%
\providecommand \natexlab [1]{#1}%
\providecommand \enquote  [1]{``#1''}%
\providecommand \bibnamefont  [1]{#1}%
\providecommand \bibfnamefont [1]{#1}%
\providecommand \citenamefont [1]{#1}%
\providecommand \href@noop [0]{\@secondoftwo}%
\providecommand \href [0]{\begingroup \@sanitize@url \@href}%
\providecommand \@href[1]{\@@startlink{#1}\@@href}%
\providecommand \@@href[1]{\endgroup#1\@@endlink}%
\providecommand \@sanitize@url [0]{\catcode `\\12\catcode `\$12\catcode
  `\&12\catcode `\#12\catcode `\^12\catcode `\_12\catcode `\%12\relax}%
\providecommand \@@startlink[1]{}%
\providecommand \@@endlink[0]{}%
\providecommand \url  [0]{\begingroup\@sanitize@url \@url }%
\providecommand \@url [1]{\endgroup\@href {#1}{\urlprefix }}%
\providecommand \urlprefix  [0]{URL }%
\providecommand \Eprint [0]{\href }%
\providecommand \doibase [0]{https://doi.org/}%
\providecommand \selectlanguage [0]{\@gobble}%
\providecommand \bibinfo  [0]{\@secondoftwo}%
\providecommand \bibfield  [0]{\@secondoftwo}%
\providecommand \translation [1]{[#1]}%
\providecommand \BibitemOpen [0]{}%
\providecommand \bibitemStop [0]{}%
\providecommand \bibitemNoStop [0]{.\EOS\space}%
\providecommand \EOS [0]{\spacefactor3000\relax}%
\providecommand \BibitemShut  [1]{\csname bibitem#1\endcsname}%
\let\auto@bib@innerbib\@empty
\bibitem [{\citenamefont {Bier}\ \emph {et~al.}(2017)\citenamefont {Bier},
  \citenamefont {Gavish}, \citenamefont {Uecker},\ and\ \citenamefont
  {Yochelis}}]{app_gavish2017PRE}%
  \BibitemOpen
  \bibfield  {author} {\bibinfo {author} {\bibfnamefont {S.}~\bibnamefont
  {Bier}}, \bibinfo {author} {\bibfnamefont {N.}~\bibnamefont {Gavish}},
  \bibinfo {author} {\bibfnamefont {H.}~\bibnamefont {Uecker}},\ and\ \bibinfo
  {author} {\bibfnamefont {A.}~\bibnamefont {Yochelis}},\ }\bibfield  {title}
  {\bibinfo {title} {From bulk self-assembly to electrical diffuse layer in a
  continuum approach for ionic liquids: The impact of anion and cation size
  asymmetry},\ }\href {https://doi.org/10.1103/PhysRevE.95.060201} {\bibfield
  {journal} {\bibinfo  {journal} {Phys. Rev. E}\ }\textbf {\bibinfo {volume}
  {95}},\ \bibinfo {pages} {060201} (\bibinfo {year} {2017})}\BibitemShut
  {NoStop}%
\bibitem [{\citenamefont {Rackauckas}\ and\ \citenamefont
  {Nie}(2017)}]{app_rackauckas2017differentialequations}%
  \BibitemOpen
  \bibfield  {author} {\bibinfo {author} {\bibfnamefont {C.}~\bibnamefont
  {Rackauckas}}\ and\ \bibinfo {author} {\bibfnamefont {Q.}~\bibnamefont
  {Nie}},\ }\bibfield  {title} {\bibinfo {title} {Differential{E}quations.jl--a
  performant and feature-rich ecosystem for solving differential equations in
  {J}ulia},\ }\href@noop {} {\bibfield  {journal} {\bibinfo  {journal} {Journal
  of Open Research Software}\ }\textbf {\bibinfo {volume} {5}} (\bibinfo {year}
  {2017})}\BibitemShut {NoStop}%
\bibitem [{\citenamefont {Celora}\ \emph {et~al.}(2023)\citenamefont {Celora},
  \citenamefont {Blossey}, \citenamefont {M\"{u}nch},\ and\ \citenamefont
  {Wagner}}]{app_celora2023}%
  \BibitemOpen
  \bibfield  {author} {\bibinfo {author} {\bibfnamefont {G.~L.}\ \bibnamefont
  {Celora}}, \bibinfo {author} {\bibfnamefont {R.}~\bibnamefont {Blossey}},
  \bibinfo {author} {\bibfnamefont {A.}~\bibnamefont {M\"{u}nch}},\ and\
  \bibinfo {author} {\bibfnamefont {B.}~\bibnamefont {Wagner}},\ }\bibfield
  {title} {\bibinfo {title} {Counterion-controlled phase equilibria in a
  charge-regulated polymer solution},\ }\href
  {https://doi.org/10.1063/5.0169610} {\bibfield  {journal} {\bibinfo
  {journal} {The Journal of Chemical Physics}\ }\textbf {\bibinfo {volume}
  {159}},\ \bibinfo {pages} {184902} (\bibinfo {year} {2023})}\BibitemShut
  {NoStop}%
\bibitem [{\citenamefont {Avni}\ \emph {et~al.}(2020)\citenamefont {Avni},
  \citenamefont {Podgornik},\ and\ \citenamefont {Andelman}}]{app_Avni2020}%
  \BibitemOpen
  \bibfield  {author} {\bibinfo {author} {\bibfnamefont {Y.}~\bibnamefont
  {Avni}}, \bibinfo {author} {\bibfnamefont {R.}~\bibnamefont {Podgornik}},\
  and\ \bibinfo {author} {\bibfnamefont {D.}~\bibnamefont {Andelman}},\
  }\bibfield  {title} {\bibinfo {title} {Critical behavior of charge-regulated
  macro-ions},\ }\href {https://doi.org/10.1063/5.0011623} {\bibfield
  {journal} {\bibinfo  {journal} {The Journal of Chemical Physics}\ }\textbf
  {\bibinfo {volume} {153}},\ \bibinfo {pages} {024901} (\bibinfo {year}
  {2020})}\BibitemShut {NoStop}%
\bibitem [{\citenamefont {Luo}\ \emph {et~al.}(2024)\citenamefont {Luo},
  \citenamefont {Hess}, \citenamefont {Aierken}, \citenamefont {Qiang},
  \citenamefont {Joseph},\ and\ \citenamefont
  {Zwicker}}]{app_luo2024condensatesizecontrolcharge}%
  \BibitemOpen
  \bibfield  {author} {\bibinfo {author} {\bibfnamefont {C.}~\bibnamefont
  {Luo}}, \bibinfo {author} {\bibfnamefont {N.}~\bibnamefont {Hess}}, \bibinfo
  {author} {\bibfnamefont {D.}~\bibnamefont {Aierken}}, \bibinfo {author}
  {\bibfnamefont {Y.}~\bibnamefont {Qiang}}, \bibinfo {author} {\bibfnamefont
  {J.~A.}\ \bibnamefont {Joseph}},\ and\ \bibinfo {author} {\bibfnamefont
  {D.}~\bibnamefont {Zwicker}},\ }\href@noop {} {\bibinfo {title} {Condensate
  size control by charge asymmetry}} (\bibinfo {year} {2024}),\ \Eprint
  {https://arxiv.org/abs/2409.15599} {arXiv:2409.15599} \BibitemShut {NoStop}%
\end{thebibliography}%

\end{document}


\preprint{}

\title{Supplementary material for ``The diffusive dynamics and electrochemical regulation of weak polyelectrolytes across liquid interfaces"}

\author{Giulia L Celora}
\email[]{giulia.celora@maths.ox.ac.uk}
\affiliation{University College London, Department of Mathematics, 25 Gordon Street, London, WC1H 0AY, UK}
\affiliation{Mathematical Institute, University of Oxford, Andrew Wiles Building, Woodstock Road, Oxford OX2 6GG, UK}

\author{Ralf Blossey}
\email[]{ralf.blossey@univ-lille.fr}
\affiliation{University of Lille, Unité de Glycobiologie Structurale et Fonctionnelle (UGSF), CNRS UMR8576, F-59000 Lille, France}

\author{Andreas M\"unch}
\email[]{muench@maths.ox.ac.uk}
\affiliation{Mathematical Institute, University of Oxford, Andrew Wiles Building, Woodstock Road, Oxford OX2 6GG, UK}

\author{Barbara Wagner}
\email[]{barbara.wagner@wias-berlin.de}
\affiliation{Weierstrass Institute, Mohrenstr. 39, 10117 Berlin, Germany}

\date{\today}

\maketitle

\tableofcontents

\section{Non-dimensional form of the modelling equations}
\label{SM:non-dimensional}
We here present the procedure used to non-dimensionalise the model presented in Section II of the main text specialised to a 1D geometry, $\Omega=[0,L]$. To do so, we introduce the following dimensionless variables
\begin{widetext}
\begin{equation}
\begin{aligned}
	\phi_{m}=\nu_m c_{m},\quad \hat{j}_{m}=j_{m} \frac{L_\psi\nu_m}{D},\quad  \tau= \frac{D}{L_\psi^2} t,\quad \hat{x}= \frac{x}{L_\psi}, \quad\hat{f}_{\eff}=\frac{\bar{f}_{\eff}\nu}{k_BT},
    \quad \hat{\mu}^{\text{el}}_\beta=\frac{\mu^{\text{el}}_\beta\nu}{k_BT\nu_\beta}, \quad \hat{P}=\frac{(P-\bar{P})\nu}{k_BT},\\\hat{u}^1_{\CR,s}=\frac{u^1_{\CR,s}-u^0_{\CR,s}}{k_BT},\quad \hat{u}^q_{\CR,p}=\frac{u^q_{\CR,p}}{k_BT}-\hat{u}^1_{\CR,s},\quad    \hat{\mu}^{\text{el}}_H=\frac{\mu_H}{k_BT}-\hat{u}^1_{\CR,s}, \quad \hat{\lambda}_{\CR}=\lambda_{\CR}+\frac{e\psi}{k_BT}+\hat{u}^1_{\CR,s},\\
\hat{\psi}=\frac{\psi e}{k_BT}, \quad \hat{\rho}=\rho\nu, \quad \hat{q}_p=\frac{q_p}{N},	\quad L_\psi=\sqrt{\frac{\epsilon \nu k_BT}{e^2}},\end{aligned}\label{eq_supp:scaling}
\end{equation}
\end{widetext}
and reformulate the modelling equations on the interval $\hat{x}\in(0,\ell_\Omega)$
\begin{subequations}\label{eq:non-dimensional-evolution-equations}
\begin{align}
\partial_\tau\phi_m -\partial_{\hat{x}}\cdot\left(\sum_{K}\hat{\mathcal{M}}_{mK}\partial_{\hat{x}}\hat{\mu}^{\text{el}}_K\right)=0,&\quad m\in\mathcal{I}_H\\
-\partial^2_{\hat{x}}\hat{\psi}=\phi_H+\phi_+-\phi_-,&\label{eq:non-dimensional poisson}
\end{align}
\end{subequations}
with the rescaled 
mobility matrix
\begin{subequations}\label{eq:rescaled_chem_potentials}

\begin{align}
    \mathcal{M}=\begin{bmatrix} \phi_p &0 & \hat{q}_p\phi_p &0 &0\\[4pt]
    0 &\phi_s& q_s\phi_s &0 &0\\[2pt]
     \hat{q}_p\phi_p  & q_s\phi_s & \langle q^2\rangle_s \phi_s +  \frac{\langle q^2 \rangle_p}{N^2}\phi_p &0 &0\\[4pt]
     0 & 0 &0 & \phi_+ &0\\[2pt]
    0 & 0 &0 & 0 &\phi_-\\
\end{bmatrix}
\end{align}
and the rescaled electrochemical potentials
\begin{align}
\hat{\mu}^{\text{el}}_{\pm}&=\hat{P}\pm\hat{\psi}+\frac{\partial\hat{f}_{\text{eff}}}{\partial \phi_\pm},\\
\hat{\mu}^{\text{el}}_s&=\hat{P}+q_s \hat{\psi}+\frac{\partial \hat{f}_{\eff}}{\partial \phi_s},\\
\hat{\mu}^{\text{el}}_{p}&=\hat{P}+\hat{q}_p\hat{\psi}+\frac{\partial\hat{f}_{\text{eff}}}{\partial \phi_p}-\ell_\kappa^2\partial_{\hat{x}}^2\phi_p,\\
\hat{\mu}^{\text{el}}_H&=-\hat{\lambda}_{\CR}+\hat{\psi},\label{eq:muHdef}
\end{align}
where the rescaled thermodynamic pressure and free energy are respectively
    \begin{align}
        \hat{P}&=\hat{f}_{\eff} - \sum_{i\in\mathcal{I}}\phi_i\frac{\partial \hat{f}_{\eff}}{\partial \phi_i}+\frac{1}{2}\left(\partial_{\hat{x}}\hat{\psi}\right)^2+\ell_\kappa^2\left(\phi_p\partial_{\hat{x}}^2\phi_p-\frac{(\partial_{\hat{x}}\phi_p)^2}{2}\right),\\
        \hat{f}_{\eff}&=\phi_+\ln\phi_++\phi_-\ln\phi_-+\frac{\phi_p}{N}\left(\ln\left(\frac{\phi_p}{\mathcal{Z}_p(\hat{\lambda}_H)}\right)+\hat{q}_p\hat{\psi}\right)-\phi_s\left(\ln\left(\frac{\phi_s}{\mathcal{Z}_s(\hat{\lambda}_H)}\right)+q_s\hat{\psi}\right)-\phi_H\hat{\lambda}_{H}.
    \end{align}
The local mean charge on the polymer and solvent phases are
\end{subequations}
\begin{subequations}
    \begin{align}
\hat{q}_{p}&=-\frac{1}{N}\frac{d}{d\hat{\lambda}_H}\ln\mathcal{Z}_{p}(\hat{\lambda}_H),\quad q_s=-\frac{d}{d\hat{\lambda}_H}\ln\mathcal{Z}_{s}(\hat{\lambda}_H)\\
\mathcal{Z}_s&=\sum\limits_{\zeta=0}^{1}\exp\left(-\hat{\lambda}_{\CR} \zeta \right),\quad
\mathcal{Z}_p=\sum\limits_{\zeta=0}^{Q_p}\exp\left(-(\hat{u}_{\CR,p}^\zeta+\hat{\lambda}_{\CR} \zeta) \right),
\end{align}
with the Lagrange multiplier $\hat{\lambda}_{\CR}$ being defined by the constraint
\begin{align}
\phi_H-\hat{q}_p\phi_p-q_s\phi_s=0.
\end{align}
\end{subequations}
The problem is then closed by imposing boundary conditions
\begin{subequations}\label{eq:non-dimensional-BC-and-BI-conditions}
\begin{align}
    \partial_{\hat{x}}\hat{\psi}(0,t)=\partial_{\hat{x}}\hat{\psi}(\ell_\Omega,t)&=0,\\\hat{j}_{m}(0,t)=\hat{j}_{m}(\ell_\Omega,t)&=0,\quad m\in \mathcal{I}_H,\\
    \int_0^{\ell_\Omega}\hat{\psi}(x,t) \,dx&=0,
    \end{align}
and appropriate initial conditions    
    \begin{align}
    \phi_{m}(x,0)=\phi^0_{m}(x), \quad x\in[0,\ell_\Omega],\quad m\in\mathcal{I}_H,
    \end{align}
\end{subequations}
where the initial data must be $\phi_i^0(x)\in (0,1)$ and satisfy global electroneutrality, \[\int_0^{\ell_\Omega} (\phi^0_H+\phi^0_+-\phi_-^0)\,dx=0,\] and the local no-void conditions \[\phi^0_s(x)=1-\phi^0_p(x)-\phi^0_+(x)-\phi^0_-(x).\]
The rescaled version of the model highlights the presence of a key non-dimensional parameter
\[\ell_\kappa=\frac{1}{L_\psi}\sqrt{\frac{\kappa}{k_BT\nu N^2}}.\] 
which is a measure of the competition between the short- (\emph{i.e.}, interfacial) and
long-range (\emph{i.e.}, Coulombic) interactions. Note that the choice of scaling the spatial dimension by the Debye-length $L_\psi$ follows the literature on dilute electrolytes~\cite{app_gavish2017PRE}. However, the screening length for weak polyelectrolytes differs as investigated in~\Cref{SM:stability}.
\section{Dynamical simulations}\label{app:numerics}
We solve the non-dimensional form of the modelling equations, Eqs.~(\ref{eq:non-dimensional-evolution-equations})-(\ref{eq:non-dimensional-BC-and-BI-conditions}) using the method of lines. We first discretise Eqs.~(\ref{eq:non-dimensional-evolution-equations}) in space using a finite volume scheme over a stencil of $N_x=500$ points to obtain a system of $4N_x$ ODEs, which are coupled to a system of $N_x$ algebraic equations for $\lambda_{\CR}$. The differential-algebraic system of equations is solved over time using \texttt{DifferentialEquations.jl} package \cite{app_rackauckas2017differentialequations} in \texttt{Julia}. The values of the model parameters used to generate the Figures are listed in~\Cref{tab:table_par}.

For the charge-regulation dynamics, we use the standard linear-quadratic form for the charge regulation energy as in \cite{app_celora2023,app_Avni2020},
\begin{align}
   \hat{u}^q_{\CR,p}= \alpha q +\frac{\eta}{2 Q_p} q^2 -\ln \left[\binom{Q_p}{q}\right],\label{eq:f_CR}
\end{align}
where we take $\alpha<0$ and large ($|\alpha|\gg 1$), so that it is energetically favourable for a proton to bind to the polymer, and set $\eta \geq 0$ to capture the repulsive short-range attraction between charges bounded to the polymer. As shown in~\cite{app_celora2023}, the equilibrium behaviour of the system is qualitatively independent of the precise value of $\eta$ as long as $\eta\geq0$. 

\begingroup
    \begin{table*}[htb]
        \caption{Default value(s) of the non-dimensional model parameters used in the dynamical simulations of Eqs.~(\ref{eq:non-dimensional-evolution-equations})-(\ref{eq:non-dimensional-BC-and-BI-conditions}) and scaling factors to covert variables to dimensional units.}
    \label{tab:table_par}
    \centering
    \begin{ruledtabular}
    \begin{tabular}{c p{125mm} c}
         Parameter & Description &Value  \\
         \hline\addlinespace[2pt]
         $\ell_\kappa$ & ratio of the interfacial and Debye length $L_\psi$ &$10$\\
         $\ell_\Omega$ & Domain size relative to the Debye length $L_\psi$& $500$\\
         $\alpha$ & binding energy of protons to ionizable residues relative to thermal energy  &-7.0\\
         $\eta$ & interaction energy of protons bound to the same polymer chain relative to thermal energy  &3.\\
         $\chi$ & Flory-Huggins constant & 1.5\\
         $Z$& number of ionizable monomers for polymer chain &4\\[2pt]
         $N$& number of monomers for polymer~\cite{app_luo2024condensatesizecontrolcharge}&10\\[2pt]
         $\phi_P^0$ & spatially averaged polymer volume fraction~\cite{app_luo2024condensatesizecontrolcharge} &0.1\\[2pt]
         $\phi_+^0$ & spatially averaged salt volume fraction& 0.001\\
          $\phi_H^0$ & spatially averaged protonised molecules volume fraction&0.0001--0.4\\[2pt]
           $L_\psi$ & Debye length scale of water at room temperature&$\approx0.1-10$ nm\\[2pt]
         $T=L^2_\psi \ell^2_\Omega D^{-1}$ & Characteristic temporal scale, taken to be the timescale associated with diffusion & $\approx10^{-4}-10^{-2}$ [s]\\
         $D$ & Diffusion coefficient of ions/solvent & $10^{-9}$ [m\textsuperscript{2}/s]\\
         $(\nu N_A)^{-1}$& conversion from volume fraction to moles &$55.35$ [M]\\
          $\nu$ &molecular volume of water& $0.03$ nm\textsuperscript{3}\\
         $N_A$ &Avogadro Number& $6.022\times 10^{23}$ [mol\textsuperscript{-1}]\\
        \end{tabular}
    \end{ruledtabular}
\end{table*}
\endgroup
\section{Stability analysis}
\label{SM:stability}
We here present the derivation of the dispersion relation for the model used in generating Figures 4-5 in the main text.
\subsection{Dispersion relation}
While the conservative formulation of the model in~\Cref{SM:non-dimensional} is the most appropriate for dynamical simulations, it leads to unnecessary complications in the derivation of the dispersion relation. To simplify this analysis, we rewrite the model by replacing the mass conservation equation for $\phi_H$ with a dynamic equation for $q_s$. In other words, we now consider the fraction of charges on the solvent molecules as an unknown variable, while $\phi_H$ is computed a posteriori. Note that this transformation is possible despite $q_s$ being a non-linear function of $\phi_H$. As mentioned in the main text, the mapping between the two variables is implicitly defined by 
\begin{equation}
\mathcal{C}(q_s)=q_s\phi_s-\phi_H+\hat{q}_p(q_s)\phi_p=0, \quad q_s,\phi_s,\phi_H,\phi_p\in (0,1)^4.\label{eq:implicit_definition_rho_s}
\end{equation}
When taking the derivative of the function $\mathcal{C}(q_s)$ with respect to its argument $q_s$, the latter is positive for any value of $q_s\in(0,1)$. Invoking the Implicit Function Theorem, we therefore have that $q_s(\phi_H):(0,1)\rightarrow(0,1)$ is invertible. Therefore, we can equivalently define our system in terms of the conserved variables $(\phi_p,\phi_+,\phi_-,\phi_s,\phi_H)$ or the transformed set of variables $(
\phi_p,\phi_+,\phi_-,\phi_s,q_s)$. This choice of variables helps with the algebraic derivation of the dispersion relation because $\hat{q}_p$ can be expressed directly as a function of $q_s$ alone. By applying this change of variables, the problem can be rewritten in terms of the psedu-free energy of the system, which then reads:
\begin{equation}
\begin{aligned}
\bar{g}(\phi_p,\phi_+,\phi_-,\phi_s) = \phi_+\ln\phi_+ + \phi_-\ln\phi_- + \phi_s\ln\phi_s +\frac{\phi_p}{N}\ln\phi_p +\chi \phi_s\phi_p,
\end{aligned}
\end{equation}
which is equivalent to the mixing free energy of a 4-component mixture. We note that here the introduction of the function $\bar{g}$ is purely for the purpose of easing the analysis. The transport equations for the different components then take the standard form:
\begin{subequations}
    \begin{align}
    \partial_t \phi_m=\nabla\cdot \left(\phi_m\left(\nabla\left(\frac{\partial \bar{g}}{\partial \phi_m}\right) +\ell_\kappa^2\left(\phi_p-\delta_{mp}\right)\nabla^3\phi_p+(q_m-\varrho)\nabla\psi\right)\right),
\end{align}
and the electric field $\psi$ satisfies 
\begin{equation}
    -\nabla^2\psi=q_s\phi_s+q_p(q_s)\phi_p+\phi_+-\phi_-.
\end{equation}
Note that for simplicity, we have dropped the hat notation from $q_p$ to indicate the rescaled polymer charge $\hat{q}_p$ (see Eq.~(\ref{eq_supp:scaling}).
When comparing to the previous formulation, here the cross-diffusion term with $\phi_H$ in the equations for $\phi_p$ and $\phi_s$ vanishes being substituted by an electric current due to the average charge on each of the two charged species ($q_{p,s}$). Importantly, we now require an explicit equation for the evolution of $q_s$, which can be derived using Eq.~(\ref{eq:implicit_definition_rho_s})
\begin{align}
\begin{aligned}
\partial_t q_s &= \frac{\partial_t \phi_H -q_p(q_s)\partial_t\phi_p-q_s\partial_t\phi_s}{\phi_s+\phi_pq_p'(q_s)}\\
&=\frac{\sigma_s}{\phi_s\sigma_s+\phi_p\sigma_pN}\nabla\cdot\left((\phi_s\sigma_s+\phi_p\sigma_p)\nabla\left(\ln\frac{q_s}{1-q_s} +\psi\right)\right)-\frac{\sigma_s\vec{j}_s+\sigma_pN\vec{j}_p}{\phi_s\sigma_s+\phi_p\sigma_pN}\cdot \nabla q_s.
\end{aligned}\label{eq:time_dependent_rho_s}
\end{align}\label{sys:eq_with_rho_s}%
\end{subequations}
where $\sigma_s=q_s(1-q_s) \in (0,1)$ and $\sigma_p=\langle q^2\rangle_p/N^2-q_p^2$ can be interpreted as the variance of the charge distribution on the solvent and polymer molecules, respectively. Note again that here $\sigma_p$ is normalised by the total number of monomers on a polymer chain $N$. When considering the linear stability analysis, the function $q_s$ can be expressed as
\begin{equation}
q_s=q_s^0+\delta \sum_{r\in\mathbb{N}} U^r_\varrho \cos(r\pi x)e^{\omega t},  
\end{equation}
where $U^r_\varrho$ is connected to the amplitude of the perturbations in the original variables by
\[U^r_\varrho=\sigma_s^0\frac{U_H^r-q_p^0U_p^r-q_s^0U^r_s}{\sigma_s^0\phi_s^0+\phi_p^0\sigma_p^0N}.\]

\noindent Using the above ansatz yields a linear 6-dimensional algebraic system for $\vec{U}^r=(U^r_p,U^r_+,U^r_-,U^r_s,U^r_\varrho,U^r_\psi)$, dependent on the wavenumber $r$ and the corresponding growth rate $\omega$. We can use the no-void condition and the Poisson equation to explicitly solve for $U_s^r$ and $U_s^r$
\begin{align}
U^r_s&=-U^r_p-U^r_+-U^r_-,\\
(r\pi)^2 U_\psi^r&=(q^0_p-q^0_s) U_p^r  +(1-q_s^0)U_+^r-(1+q_s^0)U_-^r+\frac{\phi_p^0\sigma^0_pN+\sigma^0_s\phi^0_s}{\sigma_s^0} U_\varrho=\sum_{j}q^*_j U^r_j,\label{eq:def_U_psi}
\end{align}
and reduce the problem to a linear 4-dimensional algebraic system that can be written as
\begin{subequations}\label{sys:linear_stability_analysis_reduced}
	\begin{align}
 \omega\left[ \left(\frac{1}{\phi^0_p}+\frac{1}{\phi^0_s}\right)U_p^r+ \frac{U_+^r}{\phi^0_s}+\frac{U_-^r}{\phi^0_s}\right]&=-(r\pi)^2\sum_{i} G^0_{pi} U^r_i-q^*_p\sum_jq^*_jU^r_j-\ell_\kappa^2(r\pi)^4U_p^r,\\
  \omega \left[\frac{1}{\phi^0_s}U_p^r+  \left(\frac{1}{\phi^0_+}+\frac{1}{\phi^0_s}\right)U_+^r+\frac{U_-^r}{\phi^0_s}\right]&=-(r\pi)^2\sum_{i} G^0_{+i} U^r_i-q^*_+\sum_jq^*_j U^r_j,\\
  \omega \left[\frac{1}{\phi^0_s}U_p^r+\frac{U_+^r}{\phi^0_s}+ \left(\frac{1}{\phi^0_-}+\frac{1}{\phi^0_s}\right)U_-^r\right]&=-(r\pi)^2\sum_{i}G^0_{-i} U^r_i-q^*_-\sum_jq^*_j U^r_j,\\
\omega \frac{(q^*_\varrho)^2}{\phi_p^0\sigma^0_p+\sigma^0_s\phi_s^0}U^r_\varrho&=-(r\pi)^2\frac{q^*_\varrho}{\sigma_s^0}U^r_\varrho-q^*_\varrho\sum_jq^*_jU^r_j.
\end{align}
where the index $i$ only sums over $+,-,p$, while $j$ over $+,-,p,\varrho$, and $G_{ki}$ indicate the second order partial derivatives of the function $G$ 
\[G(\phi_p,\phi_+,\phi_-)=
\bar{g}(\phi_p,\phi_+,\phi_-,1-\phi_p-\phi_+-\phi_-),\]
\end{subequations}
with respect to its arguments $\phi_k$ and $\phi_i$  
The system~(\ref{sys:linear_stability_analysis_reduced}) can be written in matrix form as a generalised eigenvalue problem
\[\tens{A}_r\vec{U}^r=\omega \tens{B}\vec{U}^r,\]
where the mass matrix 
\begin{equation}
    \tens{B}=\frac{1}{\phi^0_s}\begin{bmatrix}
        \frac{\phi^0_p+\phi^0_s}{\phi^0_p} & 1 & 1 &0\\
        1& \frac{\phi^0_++\phi^0_s}{\phi^0_+} & 1 &0\\
        1 & 1&\frac{\phi^0_-+\phi^0_s}{\phi^0_-} &0\\
        0 &0&0 &\frac{(q^*_\varrho)^2\phi_s^0}{\phi_p^0\sigma^0_p+\sigma^0_s\phi_s^0}
    \end{bmatrix}
\end{equation}
is symmetric and positive definite for all physical steady states, and the symmetric matrix $\tens{A}_r$ is given by
\begin{equation}
    \tens{A}_r=-\begin{bmatrix}
       (r\pi)^2 G^0_{pp}+\ell_\kappa^2 (r\pi)^4+(q^*_p)^2&  (r\pi)^2 G^0_{p+}+q^*_pq^*_+ &  (r\pi)^2 G^0_{p-}+q^*_pq^*_-  &q^*_pq^*_\varrho\\[4pt]
        \star& (r\pi)^2 G^0_{++}+(q^*_+)^2&(r\pi)^2 G^0_{+-}+q^*_+q^*_-  &q^*_+q^*_\varrho\\[4pt]
        \star&\star&(r\pi)^2 G^0_{--}+(q^*_-)^2 &q^*_-q^*_\varrho\\[4pt]
        \star&\star&\star&(r\pi)^2 \frac{q^*_\varrho}{\sigma^0_s}+(q^*_\varrho)^2
    \end{bmatrix}
\end{equation}
where the $\star$ symbols are used to avoid rewriting the entries already defined by the symmetry of the matrix $\tens{A}_r$.
\subsection{Computation of the spinodal}
To compute the boundary of the stability region of homogeneous solutions, we compute the values of the wavenumber $r$ for which $\det(\tens{A}_r)=0$, which is equivalent to finding conditions for which it exists a non-trivial solution to the following 4D homogenous algebraic system
\begin{subequations}
	\begin{align}
 0&=-(r\pi)^2\sum_{i} G^0_{pi} U^r_i-q^*_p\sum_jq^*_jU^r_j-\ell_\kappa^2(r\pi)^4U_p^r,\\
 0&=-(r\pi)^2\sum_{i} G^0_{\pm i} U^r_i-q^*_\pm\sum_jq^*_j U^r_j,\\
0&=-(r\pi)^2\frac{q^*_\varrho}{\sigma_s^0}U^r_\varrho-q^*_\varrho\sum_jq^*_jU^r_j.\label{eq:LSAU_rho}
\end{align}
\end{subequations}
Using Eqs.~(\ref{eq:LSAU_rho}) and~(\ref{eq:def_U_psi}), we find that there is a simple proportional relationship between the solvent mean charge and the electric field fluctuations; namely, $U^r_\varrho=\sigma_s^0 U^r_\psi$. We can, therefore, rewrite the above system in the form
\begin{subequations}\label{sys:LSA_v2}
	\begin{align}
 0&=-(r\pi)^2\left[\sum_{i} G^0_{pi} U^r_i+q^*_pU_\psi+ (\ell_\kappa r\pi)^2U_p^r\right],\label{eq:LSAU_p_v2}\\
 0&=-(r\pi)^2\left[\sum_{i} G^0_{\pm i} U^r_i+q^*_\pm U_\psi^r\right],\label{eq:LSAU_pm}\\
0&=-( r\pi)^2U^r_\psi+q^*_p U_p^r+q^*_+U_+^r+q^*_-U_-^r+q^*_\varrho\sigma_s^0U_\psi.\label{eq:LSAU_psi_v2}
\end{align}
\end{subequations}
Since the system is conservative, we only consider perturbations with positive wavenumbers $r>0$. Therefore, Eq.~(\ref{eq:LSAU_pm}) can be used to directly express $U_\pm^r$ in terms of $U_\psi$ and $U_p$ independently of the wavenumber $r$:
\[U_\pm=a_{\pm p}U_p+a_{\pm\psi}U_\psi,\]
where 
\begin{subequations}
    \begin{align}
        a^0_{+p}=\frac{-G^0_{--}G^0_{p+}+G^0_{+-}G^0_{p-}}{G^0_{++}G^0_{--}-(G^0_{+-})^2}, \quad a^0_{+\psi}=\frac{-G^0_{--}\hat{q}_++G^0_{+-}\hat{q}_-}{G^0_{++}G^0_{--}-(G^0_{+-})^2},\\
        a^0_{-p}=\frac{-G^0_{++}G^0_{p-}+G^0_{+-}G^0_{p+}}{G^0_{++}G^0_{--}-(G^0_{+-})^2}, \quad a^0_{-\psi}=\frac{-G^0_{++}\hat{q}_-+G^0_{+-}\hat{q}_+}{G^0_{++}G^0_{--}-(G^0_{+-})^2}.
     \end{align}
\end{subequations}
Substituting the expression of $U_\pm$ into Eq.~(\ref{eq:LSAU_p_v2}) and Eq.~(\ref{eq:LSAU_psi_v2}), we obtain the following 2D algebraic system
\begin{equation}\label{sys:LSA_v3}
\begin{bmatrix}
    0\\0
\end{bmatrix}=\underbrace{\begin{bmatrix}
    G^0_{pp}+a_{p+}G_{p+}^0+a_{p-}G_{p-}^0+(\ell_{\kappa}r\pi)^2 &\hat{q}_p+a^0_{+\psi}G_{p+}+a^0_{-\psi}G_{p-}\\
    \hat{q}_p +a_{p+}\hat{q}_++a_{p-}\hat{q}_-&\hat{q}_\varrho\sigma_s^0+a_{\psi+}\hat{q}_++a_{\psi-}\hat{q}_--( r\pi)^2
\end{bmatrix}}_{=\tens{G}_r}\begin{bmatrix}
    U_p\\U_\psi
\end{bmatrix}.
\end{equation}
The above system has non-trivial solution if $\det(\tens{G}_r)=0$, which leads to a forth-order polynomial in $r\pi$
\begin{subequations}\label{has nol_eigenvalue_r}
\begin{align}
\ell^2_\kappa(r\pi)^4-2\left(\frac{\ell_\kappa^2\tens{G}^{22}_{0}-\tens{G}^{11}_0}{2}\right)(r\pi)^2-\det(\tens{G}_0)=0.\label{eq_app:polynomial_r}
\end{align}
We note that the entries of the matrix $\tens{G}_0$ are equivalent to those of the Hessian matrix of the free energy $\tilde{F}$ associated with stationary problem describing non-linear localised solutions, see Eqs.~(\ref{eqs:equilibria_solution_salt}) in the main text. This is not surprising since solving for the zero-th growth eigenmodes corresponds to solving the linearised stationary problem. The entries of the matrix $\tens{G}_0$ can then can be easily computed as
\begin{align}
\tens{G}^{22}_0=\tilde{F}^0_{\psi\psi}&= \frac{(q^0_p\phi_p^0)^2}{(1-\phi_p^0)}-\phi_+^0-\phi_-^0-q^0_s\phi_{s}^0-\phi_p^0N\sigma^0_z,\\
    \tens{G}^{11}_0=\tilde{F}^0_{\phi\phi}&=\frac{1}{N\phi_p^0}+\frac{(1-\chi\phi_s^0)^2}{1-\phi_p^0}-\chi^2\phi_s^0,\\
    \tens{G}^{12}_0=\tens{G}^{21}_0=\tilde{F}^0_{\psi\phi}&=\frac{q^0_p}{1-\phi_p^0}(1-\chi\phi_p^0\phi_s^0)-\chi q_s^0\phi_{s}^0.
\end{align}\label{eq_app:coeff_free_energy_hessian}%
\end{subequations}

We discussed the number and explicit form of the roots to Eq.~(\ref{has nol_eigenvalue_r}) in Appendix B. Here, we focus instead on the connections between the roots to Eq.~(\ref{has nol_eigenvalue_r}) and the temporal stability of homogeneous solutions. By continuity of the growth rate $\omega$ with respect to the model parameters, we know that changes in the temporal stability of a given mode $r$ will imply $r$ being a real root of Eq.~(\ref{has nol_eigenvalue_r}). If Eq.~(\ref{has nol_eigenvalue_r}) has no solution, then we expect all wavenumbers $r$ to have the same stability properties. Based on the explicit computation of the dispersion relation , we know that the interval of unstable modes is bounded, which implies that if Eq.~(\ref{has nol_eigenvalue_r}) has no solution then the homogenous solutions is stable to all positive wavenumber $r$, \emph{i.e.}, the growth rates $\omega$ associated with the eigenmode $r$ are negative. 
When Eq.~(\ref{has nol_eigenvalue_r}) has only two real roots, by symmetry, these are of the form $r=\pm r_{1}$. Following the same reasoning as above, we therefore conclude that all the homogeneous state is temporally unstable to modes in the interval $r\in(0,r_{1})$. Finally, if Eq.~(\ref{eq_app:polynomial_r}) has four real eigenvalues, then these are of the form $r=\pm r_{1,2}$ with $r_1>r_2$, and we expect the homogeneous solution to be unstable to perturbation of wavenumber in the interval $r\in(r_2,r_1)$. More details on the nature of transition in the temporal stability of homogeneous states are given in Appendix B.

\section{Spatial Dynamics.} 
\label{SM:spatial dynamics}
In this section we detail the derivation of the system~(\ref{eq:equilibria_spatial_dynamics}) that defines steady solution of Eqs.~(\ref{eq:non-dimensional-evolution-equations})-(\ref{eq:non-dimensional-BC-and-BI-conditions}). For simplicity, we denote stationary solutions of the problem with capital letters, \emph{i.e.}, $\Phi_m(x)$ and $\Psi(x)$ correspond, respectively, to the stationary spatial profile of the volume fraction for the component $m$ of the mixture and the rescaled electric potential. To find the stationary solution, we set all time derivatives in Eq.~(\ref{sys:eq_with_rho_s}) to zero. Because of the no flux boundary condition and the 1D geometry, this is equivalent to setting the fluxes of each component of the mixture to zero. 
In line with the logarithmic form of the entropy of mixing, we here consider solutions with support on the whole domain $[0,L_\Omega]$, \emph{i.e.}, $\Phi_m,q_s\in(0,1)$ for all $x\in[0,L_\Omega]$. Under these assumptions, at steady state, Eq.~(\ref{eq:time_dependent_rho_s}) enables to solve explicitly for the charge density $q_s$
\[q_s=\frac{e^{\mu^\Delta_H-\Psi}}{1+e^{\mu^\Delta_H-\Psi}}=-\frac{\partial}{\partial \Psi}\left[\ln\left(1+e^{\mu^\Delta_H-\Psi}\right)\right]\]
where $\mu^0_H$ is an unknown constant characterising the equilibrium chemical potential of protons $\mu_H$ (see Eq.~(\ref{eq:muHdef})). Since $q_s$ is a unique function of the electric field, the same holds for the mean charge on the polymer $q_p$
\[\mathcal{Z}_{\beta,Q}(\Psi)= \sum_{q=0}^{Q_p} e^{-\hat{u}^q_{\CR,p}-q(\Psi-\mu^\Delta_H)}, \quad q_p=-\frac{1}{N}\frac{\partial \ln \mathcal{Z}_{\beta,Q}(\Psi) }{\partial \Psi},\]
where we have used Eq.~(\ref{eq:moments charge distribution}) to write the moments of the mean charge $q_p$ in terms of the partition function associated with the polymer charge distribution. We can further solve for the volume fractions of the salt ions $\Phi_\pm$ by considering a linear combination of zero flux conditions, namely $j_\pm\Phi_s-\Phi_\pm j_s=0$ and coupling these with the no-void condition, we obtain
\begin{equation}
    \begin{cases}
        \Phi_+(\Phi_p,\Psi)=\gamma_s(\Phi_p,\Psi)\,e^{-\Psi+\mu^\Delta_++\chi\Phi_P}(1-\Phi_p),\\
        \Phi_-(\Phi_p,\Psi)=\gamma_s(\Phi_p,\Psi)\, e^{\Psi+\mu^\Delta_-+\chi\Phi_P}(1-\Phi_p),\\
        \Phi_s(\Phi_p,\Psi)=\gamma_s(\Phi_p,\Psi)\left(1+e^{-\Psi+\mu^\Delta_H}\right)(1-\Phi_p),
    \end{cases}\label{eqs:equilibria_solution_salt}
    \end{equation}
    where the constant $\mu^\Delta_\pm$ are again unknown constants which specify the difference in chemical potential between each salt ion and the solvent, and $\gamma_s$ is defined as
    \begin{equation}
         \gamma_s(\phi,\psi)=\dfrac{1}{1+e^{-\psi+\mu^\Delta_++\chi\phi}+e^{-\psi+\mu^\Delta_H}+e^{\psi+\mu^\Delta_-+\chi\phi}}.
    \end{equation}
    Substituting Eq.~(\ref{eqs:equilibria_solution_salt}) into Eq.~(\ref{eq:non-dimensional poisson}), we find that the steady spatial profile for the electric potential $\Psi$ is defined by
\begin{equation}
    -\partial_{xx} \Psi= \frac{1-\Phi_p}{\gamma_s(\Phi_p,\Psi)}\frac{\partial \gamma_s(\Phi_p,\Psi)}{\partial \Psi}-\frac{\Phi_p}{N}\frac{\partial \ln\mathcal{Z}_{\beta,Q}(\Psi)}{\partial \Psi},\label{eq:Psi_eq_I}
\end{equation}
Finally, the equation defining the steady spatial profiles for the polymer volume fraction $\Phi_p$ is obtained by considering the linear combination of the no-flux conditions $j_p\Phi_s-\Phi_pj_s=0$, which leads to the following second order differential equation
\begin{equation}
\begin{aligned}
\ell_\kappa^2\partial_{xx}\Phi_p &= \frac{1}{N}\left[\ln\left(\frac{\Phi_p}{\mathcal{Z}_{\beta,Q}}\right)+1\right]-\left[\ln(\gamma_s(1-\Phi_p))+1\right]+\chi\left((1-\Phi_p)\gamma_s(1+e^{\mu^\Delta_H-\Psi})-\Phi_p\right)+\mu^\Delta_p\\
&=\frac{\partial}{\partial \Phi_p} \left[(1-\Phi_p)\ln(\gamma_s(\Phi_p,\Psi)(1-\Phi_p))+\frac{\Phi_p}{N}\ln\left(\frac{\Phi_p}{\mathcal{Z}_{\beta,Q}(\Psi)}\right)+\chi\Phi_p(1-\Phi_p)+\mu^\Delta_p\Phi_p\right],
\end{aligned}\label{eq:Phip_eq_I}
\end{equation}
where $\mu_p^\Delta$ is an additional unknown constant that characterises the difference in the polymer and solvent chemical potentials at steady state. To obtain Eqs.~(\ref{eq:equilibria_spatial_dynamics}) in the main text, we notice that the right-hand sides of Eqs.~(\ref{eq:Psi_eq_I})-(\ref{eq:Phip_eq_I}) can be written as the partial derivative of the rescaled free-energy:
\begin{equation}
\begin{aligned}
\tilde{F}
&=\hat{f}_{\eff}+e\psi(q_s\Phi_s+q_p(q_s)\Phi_p-\Phi_-+\Phi_+)+\sum_{m\in\mathcal{I}_H\setminus\left\{s\right\}}\mu^\Delta_m\Phi_m\\&=(1-\Phi_p)\ln(\gamma_s\,(1-\Phi_p))+\frac{\Phi_p}{N}\ln\left( \gamma_p\,\Phi_p\right)+\chi\Phi_p(1-\Phi_p),
\end{aligned}
\end{equation}
where the activity coefficients $\gamma_s$ and $\gamma_p$ take the form:
\begin{align}
    \gamma_s&=\dfrac{1}{1+e^{-\psi+\mu^\Delta_++\chi\phi}+e^{-\psi+\mu^\Delta_H}+e^{\psi+\mu^\Delta_-+\chi\phi}},\\
    \gamma_p&=\frac{e^{N\mu^\Delta_p}}{\mathcal{Z}_{\beta,Q}(\psi)}=\dfrac{e^{N\mu^\Delta_p}}{\sum\limits_{q=0}^{Q_p} e^{-\beta u_{\CR}^q-q (\psi-\mu^\Delta_H)}}.\label{eq_app:gamma_p}
\end{align}
To uniquely identify the stationary solutions, it remains to specify the chemical potentials $\mu^\Delta_m$ for $m\in\mathcal{I}_H\setminus\left\{s\right\}$. These four degrees of freedom are associated with mass conservations of ions, protons and polymers. One of this degree of freedom is constrained by the overall electroneutrality of the mixture
\[\int_0^{\ell_\Omega} (q_s\Phi_s+q_p\Phi_p+\Phi_+-\Phi_-)\,dx=0.\]
We constrain the remaining three degrees of freedom by prescribing the spatially-averaged volume fraction of polymers, protons and positive salt ions:
\begin{subequations}
    \begin{align}
        \phi_p^0&=\frac{1}{\ell_\Omega}\int_0^{\ell_\Omega}\Phi_p(x) \, dx,\\
        \phi_H^0&=\frac{1}{\ell_\Omega}\int_0^{\ell_\Omega}q_p(x)\Phi_p(x) +q_s(x)\Phi_s(x)\, dx,\\
         \phi_+^0&=\frac{1}{\ell_\Omega}\int_0^{\ell_\Omega}\Phi_+(x)\, dx,
    \end{align}
\end{subequations}
where $\phi_p^0$, $\phi_+^0$ and $\phi_H^0$ are arbitrary constant that parametrise the stationary solutions. 

A similar formulation holds for the case of a polymer with fixed charge $Q_p$, except for the form of the $\gamma_p$ term, which reduces to \[\gamma_p=\exp\left[N\mu^\Delta_p+Q_p\psi\right]\] and can take both values above and below one.

\section{Additional results: ensemble simulations}
\begin{figure}[htb]
\centering
\includegraphics[width=0.875\textwidth]{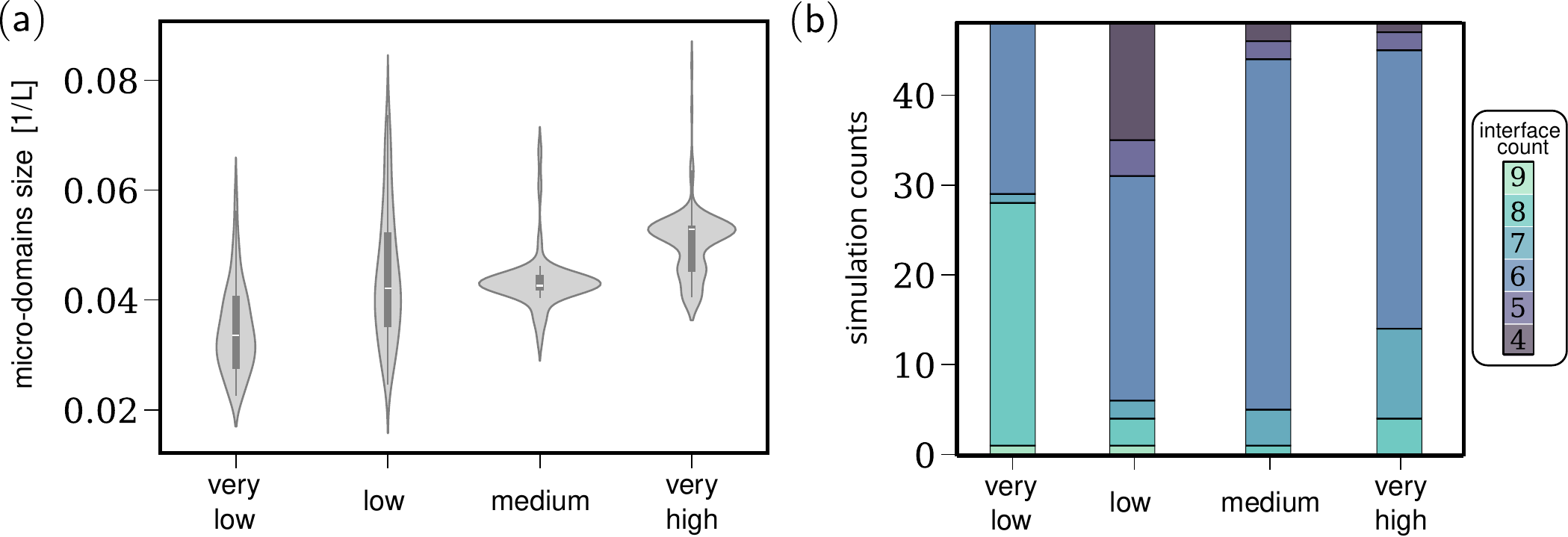}
\caption{Plots quantifying the level of variability in the pattern emerging from ensemble simulations of spinoidal decomposition for different values of the average proton concentration $c_H^0$. (a) Size distribution of the micro-domain size at simulation time $t=1000T$; (b) Cumulative barplot illustrating the distribution of interface number at simulation time $t=1000T$. The values of $c_H^0$ and all parameter are as in Figure 3 of the main text.}
\label{fig:ensemble_sim}
\end{figure}

Here, we present additional results from numerical simulations of phase separation for different values of the average proton concentration $c_H^0$. In Figure 3 of the main text, we show the evolution of the polymer concentration for one specific noisy initial condition. To study how stable the long-term pattern is, we simulate the model for an ensemble of 50 noisy initial conditions (for a fixed noise level of noise). To characterise the emergent pattern, we run the simulation up to a time $t=1000T$ and then estimate the distribution of two metrics across the simulation ensemble: the average size of condensed micro-domains, and the number of liquid interfaces in the domain, which, when halved, gives a proxy for the number of coacervates. Results are illustrated in~\Cref{fig:ensemble_sim}. We find that the variability in the number and size of condensed micro-domains is the largest in proton-dilute ($c_H^0$ very low) conditions, suggesting the pattern observed in the simulations is not robust to perturbations; in contrast, for medium levels of $c_H^0$, we find minimal variations both in the number and size of the micro-domains suggesting the pattern is more stable than in dilute conditions and the characteristic length scale of the micro-domains is characteristic of the problem. This is in line with the linear stability and steady state analyses presented in the main text. 

\nocite{*}

\bibliography{Supplementary}